\documentclass[letterpaper,twocolumn,10pt]{article}\usepackage[]{graphicx}\usepackage[]{color}
\makeatletter
\def\maxwidth{ %
  \ifdim\Gin@nat@width>\linewidth
    \linewidth
  \else
    \Gin@nat@width
  \fi
}
\makeatother

\definecolor{fgcolor}{rgb}{0.345, 0.345, 0.345}

\usepackage{framed}
\makeatletter
 {\par\unskip\endMakeFramed%
 \at@end@of@kframe}
\makeatother

\definecolor{shadecolor}{rgb}{.97, .97, .97}
\definecolor{messagecolor}{rgb}{0, 0, 0}
\definecolor{warningcolor}{rgb}{1, 0, 1}
\definecolor{errorcolor}{rgb}{1, 0, 0}

\usepackage{alltt}
\usepackage{usenix}

\usepackage[letterpaper]{geometry}
\usepackage{epsfig,endnotes,url}

\usepackage{xspace}

\usepackage[HideComments]{mycomment}

\usepackage[dvipsnames,table]{xcolor}

\usepackage[disable,colorinlistoftodos]{todonotes}   
\usepackage{amsmath, amsthm, amssymb, latexsym} 
\usepackage{ulem}
\usepackage{siunitx}

\usepackage{graphicx}

\usepackage{multirow}
\usepackage{booktabs}
\usepackage{longtable}
\usepackage{rotating}

\usepackage{wasysym}

\usepackage{siunitx}
\usepackage{pdfpages}

\usepackage{paralist}

\usepackage{balance}
\usepackage{adjustbox}

\usepackage{versions}
\excludeversion{DocumentVersionConference}
\excludeversion{AnonymizedAuthors}
\includeversion{NamedAuthors}
\includeversion{DocumentVersionTR}
\excludeversion{DocumentVersionLNCS}
\includeversion{QualitativeContent}
\includeversion{CaptionExplanations}
\excludeversion{RedundantContent}
\includeversion{Supplements}

\newcommand{\explain}[1]{\protect{\processifversion{CaptionExplanations}{\protect{#1}}}}

\usepackage[caption=false]{subfig}

\usepackage{hyperref}

\hypersetup{
bookmarks=true,
colorlinks=false,
pdftitle={Revisiting the Factor Structure of IUIPC-10}
}

\newcommand{\defterm}[2]{\hypertarget{#1}{\textit{#2}}}
\newcommand{\refterm}[2]{\hyperlink{#1}{#2}}

\newtheorem{researchquestion}{RQ}
\theoremstyle{definition} 
\newtheorem{remark}{Remark}  

\newcommand{\vari}[1]{\ensuremath{\mathit{#1}\xspace}}
\newcommand{\const}[1]{\ensuremath{\mathsf{#1}\xspace}}

\newcommand{\CASCAde}{ERC Starting Grant CASCAde (GA n\textsuperscript{o}716980)}
\IfFileExists{upquote.sty}{\usepackage{upquote}}{}
\begin{document}

\newcommand{\inputCorSDBV}{
\begin{table*}[p]
\centering\caption{Correlations and Standard Deviations of samples \textsf{B} and \textsf{V} used in the study (before outlier treatment)}
\label{tab:inputCorSDBV}
\captionsetup{position=top}
\subfloat[Sample \textsf{B}]{
\label{tab:inputCorSDBraw}
\centering
\begingroup\footnotesize
\begin{tabular}{rrrrrrrrrrr}
  \toprule
 & 1 & 2 & 3 & 4 & 5 & 6 & 7 & 8 & 9 & 10 \\ 
  \midrule
1. ctrl1 &  &  &  &  &  &  &  &  &  &  \\ 
  2. ctrl2 & 0.617 &  &  &  &  &  &  &  &  &  \\ 
  3. ctrl3 & 0.238 & 0.274 &  &  &  &  &  &  &  &  \\ 
  4. awa1 & 0.230 & 0.261 & 0.292 &  &  &  &  &  &  &  \\ 
  5. awa2 & 0.265 & 0.308 & 0.321 & 0.653 &  &  &  &  &  &  \\ 
  6. awa3 & 0.224 & 0.187 & 0.256 & 0.311 & 0.286 &  &  &  &  &  \\ 
  7. coll1 & 0.017 & 0.028 & 0.263 & 0.062 & 0.106 & 0.329 &  &  &  &  \\ 
  8. coll2 & 0.071 & 0.062 & 0.296 & 0.240 & 0.251 & 0.274 & 0.650 &  &  &  \\ 
  9. coll3 & 0.130 & 0.079 & 0.309 & 0.139 & 0.188 & 0.399 & 0.765 & 0.692 &  &  \\ 
  10. coll4 & 0.165 & 0.113 & 0.315 & 0.268 & 0.244 & 0.481 & 0.717 & 0.613 & 0.810 &  \\ 
  SD & 1.018 & 1.024 & 1.008 & 0.593 & 0.581 & 0.852 & 1.380 & 1.122 & 1.290 & 1.278 \\ 
   \bottomrule
\multicolumn{11}{c}{\emph{Note:} $N_\mathsf{B} = 379$}\\
\end{tabular}
\endgroup
}

\subfloat[Sample \textsf{V}]{%
\label{tab:inputCorSDVraw}
\centering
\begingroup\footnotesize
\begin{tabular}{rrrrrrrrrrr}
  \toprule
 & 1 & 2 & 3 & 4 & 5 & 6 & 7 & 8 & 9 & 10 \\ 
  \midrule
1. ctrl1 &  &  &  &  &  &  &  &  &  &  \\ 
  2. ctrl2 & 0.614 &  &  &  &  &  &  &  &  &  \\ 
  3. ctrl3 & 0.276 & 0.270 &  &  &  &  &  &  &  &  \\ 
  4. awa1 & 0.278 & 0.291 & 0.385 &  &  &  &  &  &  &  \\ 
  5. awa2 & 0.263 & 0.257 & 0.350 & 0.589 &  &  &  &  &  &  \\ 
  6. awa3 & 0.171 & 0.210 & 0.248 & 0.326 & 0.389 &  &  &  &  &  \\ 
  7. coll1 & 0.027 & 0.075 & 0.249 & 0.127 & 0.190 & 0.281 &  &  &  &  \\ 
  8. coll2 & 0.137 & 0.162 & 0.270 & 0.161 & 0.227 & 0.346 & 0.531 &  &  &  \\ 
  9. coll3 & 0.117 & 0.102 & 0.303 & 0.221 & 0.270 & 0.348 & 0.713 & 0.599 &  &  \\ 
  10. coll4 & 0.085 & 0.137 & 0.319 & 0.263 & 0.287 & 0.330 & 0.595 & 0.511 & 0.757 &  \\ 
  SD & 1.112 & 1.100 & 1.056 & 0.756 & 0.721 & 1.063 & 1.400 & 1.063 & 1.231 & 1.217 \\ 
   \bottomrule
\multicolumn{11}{c}{\emph{Note:} $N_\mathsf{V} = 433$}\\
\end{tabular}
\endgroup
}
\end{table*}
}

\newcommand{\histograms}{
\begin{figure}[p]
\centering\caption{Histograms and density plots of all samples}
\captionsetup{position=top}
\label{fig:histograms}
\subfloat[Sample \textsf{A}]{
\label{fig:histA}
\centering\includegraphics[keepaspectratio,width=.8\maxwidth]{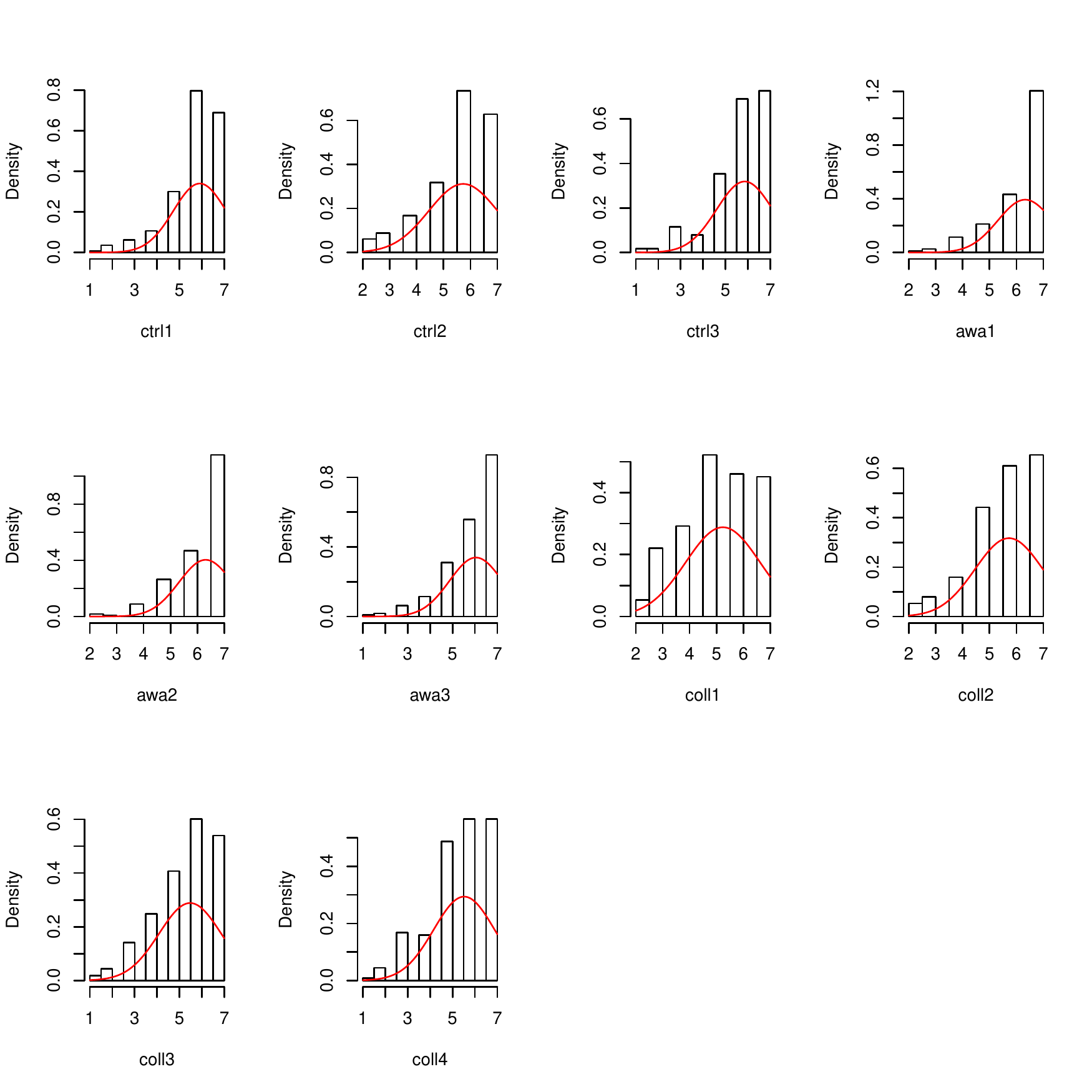}}

\subfloat[Sample \textsf{B}]{
\label{fig:histB}
\centering\includegraphics[keepaspectratio,width=.8\maxwidth]{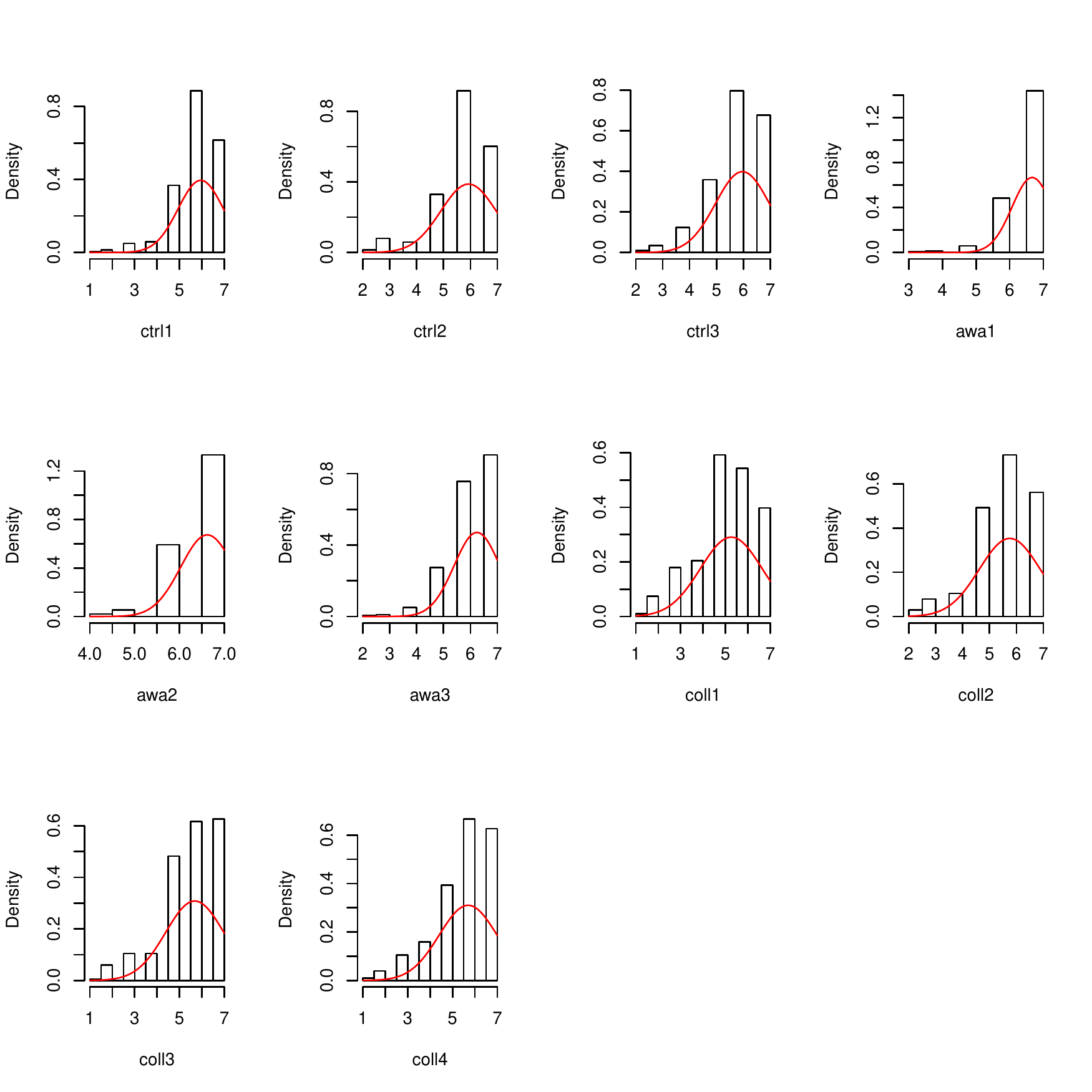}}

\subfloat[Sample \textsf{V}]{%
\label{fig:ihistV}
\centering\includegraphics[keepaspectratio,width=.8\maxwidth]{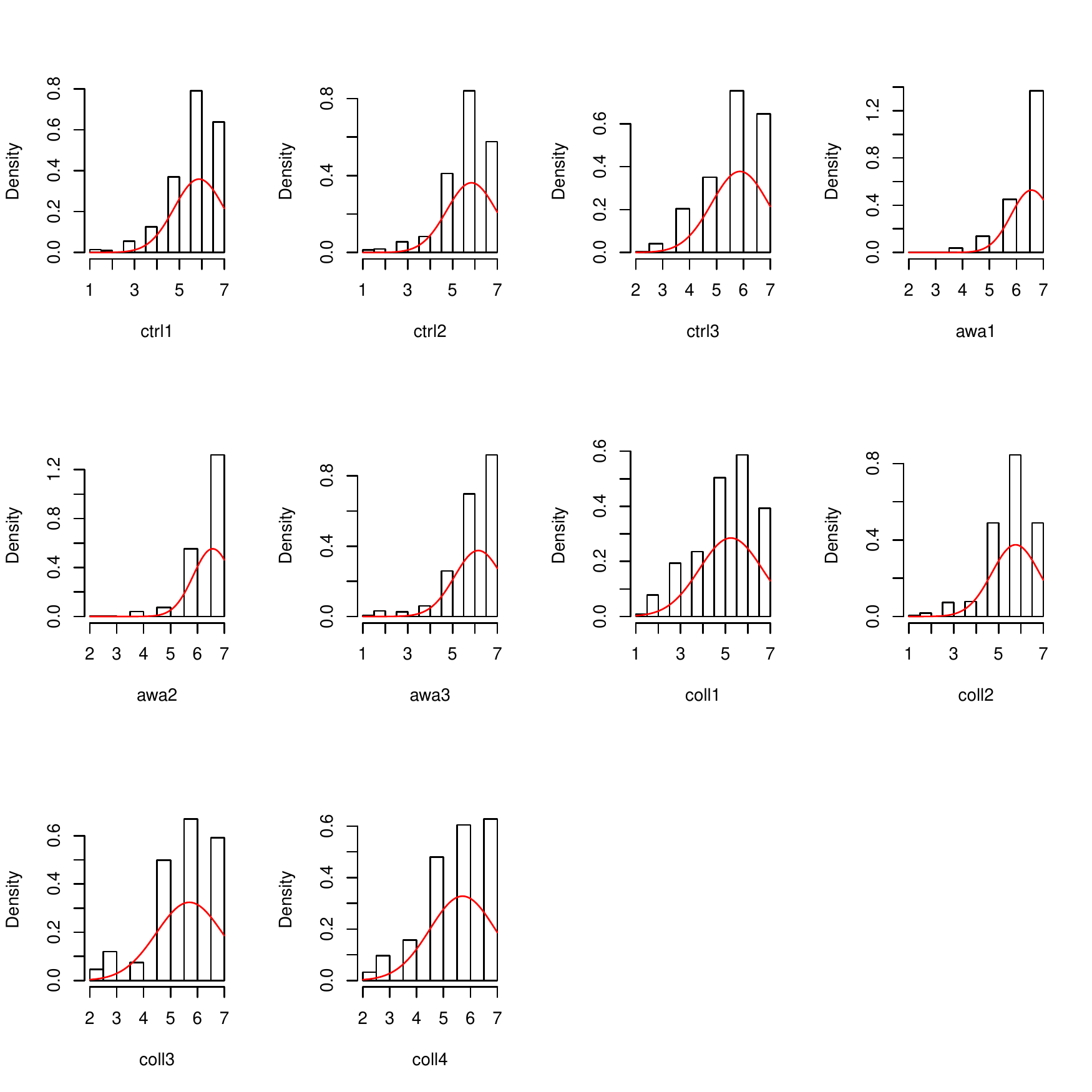}}
\end{figure}
}
\newcommand{\inputCorSDB}{
\begin{table*}[ht]
\centering
\caption{Correlations and Standard Deviations of Sample \textsf{B}} 
\label{tab:inputCorSDB}
\begingroup\footnotesize
\begin{tabular}{rrrrrrrrrrr}
  \toprule
 & 1 & 2 & 3 & 4 & 5 & 6 & 7 & 8 & 9 & 10 \\ 
  \midrule
1. ctrl1 &  &  &  &  &  &  &  &  &  &  \\ 
  2. ctrl2 & 0.56 &  &  &  &  &  &  &  &  &  \\ 
  3. ctrl3 & 0.25 & 0.27 &  &  &  &  &  &  &  &  \\ 
  4. awa1 & 0.25 & 0.23 & 0.25 &  &  &  &  &  &  &  \\ 
  5. awa2 & 0.32 & 0.32 & 0.30 & 0.62 &  &  &  &  &  &  \\ 
  6. awa3 & 0.23 & 0.19 & 0.26 & 0.32 & 0.31 &  &  &  &  &  \\ 
  7. coll1 & 0.05 & 0.05 & 0.26 & 0.05 & 0.11 & 0.33 &  &  &  &  \\ 
  8. coll2 & 0.10 & 0.06 & 0.28 & 0.22 & 0.24 & 0.30 & 0.66 &  &  &  \\ 
  9. coll3 & 0.16 & 0.09 & 0.32 & 0.16 & 0.22 & 0.40 & 0.76 & 0.72 &  &  \\ 
  10. coll4 & 0.19 & 0.10 & 0.31 & 0.23 & 0.23 & 0.47 & 0.71 & 0.62 & 0.81 &  \\ 
  SD & 0.93 & 0.93 & 1.00 & 0.55 & 0.56 & 0.82 & 1.36 & 1.11 & 1.24 & 1.22 \\ 
   \bottomrule
\multicolumn{11}{c}{\emph{Note:} $N_\mathsf{B} = 370$}\\
\end{tabular}
\endgroup
\end{table*}
}
\newcommand{\inputCorSD}{
\begin{table*}[p]
\centering\caption{Correlations and Standard Deviations of samples used in the study}
\captionsetup{position=top}
\label{tab:inputCorSD}
\subfloat[Sample \textsf{A}]{
\label{tab:inputCorSDA}
\centering
\begingroup\footnotesize
\begin{tabular}{rrrrrrrrrrr}
  \toprule
 & 1 & 2 & 3 & 4 & 5 & 6 & 7 & 8 & 9 & 10 \\ 
  \midrule
1. ctrl1 &  &  &  &  &  &  &  &  &  &  \\ 
  2. ctrl2 & 0.531 &  &  &  &  &  &  &  &  &  \\ 
  3. ctrl3 & 0.298 & 0.420 &  &  &  &  &  &  &  &  \\ 
  4. awa1 & 0.238 & 0.456 & 0.393 &  &  &  &  &  &  &  \\ 
  5. awa2 & 0.313 & 0.367 & 0.306 & 0.494 &  &  &  &  &  &  \\ 
  6. awa3 & 0.306 & 0.282 & 0.142 & 0.130 & 0.062 &  &  &  &  &  \\ 
  7. coll1 & 0.094 & 0.072 & 0.092 & 0.162 & 0.128 & 0.268 &  &  &  &  \\ 
  8. coll2 & 0.129 & 0.134 & 0.236 & 0.194 & 0.225 & 0.273 & 0.567 &  &  &  \\ 
  9. coll3 & 0.105 & 0.184 & 0.236 & 0.304 & 0.214 & 0.170 & 0.674 & 0.608 &  &  \\ 
  10. coll4 & 0.167 & 0.182 & 0.261 & 0.309 & 0.292 & 0.313 & 0.580 & 0.393 & 0.660 &  \\ 
  SD & 1.031 & 1.216 & 1.201 & 0.942 & 0.871 & 1.097 & 1.357 & 1.273 & 1.371 & 1.310 \\ 
   \bottomrule
\multicolumn{11}{c}{\emph{Note:} $N_{\mathsf{A}}^\prime = 201$}\\
\end{tabular}
\endgroup
}

\subfloat[Sample \textsf{B}]{
\label{tab:inputCorSDB}
\centering
\begingroup\footnotesize
\begin{tabular}{rrrrrrrrrrr}
  \toprule
 & 1 & 2 & 3 & 4 & 5 & 6 & 7 & 8 & 9 & 10 \\ 
  \midrule
1. ctrl1 &  &  &  &  &  &  &  &  &  &  \\ 
  2. ctrl2 & 0.557 &  &  &  &  &  &  &  &  &  \\ 
  3. ctrl3 & 0.253 & 0.275 &  &  &  &  &  &  &  &  \\ 
  4. awa1 & 0.253 & 0.232 & 0.253 &  &  &  &  &  &  &  \\ 
  5. awa2 & 0.316 & 0.325 & 0.296 & 0.619 &  &  &  &  &  &  \\ 
  6. awa3 & 0.226 & 0.192 & 0.262 & 0.321 & 0.312 &  &  &  &  &  \\ 
  7. coll1 & 0.053 & 0.050 & 0.264 & 0.050 & 0.107 & 0.325 &  &  &  &  \\ 
  8. coll2 & 0.102 & 0.058 & 0.281 & 0.219 & 0.236 & 0.297 & 0.656 &  &  &  \\ 
  9. coll3 & 0.157 & 0.087 & 0.321 & 0.163 & 0.220 & 0.401 & 0.756 & 0.717 &  &  \\ 
  10. coll4 & 0.194 & 0.099 & 0.310 & 0.234 & 0.234 & 0.470 & 0.706 & 0.618 & 0.812 &  \\ 
  SD & 0.931 & 0.933 & 0.996 & 0.549 & 0.563 & 0.819 & 1.358 & 1.106 & 1.242 & 1.218 \\ 
   \bottomrule
\multicolumn{11}{c}{\emph{Note:} $N_{\mathsf{B}}^\prime = 370$}\\
\end{tabular}
\endgroup
}

\subfloat[Sample \textsf{V}]{%
\label{tab:inputCorSDV}
\centering
\begingroup\footnotesize
\begin{tabular}{rrrrrrrrrrr}
  \toprule
 & 1 & 2 & 3 & 4 & 5 & 6 & 7 & 8 & 9 & 10 \\ 
  \midrule
1. ctrl1 &  &  &  &  &  &  &  &  &  &  \\ 
  2. ctrl2 & 0.534 &  &  &  &  &  &  &  &  &  \\ 
  3. ctrl3 & 0.288 & 0.309 &  &  &  &  &  &  &  &  \\ 
  4. awa1 & 0.277 & 0.339 & 0.320 &  &  &  &  &  &  &  \\ 
  5. awa2 & 0.272 & 0.310 & 0.291 & 0.549 &  &  &  &  &  &  \\ 
  6. awa3 & 0.154 & 0.232 & 0.176 & 0.210 & 0.274 &  &  &  &  &  \\ 
  7. coll1 & 0.065 & 0.126 & 0.255 & 0.120 & 0.214 & 0.295 &  &  &  &  \\ 
  8. coll2 & 0.175 & 0.218 & 0.272 & 0.127 & 0.218 & 0.366 & 0.522 &  &  &  \\ 
  9. coll3 & 0.144 & 0.143 & 0.296 & 0.183 & 0.265 & 0.339 & 0.705 & 0.581 &  &  \\ 
  10. coll4 & 0.075 & 0.148 & 0.305 & 0.226 & 0.257 & 0.294 & 0.591 & 0.493 & 0.743 &  \\ 
  SD & 1.017 & 0.998 & 1.031 & 0.691 & 0.622 & 0.982 & 1.390 & 1.036 & 1.192 & 1.179 \\ 
   \bottomrule
\multicolumn{11}{c}{\emph{Note:} $N_{\mathsf{V}}^\prime = 419$}\\
\end{tabular}
\endgroup
}
\end{table*}
}
\newcommand{\descSubScales}{
\begin{table}[htb]
\centering\caption{Means (SDs) of the parceled sub-scales of IUIPC-10}
\label{tab:descSubScales}
\begingroup\footnotesize
\begin{tabular}{rlll}
  \toprule
 & Sample \textsf{A} & Sample \textsf{B} & Sample \textsf{V} \\ 
  \midrule
\textsf{ctrl} & $5.82$ $(0.99)$ & $5.93$ $(0.78)$ & $5.86$ $(0.84)$ \\ 
  \textsf{awa} & $6.22$ $(0.78)$ & $6.51$ $(0.52)$ & $6.43$ $(0.66)$ \\ 
  \textsf{coll} & $5.48$ $(1.12)$ & $5.58$ $(1.12)$ & $5.60$ $(1.04)$ \\ 
  \textsf{iuipc} & $5.84$ $(0.75)$ & $6.00$ $(0.61)$ & $5.96$ $(0.64)$ \\ 
   \bottomrule
\end{tabular}
\endgroup
\end{table}
}

\newcommand{\densitySubScales}{
\definecolor{viriviolet}{HTML}{351042}
\definecolor{virigreen}{HTML}{317F79}
\definecolor{viriyellow}{HTML}{FCE528}
\begin{figure*}[tb]
\centering\captionsetup{position=bottom}
\begin{minipage}{0.24\textwidth}%
\subfloat[Control]{
\label{fig:densityCtrl}
\centering
\includegraphics[width=\maxwidth]{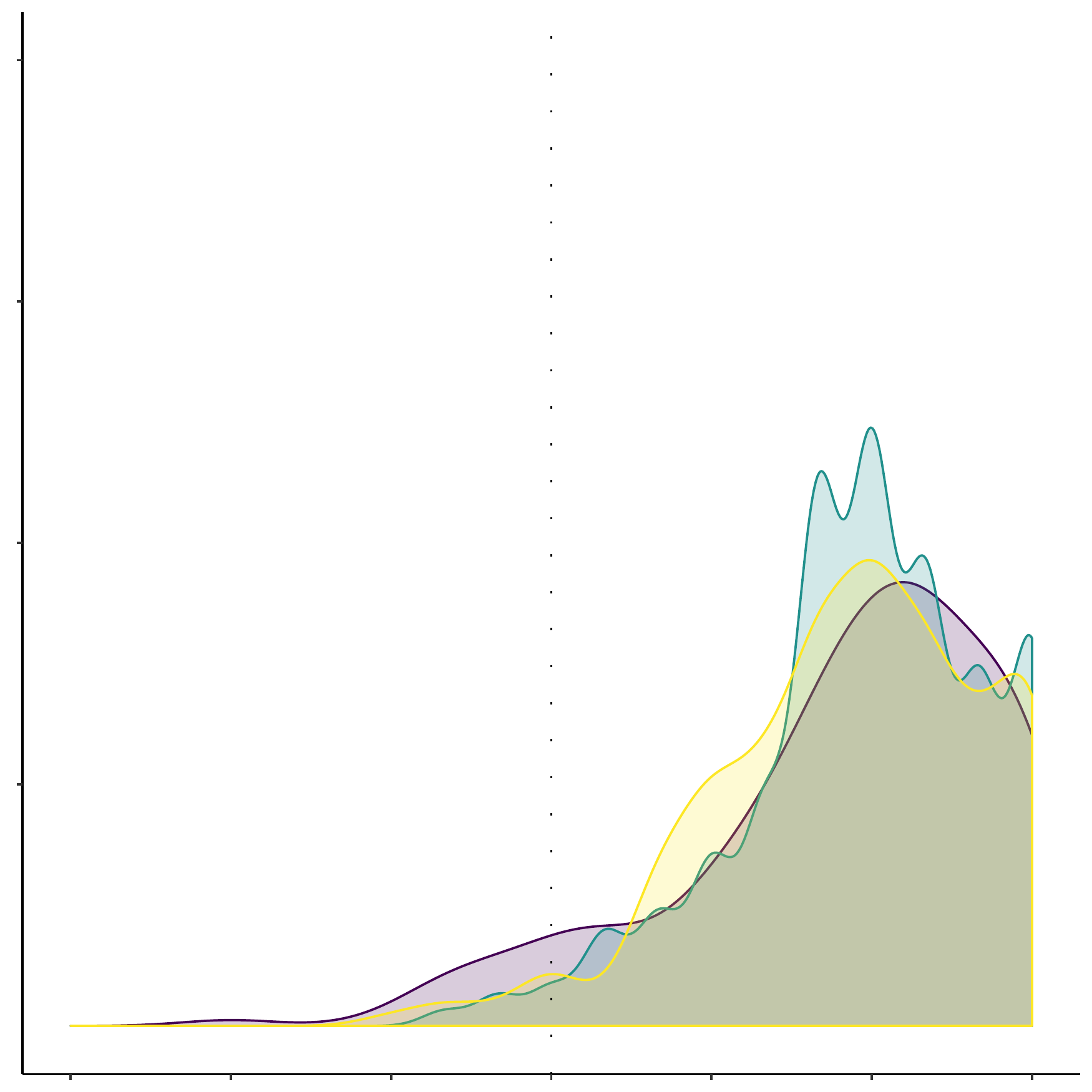} 
}
\end{minipage}~
\begin{minipage}{0.24\textwidth}%
\subfloat[Awareness]{
\label{fig:densityAwa}
\centering
\includegraphics[width=\maxwidth]{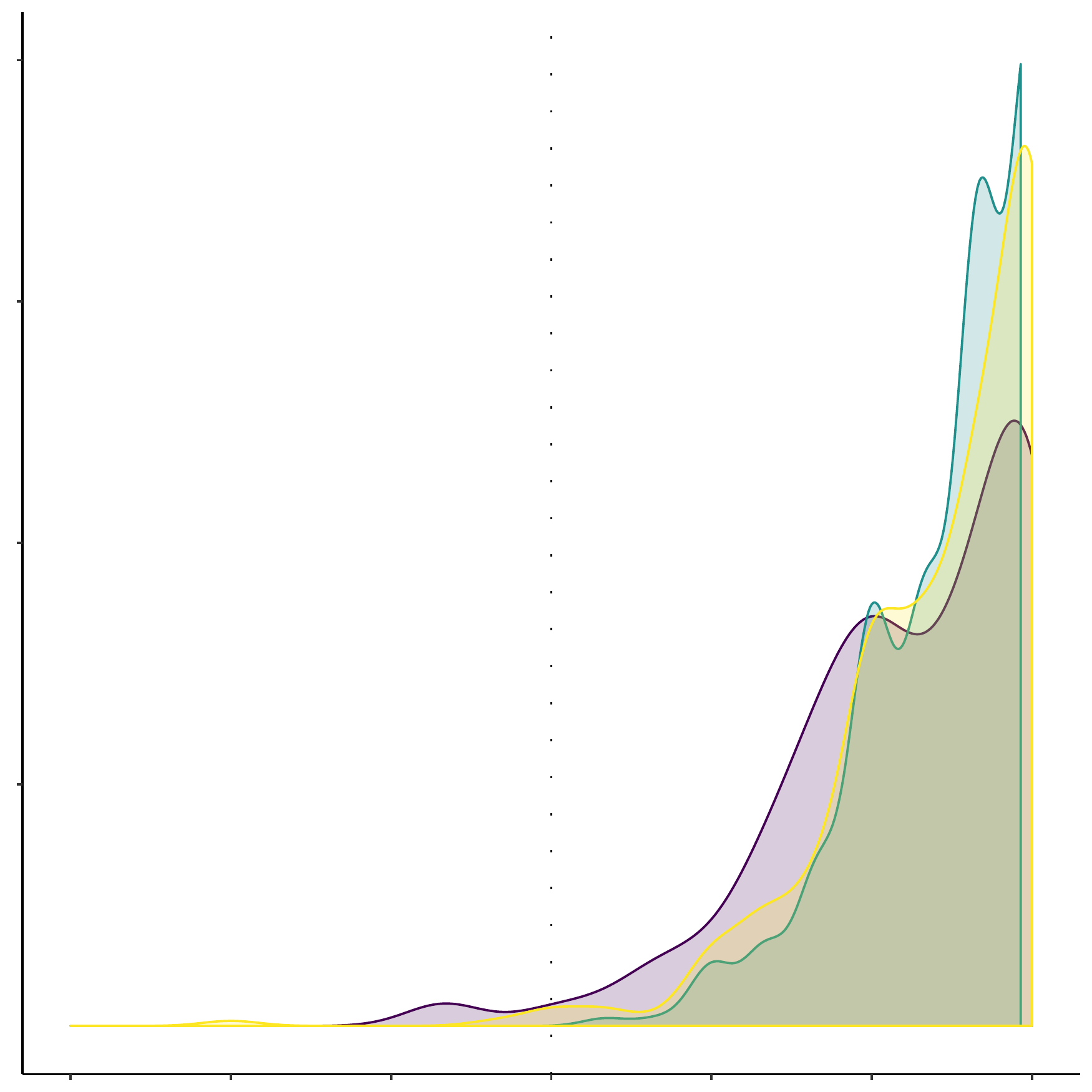} 
}
\end{minipage}~
\begin{minipage}{0.24\textwidth}%
\subfloat[Collection]{%
\label{fig:densityColl}
\centering
\includegraphics[width=\maxwidth]{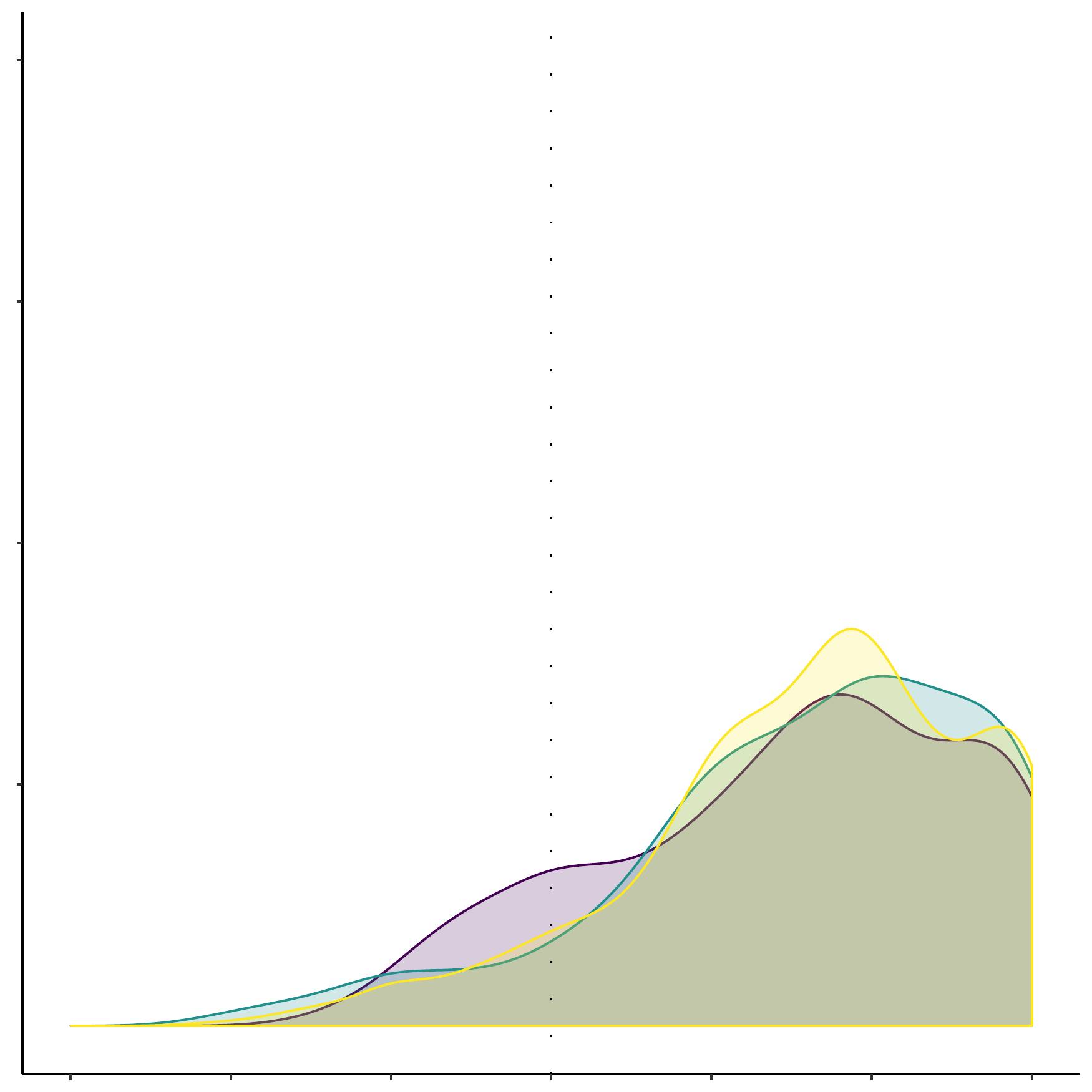} 
}
\end{minipage}
\begin{minipage}{0.24\textwidth}%
\subfloat[IUIPC-10 Overall]{%
\label{fig:densityIUIPC}
\centering
\includegraphics[width=\maxwidth]{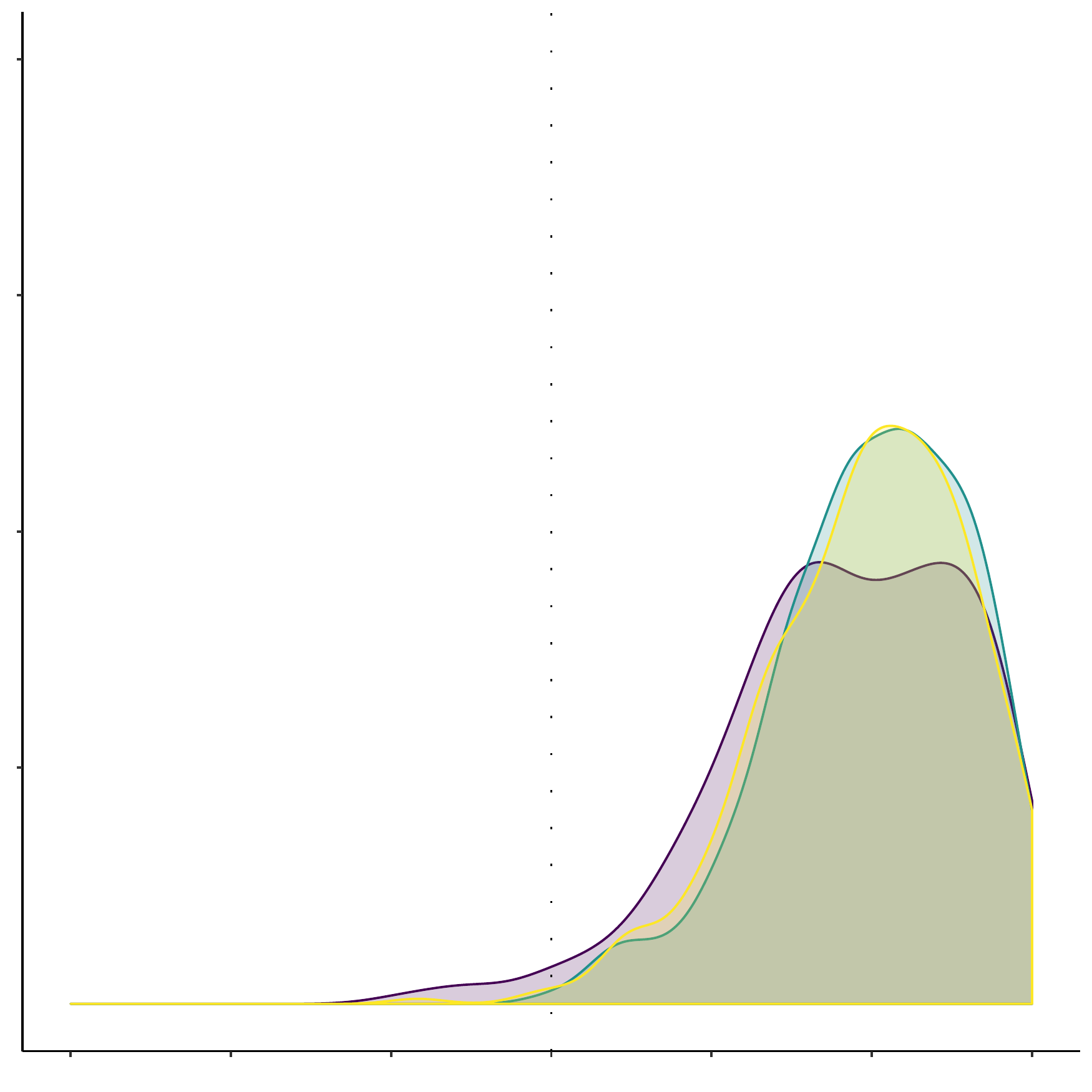} 
}
\end{minipage}
\caption{Density of IUIPC-10 subscale responses across samples (\textsf{A}: \textcolor{viriviolet}{violet}, \textsf{B}: \textcolor{virigreen}{green}, \textsf{V}: \textcolor{viriyellow}{yellow}). \emph{Note:} All graphs are on the same scale}
\label{fig:densitySubScales}
\end{figure*}
}

\newcommand{\efaloadings}{
\begin{table*}[tbp]
\centering\caption{Factor loadings of \textsf{Oblimin}-transformed EFA on auxiliary Sample~\textsf{A} and base Sample \textsf{B}}
\captionsetup{position=top}
\label{tab:efaloadings}
\subfloat[EFA on Sample \textsf{A} ($N_\mathsf{A} = 204$)]{
\label{tab:efaloadingsA}
\centering
\begingroup\footnotesize
\begin{tabular}{rrrrr}
  \toprule
 & Collection & Awareness & Control & Uniqueness \\ 
  \midrule
ctrl1 &  & 0.62 &  & 0.57 \\ 
  ctrl2 &  & 0.59 & 0.37 & 0.37 \\ 
  ctrl3 &  &  & 0.41 & 0.69 \\ 
  awa1 &  &  & 0.64 & 0.49 \\ 
  awa2 &  &  & 0.48 & 0.63 \\ 
  awa3 & 0.23 & 0.57 & -0.23 & 0.63 \\ 
  coll1 & 0.84 &  &  & 0.31 \\ 
  coll2 & 0.64 &  &  & 0.55 \\ 
  coll3 & 0.87 &  &  & 0.21 \\ 
  coll4 & 0.67 &  &  & 0.46 \\ 
   \midrule
SS loadings & 2.43 & 1.16 & 1.07 &  \\ 
  Proportion Var & 0.24 & 0.12 & 0.11 &  \\ 
  Cumulative Var & 0.24 & 0.36 & 0.47 &  \\ 
   \bottomrule
\multicolumn{5}{c}{\emph{Note:} $\chi^2(18) = 42.552, p < .001$, Variance explained $R^2=.47$}\\
\end{tabular}
\endgroup
}~\subfloat[EFA on Sample \textsf{B} ($N_\mathsf{B} = 371$)]{%
\label{tab:efaloadingsB}
\centering
\begingroup\footnotesize
\begin{tabular}{rrrrr}
  \toprule
 & Collection & Awareness & Control & Uniqueness \\ 
  \midrule
ctrl1 &  &  & 0.67 & 0.51 \\ 
  ctrl2 &  &  & 0.82 & 0.36 \\ 
  ctrl3 & 0.28 &  & 0.24 & 0.76 \\ 
  awa1 &  & 0.88 &  & 0.26 \\ 
  awa2 &  & 0.66 &  & 0.45 \\ 
  awa3 & 0.36 & 0.26 &  & 0.70 \\ 
  coll1 & 0.87 &  &  & 0.29 \\ 
  coll2 & 0.75 &  &  & 0.41 \\ 
  coll3 & 0.93 &  &  & 0.14 \\ 
  coll4 & 0.83 &  &  & 0.26 \\ 
   \midrule
SS loadings & 3.08 & 1.34 & 1.23 &  \\ 
  Proportion Var & 0.31 & 0.13 & 0.12 &  \\ 
  Cumulative Var & 0.31 & 0.44 & 0.56 &  \\ 
   \bottomrule
\multicolumn{5}{c}{\emph{Note:} $\chi^2(18) = 36.37, p = .006$, Variance explained $R^2=.56$}\\
\end{tabular}
\endgroup
}
\end{table*}
}
\newcommand{\efaloadingspoly}{
\begin{table*}[tbp]
\centering\caption{Factor loadings of \textsf{Oblimin}-transformed Polychoric EFA on Samples~\textsf{A} and~\textsf{B}}
\captionsetup{position=top}
\label{tab:efaloadingspoly}
\subfloat[EFA on Sample \textsf{A} ($N_\mathsf{A} = 204$)]{
\label{tab:efaloadingsApoly}
\centering
\begingroup\footnotesize
\begin{tabular}{rrrrr}
  \toprule
  &\multicolumn{3}{c}{Factors} & \multirow{2}{*}{Uniqueness}\\
\cmidrule(lr){2-4}
 & Collection & Awareness & Control\\
 \midrule
ctrl1 &  &  & 0.75 & 0.43 \\ 
  ctrl2 &  & 0.29 & 0.59 & 0.39 \\ 
  ctrl3 &  & 0.46 &  & 0.58 \\ 
  awa1 &  & 0.75 &  & 0.39 \\ 
  awa2 &  & 0.66 &  & 0.45 \\ 
  awa3 & 0.26 &  & 0.56 & 0.62 \\ 
  coll1 & 0.86 &  &  & 0.33 \\ 
  coll2 & 0.69 &  &  & 0.45 \\ 
  coll3 & 0.79 &  &  & 0.30 \\ 
  coll4 & 0.59 &  &  & 0.47 \\ 
   \midrule
SS loadings & 2.26 & 1.39 & 1.30 &  \\ 
  Proportion Var & 0.23 & 0.14 & 0.13 &  \\ 
  Cumulative Var & 0.23 & 0.37 & 0.50 &  \\ 
   \bottomrule
\multicolumn{5}{c}{\emph{Note:} $\chi^2(18) = 19.515, p = .361$, Variance explained $R^2=.50$}\\
\end{tabular}
\endgroup
}~\subfloat[EFA on Sample \textsf{B} ($N_\mathsf{B} = 371$)]{%
\label{tab:efaloadingsBpoly}
\centering
\begingroup\footnotesize
\begin{tabular}{rrrrr}
  \toprule
  &\multicolumn{3}{c}{Factors} & \multirow{2}{*}{Uniqueness}\\
\cmidrule(lr){2-4}
 & Collection & Awareness & Control\\
 \midrule
ctrl1 &  &  & 0.66 & 0.48 \\ 
  ctrl2 &  &  & 0.89 & 0.25 \\ 
  ctrl3 & 0.29 &  & 0.29 & 0.65 \\ 
  awa1 &  & 0.95 &  & 0.17 \\ 
  awa2 &  & 0.76 &  & 0.29 \\ 
  awa3 & 0.38 & 0.32 &  & 0.61 \\ 
  coll1 & 0.84 &  &  & 0.35 \\ 
  coll2 & 0.78 &  &  & 0.37 \\ 
  coll3 & 0.93 &  &  & 0.15 \\ 
  coll4 & 0.84 &  &  & 0.25 \\ 
   \midrule
SS loadings & 3.11 & 1.63 & 1.35 &  \\ 
  Proportion Var & 0.31 & 0.16 & 0.14 &  \\ 
  Cumulative Var & 0.31 & 0.47 & 0.61 &  \\ 
   \bottomrule
\multicolumn{5}{c}{\emph{Note:} $\chi^2(18) = 24.971, p = .126$, Variance explained $R^2=.61$}\\
\end{tabular}
\endgroup
}
\end{table*}
}
\newcommand{\biplotB}{
\begin{figure}[htb]
\centering
\includegraphics[width=\maxwidth]{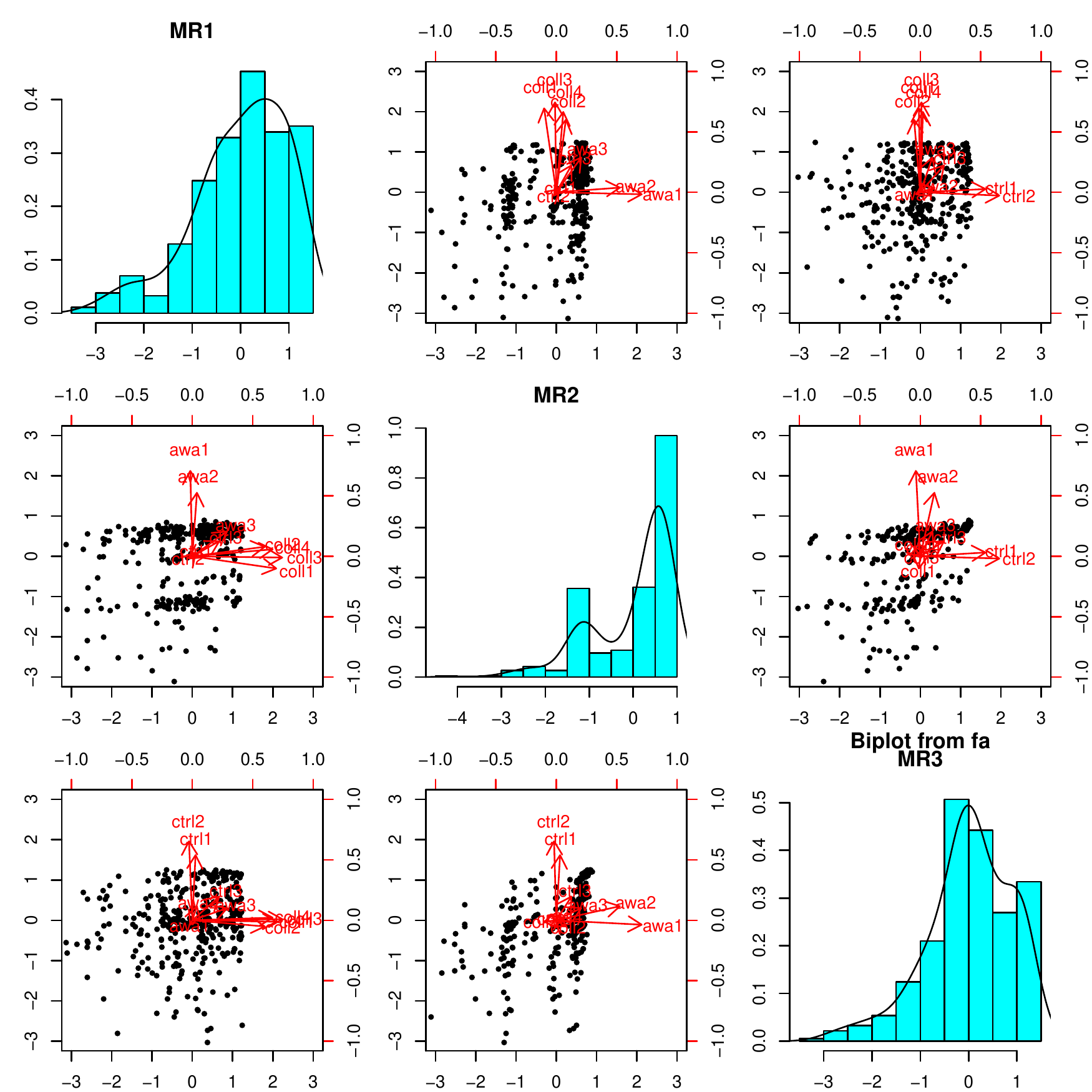} 
\caption{Biplot of the EFA of sample \textsf{B}}
\label{fig:biplotB}
\end{figure}
}
\newcommand{\efadiagramB}{
\begin{figure}[htb]
\centering
\includegraphics[width=\maxwidth]{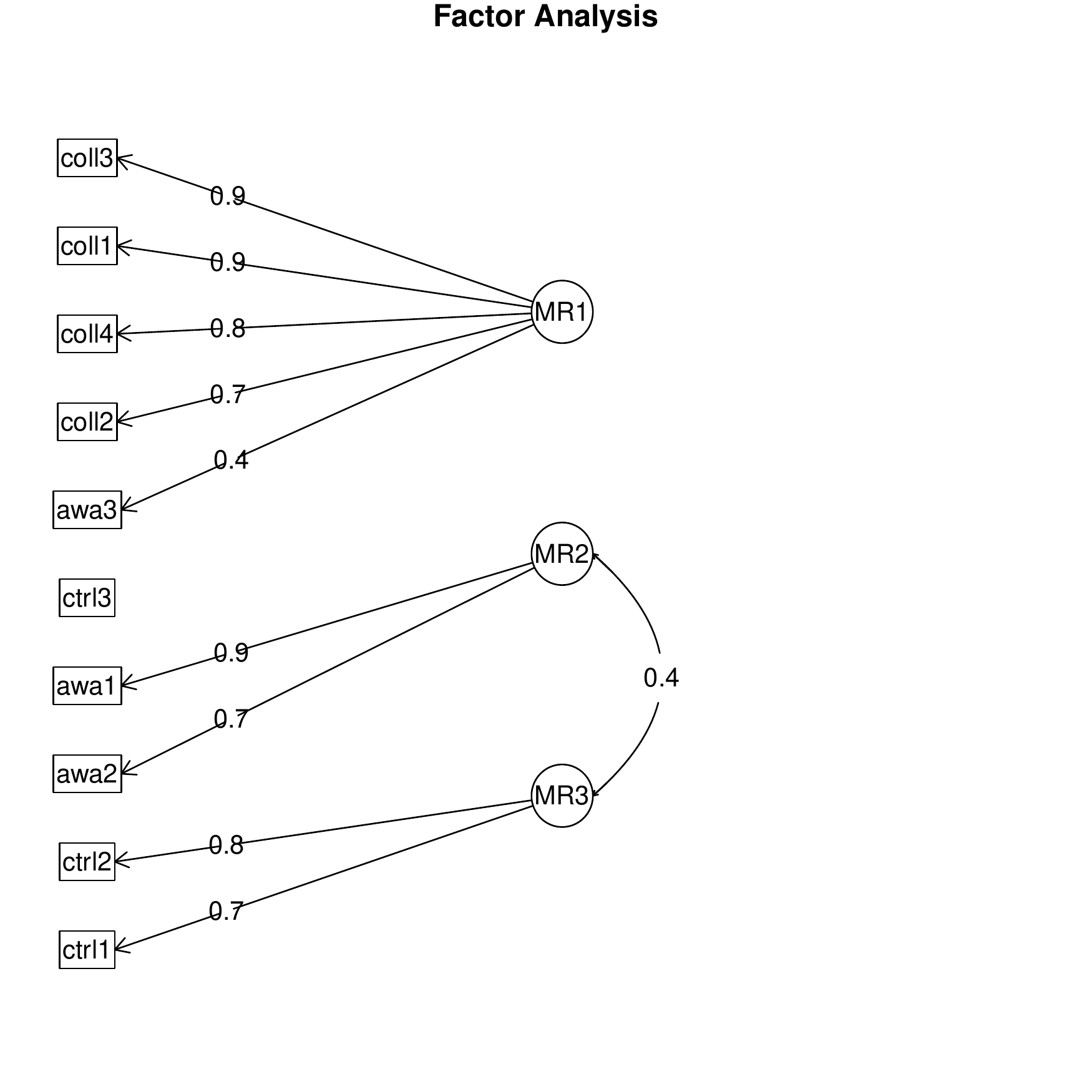} 
\caption{Path diagram of the EFA of sample \textsf{B}}
\label{fig:efadiagramB}
\end{figure}
}
\newcommand{\scatterFactorsB}{
\begin{figure}[htb]
\centering
\includegraphics[width=\maxwidth]{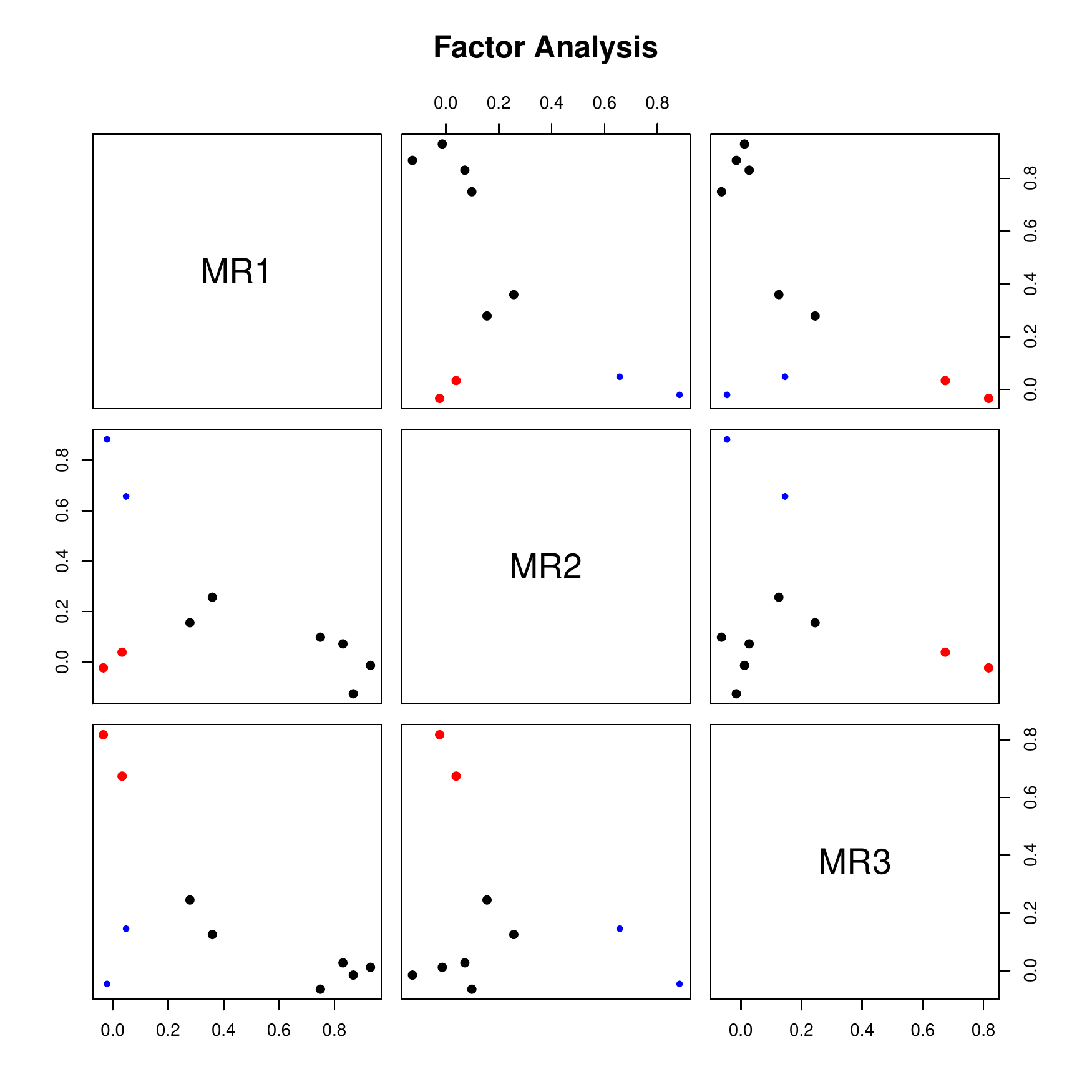} 
\caption{Item/factor scatterplot of the EFA of sample \textsf{B}}
\label{fig:scatterFactorsB}
\end{figure}
}

\newcommand{\fitmeasuresB}{
\begin{table*}[ht]
\centering
\caption{Fit measures of the MLM CFA model on Sample \textsf{B}} 
\label{tab:fitmeasuresB}
\begingroup\footnotesize
\begin{tabular}{rrrrrrrrrr}
  \toprule
$\chi^2$ & $\mathit{df}$ & $p_{\chi^2}$ & \textsf{CFI} & \textsf{TFI} & \textsf{RMSEA} & \textsf{LL} & \textsf{UL} & $p_{\mathsf{rmsea}}$ & \textsf{SRMR} \\ 
  \midrule
163.69 & 32.00 & 0.00 & 0.92 & 0.89 & 0.11 & 0.09 & 0.12 & 0.00 & 0.10 \\ 
   \bottomrule
\end{tabular}
\endgroup
\end{table*}
}
\newcommand{\fitmeasuresBwls}{
\begin{table*}[ht]
\centering
\caption{Fit measures of the WLSMVS CFA model on Sample \textsf{B}} 
\label{tab:fitmeasuresBwls}
\begingroup\footnotesize
\begin{tabular}{rrrrrrrrrr}
  \toprule
$\chi^2$ & $\mathit{df}$ & $p_{\chi^2}$ & \textsf{CFI} & \textsf{TFI} & \textsf{RMSEA} & \textsf{LL} & \textsf{UL} & $p_{\mathsf{rmsea}}$ & \textsf{SRMR} \\ 
  \midrule
213.83 & 32.00 & 0.00 & 0.98 & 0.98 & 0.12 & 0.11 & 0.14 & 0.00 & 0.10 \\ 
   \bottomrule
\end{tabular}
\endgroup
\end{table*}
}
\newcommand{\corfitB}{
\begin{table}[ht]
\centering
\caption{Implied correlations of fitted CFA model on Sample \textsf{B}} 
\label{tab:corfitB}
\begingroup\footnotesize
\begin{tabular}{rllll}
  \toprule
 & 1 & 2 & 3 & 4 \\ 
  \midrule
1. ctrl & 0.634 &  &  &  \\ 
  2. aware & 0.546 & 0.627 &  &  \\ 
  3. collect & 0.227 & 0.329 & 0.848 &  \\ 
  4. iuipc & 0.613 & 0.891 & 0.37 & 0.769 \\ 
   \bottomrule
\multicolumn{5}{c}{\emph{Note:} The diagonal contains the $\sqrt{\mathit{AVE}}$}\\
\end{tabular}
\endgroup
\end{table}
}
\newcommand{\residualsB}{
\begin{table*}[tbp]
\centering\caption{Residuals of the MLM-estimated CFA of IUIPC-10 on Sample \textsf{B}\explain{.\newline{}The highlighted distinctive residual patterns show a poor local fit. They indicate misloadings of \textsf{ctrl3} and \textsf{awa3}.}}
\label{tab:residualsB}
\captionsetup{position=top}
\subfloat[Correlation residuals]{
\label{tab:residualsBcor}
\centering
\begingroup\footnotesize
\begin{tabular}{rllllllllll}
  \toprule
 & 1 & 2 & 3 & 4 & 5 & 6 & 7 & 8 & 9 & 10 \\ 
  \midrule
1. ctrl1 & --- &  &  &  &  &  &  &  &  &  \\ 
  2. ctrl2 & 0.021 & --- &  &  &  &  &  &  &  &  \\ 
  3. ctrl3 & -0.046 & -0.022 & --- &  &  &  &  &  &  &  \\ 
  4. awa1 & -0.042 & -0.062 & 0.089 & --- &  &  &  &  &  &  \\ 
  5. awa2 & -0.01 & 0 & \textbf{0.115} & 0.018 & --- &  &  &  &  &  \\ 
  6. awa3 & 0.048 & 0.015 & \textbf{0.163} & -0.007 & -0.049 & --- &  &  &  &  \\ 
  7. coll1 & -0.082 & -0.085 & \textbf{0.189} & \textbf{-0.147} & \textbf{-0.111} & \textbf{0.206} & --- &  &  &  \\ 
  8. coll2 & -0.025 & -0.068 & \textbf{0.211} & 0.034 & 0.032 & \textbf{0.186} & 0.037 & --- &  &  \\ 
  9. coll3 & 0.001 & -0.069 & \textbf{0.234} & -0.065 & -0.031 & \textbf{0.264} & -0.007 & 0.003 & --- &  \\ 
  10. coll4 & 0.051 & -0.043 & \textbf{0.231} & 0.024 & 0.003 & \textbf{0.345} & 0.006 & -0.038 & 0.004 & --- \\ 
   \bottomrule
\multicolumn{11}{c}{\emph{Note:} Correlation residuals in absolute $> 0.1$ are marked}\\
\end{tabular}
\endgroup
}

\subfloat[Standardized residuals]{%
\label{tab:residualsBstd}
\centering
\begingroup\footnotesize
\begin{tabular}{rllllllllll}
  \toprule
 & 1 & 2 & 3 & 4 & 5 & 6 & 7 & 8 & 9 & 10 \\ 
  \midrule
1. ctrl1 & --- &  &  &  &  &  &  &  &  &  \\ 
  2. ctrl2 & \textbf{3.703} & --- &  &  &  &  &  &  &  &  \\ 
  3. ctrl3 & \textbf{-2.345} & -1.016 & --- &  &  &  &  &  &  &  \\ 
  4. awa1 & -1.585 & \textbf{-2.466} & 1.921 & --- &  &  &  &  &  &  \\ 
  5. awa2 & -0.457 & 0.003 & \textbf{2.597} & \textbf{4.618} & --- &  &  &  &  &  \\ 
  6. awa3 & 0.972 & 0.379 & \textbf{3.012} & -0.36 & \textbf{-3.891} & --- &  &  &  &  \\ 
  7. coll1 & \textbf{-2.304} & \textbf{-2.617} & \textbf{3.324} & \textbf{-4.525} & \textbf{-3.544} & \textbf{3.878} & --- &  &  &  \\ 
  8. coll2 & -0.669 & -1.843 & \textbf{4.334} & 0.706 & 0.845 & \textbf{3.971} & \textbf{2.035} & --- &  &  \\ 
  9. coll3 & 0.031 & \textbf{-2.609} & \textbf{4.142} & \textbf{-2.448} & -1.365 & \textbf{5.144} & -1.384 & 0.405 & --- &  \\ 
  10. coll4 & 1.293 & -1.429 & \textbf{4.284} & 0.688 & 0.101 & \textbf{5.912} & 0.475 & \textbf{-2.636} & 1.397 & --- \\ 
   \bottomrule
\multicolumn{11}{c}{\emph{Note:} Statistically significant residuals ($\mathsf{abs} > 1.96$) are marked}\\
\end{tabular}
\endgroup
}
\end{table*}
}
\newcommand{\residualsBml}{
\begin{table*}[tbp]
\centering\caption{Residuals of the ML-estimated CFA of IUIPC-10 on Sample \textsf{B} (with outlier treatment)}
\label{tab:residualsBml}
\captionsetup{position=top}
\subfloat[Correlation residuals]{
\label{tab:residualsBmlcor}
\centering
\begingroup\footnotesize
\begin{tabular}{rllllllllll}
  \toprule
 & 1 & 2 & 3 & 4 & 5 & 6 & 7 & 8 & 9 & 10 \\ 
  \midrule
1. ctrl1 & --- &  &  &  &  &  &  &  &  &  \\ 
  2. ctrl2 & 0.021 & --- &  &  &  &  &  &  &  &  \\ 
  3. ctrl3 & -0.046 & -0.022 & --- &  &  &  &  &  &  &  \\ 
  4. awa1 & -0.042 & -0.062 & 0.089 & --- &  &  &  &  &  &  \\ 
  5. awa2 & -0.01 & 0 & \textbf{0.115} & 0.018 & --- &  &  &  &  &  \\ 
  6. awa3 & 0.048 & 0.015 & \textbf{0.163} & -0.007 & -0.049 & --- &  &  &  &  \\ 
  7. coll1 & -0.082 & -0.085 & \textbf{0.189} & \textbf{-0.147} & \textbf{-0.111} & \textbf{0.206} & --- &  &  &  \\ 
  8. coll2 & -0.025 & -0.068 & \textbf{0.211} & 0.034 & 0.032 & \textbf{0.186} & 0.037 & --- &  &  \\ 
  9. coll3 & 0.001 & -0.069 & \textbf{0.234} & -0.065 & -0.031 & \textbf{0.264} & -0.007 & 0.003 & --- &  \\ 
  10. coll4 & 0.051 & -0.043 & \textbf{0.231} & 0.024 & 0.003 & \textbf{0.345} & 0.006 & -0.038 & 0.004 & --- \\ 
   \bottomrule
\multicolumn{11}{c}{\emph{Note:} Correlation residuals in absolute $> 0.1$ are marked}\\
\end{tabular}
\endgroup
}

\subfloat[Standardized residuals]{%
\label{tab:residualsBmlstd}
\centering
\begingroup\footnotesize
\begin{tabular}{rllllllllll}
  \toprule
 & 1 & 2 & 3 & 4 & 5 & 6 & 7 & 8 & 9 & 10 \\ 
  \midrule
1. ctrl1 & --- &  &  &  &  &  &  &  &  &  \\ 
  2. ctrl2 & \textbf{4.214} & --- &  &  &  &  &  &  &  &  \\ 
  3. ctrl3 & \textbf{-2.68} & -1.249 & --- &  &  &  &  &  &  &  \\ 
  4. awa1 & -1.676 & \textbf{-2.408} & \textbf{2.097} & --- &  &  &  &  &  &  \\ 
  5. awa2 & -0.484 & 0.003 & \textbf{2.853} & \textbf{4.646} & --- &  &  &  &  &  \\ 
  6. awa3 & 1.157 & 0.352 & \textbf{3.514} & -0.367 & \textbf{-4.015} & --- &  &  &  &  \\ 
  7. coll1 & \textbf{-2.4} & \textbf{-2.494} & \textbf{3.852} & \textbf{-4.349} & \textbf{-3.684} & \textbf{4.255} & --- &  &  &  \\ 
  8. coll2 & -0.695 & -1.864 & \textbf{4.314} & 0.955 & 0.957 & \textbf{4.03} & \textbf{2.194} & --- &  &  \\ 
  9. coll3 & 0.028 & \textbf{-2.708} & \textbf{4.91} & \textbf{-2.643} & -1.654 & \textbf{5.684} & -1.731 & 0.495 & --- &  \\ 
  10. coll4 & 1.667 & -1.386 & \textbf{4.848} & 0.818 & 0.111 & \textbf{7.349} & 0.611 & \textbf{-3.36} & 1.482 & --- \\ 
   \bottomrule
\multicolumn{11}{c}{\emph{Note:} Statistically significant residuals ($\mathsf{abs} > 1.96$) are marked}\\
\end{tabular}
\endgroup
}
\end{table*}
}
\newcommand{\residualsBwls}{
\begin{table*}[tbp]
\centering\caption{Residuals of the WLSMVS-estimated CFA of IUIPC-10 on Sample \textsf{B}}
\label{tab:residualsBwls}
\captionsetup{position=top}
\subfloat[Correlation residuals]{
\label{tab:residualsBwlscor}
\centering
\begingroup\footnotesize
\begin{tabular}{rllllllllll}
  \toprule
 & 1 & 2 & 3 & 4 & 5 & 6 & 7 & 8 & 9 & 10 \\ 
  \midrule
1. ctrl1 & --- &  &  &  &  &  &  &  &  &  \\ 
  2. ctrl2 & \textbf{0.123} & --- &  &  &  &  &  &  &  &  \\ 
  3. ctrl3 & \textbf{-0.151} & \textbf{-0.111} & --- &  &  &  &  &  &  &  \\ 
  4. awa1 & -0.003 & -0.017 & 0.015 & --- &  &  &  &  &  &  \\ 
  5. awa2 & 0.043 & 0.052 & 0.03 & 0.09 & --- &  &  &  &  &  \\ 
  6. awa3 & -0.054 & -0.075 & 0.008 & \textbf{-0.151} & \textbf{-0.223} & --- &  &  &  &  \\ 
  7. coll1 & \textbf{-0.151} & \textbf{-0.163} & 0.082 & \textbf{-0.248} & \textbf{-0.214} & 0.079 & --- &  &  &  \\ 
  8. coll2 & -0.084 & \textbf{-0.129} & \textbf{0.151} & -0.041 & -0.032 & 0.078 & 0.039 & --- &  &  \\ 
  9. coll3 & -0.053 & \textbf{-0.118} & \textbf{0.133} & \textbf{-0.152} & \textbf{-0.111} & \textbf{0.129} & -0.003 & 0.005 & --- &  \\ 
  10. coll4 & -0.007 & -0.088 & \textbf{0.138} & -0.049 & -0.076 & \textbf{0.221} & 0.008 & -0.043 & 0 & --- \\ 
   \bottomrule
\multicolumn{11}{c}{\emph{Note:} Correlation residuals in absolute $> 0.1$ are marked}\\
\end{tabular}
\endgroup
}

\subfloat[Covariance residuals]{%
\label{tab:residualsBwlsstd}
\centering
\begingroup\footnotesize
\begin{tabular}{rllllllllll}
  \toprule
 & 1 & 2 & 3 & 4 & 5 & 6 & 7 & 8 & 9 & 10 \\ 
  \midrule
1. ctrl1 & --- &  &  &  &  &  &  &  &  &  \\ 
  2. ctrl2 & 0.123 & --- &  &  &  &  &  &  &  &  \\ 
  3. ctrl3 & -0.151 & -0.111 & --- &  &  &  &  &  &  &  \\ 
  4. awa1 & -0.003 & -0.017 & 0.015 & --- &  &  &  &  &  &  \\ 
  5. awa2 & 0.043 & 0.052 & 0.03 & 0.09 & --- &  &  &  &  &  \\ 
  6. awa3 & -0.054 & -0.075 & 0.008 & -0.151 & -0.223 & --- &  &  &  &  \\ 
  7. coll1 & -0.151 & -0.163 & 0.082 & -0.248 & -0.214 & 0.079 & --- &  &  &  \\ 
  8. coll2 & -0.084 & -0.129 & 0.151 & -0.041 & -0.032 & 0.078 & 0.039 & --- &  &  \\ 
  9. coll3 & -0.053 & -0.118 & 0.133 & -0.152 & -0.111 & 0.129 & -0.003 & 0.005 & --- &  \\ 
  10. coll4 & -0.007 & -0.088 & 0.138 & -0.049 & -0.076 & 0.221 & 0.008 & -0.043 & 0 & --- \\ 
   \bottomrule
\multicolumn{11}{c}{\emph{Note:} Standardized residuals are not available for estimator WLSMVS}\\
\end{tabular}
\endgroup
}
\end{table*}
}
\newcommand{\residualsBasIUIPC}{
\begin{table*}[tbp]
\centering\caption{Residuals of the ML-estimated CFA of IUIPC-10 on Sample \textsf{B} (as done in IUIPC itself, without outlier treatment)}
\label{tab:residualsBasIUIPC}
\captionsetup{position=top}
\subfloat[Correlation residuals]{
\label{tab:residualsBasIUIPCcor}
\centering
\begingroup\footnotesize
\begin{tabular}{rllllllllll}
  \toprule
 & 1 & 2 & 3 & 4 & 5 & 6 & 7 & 8 & 9 & 10 \\ 
  \midrule
1. ctrl1 & --- &  &  &  &  &  &  &  &  &  \\ 
  2. ctrl2 & 0.013 & --- &  &  &  &  &  &  &  &  \\ 
  3. ctrl3 & -0.037 & -0.023 & --- &  &  &  &  &  &  &  \\ 
  4. awa1 & -0.047 & -0.039 & \textbf{0.155} & --- &  &  &  &  &  &  \\ 
  5. awa2 & -0.023 & -0.003 & \textbf{0.179} & 0.013 & --- &  &  &  &  &  \\ 
  6. awa3 & 0.08 & 0.031 & \textbf{0.185} & -0.009 & -0.047 & --- &  &  &  &  \\ 
  7. coll1 & -0.089 & -0.087 & \textbf{0.211} & \textbf{-0.136} & -0.099 & \textbf{0.226} & --- &  &  &  \\ 
  8. coll2 & -0.024 & -0.041 & \textbf{0.249} & 0.062 & 0.067 & \textbf{0.182} & 0.034 & --- &  &  \\ 
  9. coll3 & 0.011 & -0.049 & \textbf{0.25} & -0.083 & -0.042 & \textbf{0.284} & -0.003 & 0.001 & --- &  \\ 
  10. coll4 & 0.054 & -0.007 & \textbf{0.26} & 0.061 & 0.029 & \textbf{0.374} & -0.001 & -0.033 & 0.005 & --- \\ 
   \bottomrule
\multicolumn{11}{c}{\emph{Note:} Correlation residuals in absolute $> 0.1$ are marked}\\
\end{tabular}
\endgroup
}

\subfloat[Standardized residuals]{%
\label{tab:residualsBasIUIPCstd}
\centering
\begingroup\footnotesize
\begin{tabular}{rllllllllll}
  \toprule
 & 1 & 2 & 3 & 4 & 5 & 6 & 7 & 8 & 9 & 10 \\ 
  \midrule
1. ctrl1 & --- &  &  &  &  &  &  &  &  &  \\ 
  2. ctrl2 & \textbf{5.149} & --- &  &  &  &  &  &  &  &  \\ 
  3. ctrl3 & \textbf{-2.21} & -1.819 & --- &  &  &  &  &  &  &  \\ 
  4. awa1 & \textbf{-2.021} & -1.8 & \textbf{3.474} & --- &  &  &  &  &  &  \\ 
  5. awa2 & -1.082 & -0.155 & \textbf{4.071} & \textbf{5.177} & --- &  &  &  &  &  \\ 
  6. awa3 & 1.842 & 0.73 & \textbf{3.911} & -0.586 & \textbf{-3.71} & --- &  &  &  &  \\ 
  7. coll1 & \textbf{-2.632} & \textbf{-2.857} & \textbf{4.217} & \textbf{-4.399} & \textbf{-3.388} & \textbf{4.608} & --- &  &  &  \\ 
  8. coll2 & -0.645 & -1.17 & \textbf{4.992} & 1.744 & 1.927 & \textbf{3.919} & \textbf{2.065} & --- &  &  \\ 
  9. coll3 & 0.402 & \textbf{-2.202} & \textbf{5.12} & \textbf{-3.665} & \textbf{-2.086} & \textbf{5.905} & -0.801 & 0.078 & --- &  \\ 
  10. coll4 & 1.754 & -0.26 & \textbf{5.344} & \textbf{2.237} & 1.093 & \textbf{7.846} & -0.132 & \textbf{-2.91} & 1.505 & --- \\ 
   \bottomrule
\multicolumn{11}{c}{\emph{Note:} Statistically significant residuals ($\mathsf{abs} > 1.96$) are marked}\\
\end{tabular}
\endgroup
}
\end{table*}
}
\newcommand{\loadingsB}{
\begin{table*}[tb]
\centering
\caption{Factor loadings and their standardized solution of the MLM CFA of IUIPC-10 on Sample~\textsf{B}\explain{.\newline{}We find sub-par standardized loadings $\beta < .70$ for \textsf{ctrl3} and \textsf{awa3}, yielding a poor variance extracted $R^2 \leq .20$ and, thereby, low $\vari{AVE} < .50$ for \textsf{control} and \textsf{awareness}, indicating sub-par internal consistency. The equally sub-par construct reliability $\omega < .70$ yields a low signal-to-noise ratio less than $2$.}} 
\label{tab:loadingsB}
\begingroup\footnotesize
\begin{tabular}{lllrrrrrrrrrrrr}
  \toprule
 \multirow{2}{*}{Factor} & \multirow{2}{*}{Indicator} & \multicolumn{4}{c}{Factor Loading} & \multicolumn{4}{c}{Standardized Solution} & \multicolumn{5}{c}{Reliability} \\ 
 \cmidrule(lr){3-6} \cmidrule(lr){7-10} \cmidrule(lr){11-15} 
  & & $\lambda$ & $\vari{SE}_{\lambda}$ & $Z_{\lambda}$ & $p_{\lambda}$ & $\beta$ & $\vari{SE}_{\beta}$ & $Z_{\beta}$ & $p_{\beta}$ & $R^2$ & $\mathit{AVE}$ & $\alpha$ & $\omega$ & $\mathit{S/\!N}_\omega$ \\
  \midrule
 ctrl & ctrl1 & $1.00^+$ &  &  &  & 0.73 & 0.05 & 15.14 & $<.001$ & 0.54 & 0.40 & 0.62 & 0.66 & 1.92 \\ 
   & ctrl2 & $1.00\phantom{^+}$ & 0.11 & 8.76 & $<.001$ & 0.73 & 0.05 & 13.73 & $<.001$ & 0.53 &  &  &  &  \\ 
   & ctrl3 & $0.59\phantom{^+}$ & 0.11 & 5.36 & $<.001$ & 0.41 & 0.06 & 6.89 & $<.001$ & 0.17 &  &  &  &  \\ 
  aware & awa1 & $1.00^+$ &  &  &  & 0.74 & 0.05 & 15.36 & $<.001$ & 0.54 & 0.39 & 0.64 & 0.66 & 1.92 \\ 
   & awa2 & $1.13\phantom{^+}$ & 0.13 & 8.53 & $<.001$ & 0.81 & 0.04 & 18.45 & $<.001$ & 0.66 &  &  &  &  \\ 
   & awa3 & $0.90\phantom{^+}$ & 0.14 & 6.64 & $<.001$ & 0.44 & 0.05 & 8.83 & $<.001$ & 0.20 &  &  &  &  \\ 
  collect & coll1 & $1.00^+$ &  &  &  & 0.81 & 0.02 & 38.77 & $<.001$ & 0.66 & 0.72 & 0.91 & 0.91 & 10.13 \\ 
   & coll2 & $0.76\phantom{^+}$ & 0.05 & 14.86 & $<.001$ & 0.76 & 0.04 & 20.99 & $<.001$ & 0.58 &  &  &  &  \\ 
   & coll3 & $1.06\phantom{^+}$ & 0.04 & 23.63 & $<.001$ & 0.94 & 0.01 & 70.89 & $<.001$ & 0.88 &  &  &  &  \\ 
   & coll4 & $0.95\phantom{^+}$ & 0.05 & 18.14 & $<.001$ & 0.86 & 0.03 & 33.94 & $<.001$ & 0.74 &  &  &  &  \\ 
  iuipc & collect & $0.41\phantom{^+}$ & 0.08 & 5.42 & $<.001$ & 0.37 & 0.07 & 5.57 & $<.001$ & 0.14 &  &  &  &  \\ 
   & ctrl & $0.42\phantom{^+}$ & 0.07 & 6.05 & $<.001$ & 0.61 & 0.09 & 6.47 & $<.001$ & 0.38 &  &  &  &  \\ 
   & aware & $0.36\phantom{^+}$ & 0.06 & 6.40 & $<.001$ & 0.89 & 0.11 & 8.06 & $<.001$ & 0.79 &  &  &  &  \\ 
   \bottomrule
\multicolumn{14}{c}{\emph{Note:} $^+$ fixed parameter; the standardized solution is STDALL}\\
\end{tabular}
\endgroup
\end{table*}
}
\newcommand{\fitmeasuresBredux}{
\begin{table*}[ht]
\centering
\caption{Fit measures of the improved MLM CFA model on Sample \textsf{B}} 
\label{tab:fitmeasuresBredux}
\begingroup\footnotesize
\begin{tabular}{rrrrrrrrrr}
  \toprule
$\chi^2$ & $\mathit{df}$ & $p_{\chi^2}$ & \textsf{CFI} & \textsf{TFI} & \textsf{RMSEA} & \textsf{LL} & \textsf{UL} & $p_{\mathsf{rmsea}}$ & \textsf{SRMR} \\ 
  \midrule
54.86 & 17.00 & 0.00 & 0.97 & 0.96 & 0.08 & 0.06 & 0.10 & 0.02 & 0.03 \\ 
   \bottomrule
\end{tabular}
\endgroup
\end{table*}
}
\newcommand{\fitmeasuresBreduxwls}{
\begin{table*}[ht]
\centering
\caption{Fit measures of the improved WLSMVS CFA model on Sample \textsf{B}} 
\label{tab:fitmeasuresBreduxwls}
\begingroup\footnotesize
\begin{tabular}{rrrrrrrrrr}
  \toprule
$\chi^2$ & $\mathit{df}$ & $p_{\chi^2}$ & \textsf{CFI} & \textsf{TFI} & \textsf{RMSEA} & \textsf{LL} & \textsf{UL} & $p_{\mathsf{rmsea}}$ & \textsf{SRMR} \\ 
  \midrule
22.77 & 17.00 & 0.16 & 1.00 & 1.00 & 0.03 & 0.00 & 0.06 & 0.85 & 0.04 \\ 
   \bottomrule
\end{tabular}
\endgroup
\end{table*}
}
\newcommand{\corfitBredux}{
\begin{table}[ht]
\centering
\caption{Implied correlations of the improved CFA model on Sample \textsf{B}} 
\label{tab:corfitBredux}
\begingroup\footnotesize
\begin{tabular}{rllll}
  \toprule
 & 1 & 2 & 3 & 4 \\ 
  \midrule
1. ctrl & 0.747 &  &  &  \\ 
  2. aware & 0.477 & 0.803 &  &  \\ 
  3. collect & 0.171 & 0.265 & 0.848 &  \\ 
  4. iuipc & 0.556 & 0.859 & 0.308 & 0.825 \\ 
   \bottomrule
\multicolumn{5}{c}{\emph{Note:} The diagonal contains the $\sqrt{\mathit{AVE}}$}\\
\end{tabular}
\endgroup
\end{table}
}
\newcommand{\residualsBredux}{
\begin{table*}[p]
\centering\caption{Residuals of the MLM CFA of IUIPC-8 on Sample \textsf{B}}
\label{tab:residualsBredux}
\captionsetup{position=top}
\subfloat[Correlation residuals]{
\label{tab:residualsBreduxcor}
\centering
\begingroup\footnotesize
\begin{tabular}{rllllllll}
  \toprule
 & 1 & 2 & 3 & 4 & 5 & 6 & 7 & 8 \\ 
  \midrule
1. ctrl1 & --- &  &  &  &  &  &  &  \\ 
  2. ctrl2 & 0 & --- &  &  &  &  &  &  \\ 
  3. awa1 & 0.004 & -0.011 & --- &  &  &  &  &  \\ 
  4. awa2 & -0.007 & 0.01 & 0 & --- &  &  &  &  \\ 
  5. coll1 & -0.052 & -0.053 & -0.099 & -0.086 & --- &  &  &  \\ 
  6. coll2 & 0.003 & -0.038 & 0.08 & 0.056 & 0.037 & --- &  &  \\ 
  7. coll3 & 0.035 & -0.032 & -0.009 & -0.002 & -0.008 & 0.002 & --- &  \\ 
  8. coll4 & 0.083 & -0.009 & 0.076 & 0.03 & 0.007 & -0.037 & 0.005 & --- \\ 
   \bottomrule
\multicolumn{9}{c}{\emph{Note:} Correlation residuals in absolute $> 0.1$ are marked}\\
\end{tabular}
\endgroup
}

\subfloat[Standardized residuals]{%
\label{tab:residualsBreduxstd}
\centering
\begingroup\footnotesize
\begin{tabular}{rllllllll}
  \toprule
 & 1 & 2 & 3 & 4 & 5 & 6 & 7 & 8 \\ 
  \midrule
1. ctrl1 & --- &  &  &  &  &  &  &  \\ 
  2. ctrl2 & 0 & --- &  &  &  &  &  &  \\ 
  3. awa1 & 0.191 & -0.501 & --- &  &  &  &  &  \\ 
  4. awa2 & -0.976 & 1.345 & 0 & --- &  &  &  &  \\ 
  5. coll1 & -1.482 & -1.668 & \textbf{-2.724} & \textbf{-3.134} & --- &  &  &  \\ 
  6. coll2 & 0.082 & -1.066 & 1.584 & 1.476 & \textbf{2.033} & --- &  &  \\ 
  7. coll3 & 1.589 & -1.245 & -0.254 & -0.155 & -1.653 & 0.349 & --- &  \\ 
  8. coll4 & \textbf{2.133} & -0.299 & \textbf{1.961} & 1.17 & 0.524 & \textbf{-2.531} & 1.58 & --- \\ 
   \bottomrule
\multicolumn{9}{c}{\emph{Note:} Statistically significant residuals ($\mathsf{abs} > 1.96$) are marked}\\
\end{tabular}
\endgroup
}
\end{table*}
}
\newcommand{\residualsBreduxml}{
\begin{table*}[p]
\centering\caption{Residuals of the ML CFA of IUIPC-8 on Sample \textsf{B}}
\label{tab:residualsBreduxml}
\captionsetup{position=top}
\subfloat[Correlation residuals]{
\label{tab:residualsBreduxmlcor}
\centering
\begingroup\footnotesize
\begin{tabular}{rllllllll}
  \toprule
 & 1 & 2 & 3 & 4 & 5 & 6 & 7 & 8 \\ 
  \midrule
1. ctrl1 & --- &  &  &  &  &  &  &  \\ 
  2. ctrl2 & 0 & --- &  &  &  &  &  &  \\ 
  3. awa1 & 0.004 & -0.011 & --- &  &  &  &  &  \\ 
  4. awa2 & -0.007 & 0.01 & 0 & --- &  &  &  &  \\ 
  5. coll1 & -0.052 & -0.053 & -0.099 & -0.086 & --- &  &  &  \\ 
  6. coll2 & 0.003 & -0.038 & 0.08 & 0.056 & 0.037 & --- &  &  \\ 
  7. coll3 & 0.035 & -0.032 & -0.009 & -0.002 & -0.008 & 0.002 & --- &  \\ 
  8. coll4 & 0.083 & -0.009 & 0.076 & 0.03 & 0.007 & -0.037 & 0.005 & --- \\ 
   \bottomrule
\multicolumn{9}{c}{\emph{Note:} Correlation residuals in absolute $> 0.1$ are marked}\\
\end{tabular}
\endgroup
}

\subfloat[Standardized residuals]{%
\label{tab:residualsBreduxmlstd}
\centering
\begingroup\footnotesize
\begin{tabular}{rllllllll}
  \toprule
 & 1 & 2 & 3 & 4 & 5 & 6 & 7 & 8 \\ 
  \midrule
1. ctrl1 & --- &  &  &  &  &  &  &  \\ 
  2. ctrl2 & 0 & --- &  &  &  &  &  &  \\ 
  3. awa1 & 0.2 & -0.494 & --- &  &  &  &  &  \\ 
  4. awa2 & -1.067 & 1.292 & 0 & --- &  &  &  &  \\ 
  5. coll1 & -1.555 & -1.529 & \textbf{-2.647} & \textbf{-3.1} & --- &  &  &  \\ 
  6. coll2 & 0.084 & -1.033 & \textbf{2.073} & 1.739 & \textbf{2.192} & --- &  &  \\ 
  7. coll3 & 1.458 & -1.215 & -0.288 & -0.175 & \textbf{-2.067} & 0.432 & --- &  \\ 
  8. coll4 & \textbf{2.747} & -0.284 & \textbf{2.283} & 1.227 & 0.668 & \textbf{-3.231} & 1.683 & --- \\ 
   \bottomrule
\multicolumn{9}{c}{\emph{Note:} Statistically significant residuals ($\mathsf{abs} > 1.96$) are marked}\\
\end{tabular}
\endgroup
}
\end{table*}
}
\newcommand{\residualsBreduxwls}{
\begin{table*}[p]
\centering\caption{Residuals of the WLSMVS  CFA of IUIPC-8 on Sample \textsf{B}}
\label{tab:residualsBreduxwls}
\captionsetup{position=top}
\subfloat[Correlation residuals]{
\label{tab:residualsBreduxwlscor}
\centering
\begingroup\footnotesize
\begin{tabular}{rllllllll}
  \toprule
 & 1 & 2 & 3 & 4 & 5 & 6 & 7 & 8 \\ 
  \midrule
1. ctrl1 & --- &  &  &  &  &  &  &  \\ 
  2. ctrl2 & 0 & --- &  &  &  &  &  &  \\ 
  3. awa1 & -0.017 & 0.003 & --- &  &  &  &  &  \\ 
  4. awa2 & -0.017 & 0.032 & 0 & --- &  &  &  &  \\ 
  5. coll1 & -0.071 & -0.068 & \textbf{-0.132} & \textbf{-0.115} & --- &  &  &  \\ 
  6. coll2 & -0.005 & -0.037 & 0.071 & 0.065 & 0.034 & --- &  &  \\ 
  7. coll3 & 0.04 & -0.009 & -0.019 & 0.003 & -0.012 & 0.003 & --- &  \\ 
  8. coll4 & 0.082 & 0.015 & 0.078 & 0.034 & 0.008 & -0.036 & 0.004 & --- \\ 
   \bottomrule
\multicolumn{9}{c}{\emph{Note:} Correlation residuals in absolute $> 0.1$ are marked}\\
\end{tabular}
\endgroup
}

\subfloat[Covariance residuals]{%
\label{tab:residualsBreduxwlsraw}
\centering
\begingroup\footnotesize
\begin{tabular}{rllllllll}
  \toprule
 & 1 & 2 & 3 & 4 & 5 & 6 & 7 & 8 \\ 
  \midrule
1. ctrl1 & --- &  &  &  &  &  &  &  \\ 
  2. ctrl2 & 0 & --- &  &  &  &  &  &  \\ 
  3. awa1 & -0.017 & 0.003 & --- &  &  &  &  &  \\ 
  4. awa2 & -0.017 & 0.032 & 0 & --- &  &  &  &  \\ 
  5. coll1 & -0.071 & -0.068 & -0.132 & -0.115 & --- &  &  &  \\ 
  6. coll2 & -0.005 & -0.037 & 0.071 & 0.065 & 0.034 & --- &  &  \\ 
  7. coll3 & 0.04 & -0.009 & -0.019 & 0.003 & -0.012 & 0.003 & --- &  \\ 
  8. coll4 & 0.082 & 0.015 & 0.078 & 0.034 & 0.008 & -0.036 & 0.004 & --- \\ 
   \bottomrule
\multicolumn{9}{c}{\emph{Note:} Standardized residuals are not available for estimator WLSMVS}\\
\end{tabular}
\endgroup
}
\end{table*}
}
\newcommand{\loadingsBredux}{
\begin{table*}[tbp]
\centering
\caption{Factor loadings and their standardized solution of the MLM CFA of IUIPC-8 on Sample~\textsf{B}\explain{.\newline{}The standardized factor loadings $\beta$ are largely greater than $.70$, yielding satisfactory $\vari{AVE} > .50$. The construct reliability $\omega > .70$ passes the threshold for control and awareness yielding moderate signal-to-noise ratios; the reliability of collection is excellent.}} 
\label{tab:loadingsBredux}
\begingroup\footnotesize
\begin{tabular}{lllrrrrrrrrrrrr}
  \toprule
 \multirow{2}{*}{Factor} & \multirow{2}{*}{Indicator} & \multicolumn{4}{c}{Factor Loading} & \multicolumn{4}{c}{Standardized Solution} & \multicolumn{5}{c}{Reliability} \\ 
 \cmidrule(lr){3-6} \cmidrule(lr){7-10} \cmidrule(lr){11-15} 
  & & $\lambda$ & $\vari{SE}_{\lambda}$ & $Z_{\lambda}$ & $p_{\lambda}$ & $\beta$ & $\vari{SE}_{\beta}$ & $Z_{\beta}$ & $p_{\beta}$ & $R^2$ & $\mathit{AVE}$ & $\alpha$ & $\omega$ & $\mathit{S/\!N}_\omega$ \\
  \midrule
 ctrl & ctrl1 & $1.00^+$ &  &  &  & 0.76 & 0.06 & 13.40 & $<.001$ & 0.57 & 0.56 & 0.72 & 0.72 & 2.52 \\ 
   & ctrl2 & $0.98\phantom{^+}$ & 0.14 & 7.01 & $<.001$ & 0.74 & 0.06 & 11.58 & $<.001$ & 0.54 &  &  &  &  \\ 
  aware & awa1 & $1.00^+$ &  &  &  & 0.69 & 0.05 & 13.45 & $<.001$ & 0.48 & 0.64 & 0.76 & 0.78 & 3.55 \\ 
   & awa2 & $1.33\phantom{^+}$ & 0.18 & 7.33 & $<.001$ & 0.90 & 0.05 & 17.34 & $<.001$ & 0.80 &  &  &  &  \\ 
  collect & coll1 & $1.00^+$ &  &  &  & 0.81 & 0.02 & 39.01 & $<.001$ & 0.66 & 0.72 & 0.91 & 0.91 & 10.13 \\ 
   & coll2 & $0.76\phantom{^+}$ & 0.05 & 14.83 & $<.001$ & 0.76 & 0.04 & 20.90 & $<.001$ & 0.58 &  &  &  &  \\ 
   & coll3 & $1.06\phantom{^+}$ & 0.04 & 23.67 & $<.001$ & 0.94 & 0.01 & 70.72 & $<.001$ & 0.88 &  &  &  &  \\ 
   & coll4 & $0.95\phantom{^+}$ & 0.05 & 18.15 & $<.001$ & 0.86 & 0.03 & 33.67 & $<.001$ & 0.74 &  &  &  &  \\ 
  iuipc & collect & $0.34\phantom{^+}$ & 0.08 & 4.28 & $<.001$ & 0.31 & 0.07 & 4.31 & $<.001$ & 0.09 &  &  &  &  \\ 
   & ctrl & $0.39\phantom{^+}$ & 0.08 & 4.90 & $<.001$ & 0.56 & 0.11 & 5.01 & $<.001$ & 0.31 &  &  &  &  \\ 
   & aware & $0.33\phantom{^+}$ & 0.07 & 4.99 & $<.001$ & 0.86 & 0.14 & 6.03 & $<.001$ & 0.74 &  &  &  &  \\ 
   \bottomrule
\multicolumn{14}{c}{\emph{Note:} $^+$ fixed parameter; the standardized solution is STDALL}\\
\end{tabular}
\endgroup
\end{table*}
}

\newcommand{\loadingsV}{
\begin{table*}[tbp]
\centering
\caption{Factor loadings and their standardized solution of the MLM CFA of IUIPC-10 on validation Sample~\textsf{V}} 
\label{tab:loadingsV}
\begingroup\footnotesize
\begin{tabular}{lllrrrrrrrrrrrr}
  \toprule
 \multirow{2}{*}{Factor} & \multirow{2}{*}{Indicator} & \multicolumn{4}{c}{Factor Loading} & \multicolumn{4}{c}{Standardized Solution} & \multicolumn{5}{c}{Reliability} \\ 
 \cmidrule(lr){3-6} \cmidrule(lr){7-10} \cmidrule(lr){11-15} 
  & & $\lambda$ & $\vari{SE}_{\lambda}$ & $Z_{\lambda}$ & $p_{\lambda}$ & $\beta$ & $\vari{SE}_{\beta}$ & $Z_{\beta}$ & $p_{\beta}$ & $R^2$ & $\mathit{AVE}$ & $\alpha$ & $\omega$ & $\mathit{S/\!N}_\omega$ \\
  \midrule
 ctrl & ctrl1 & $1.00^+$ &  &  &  & 0.68 & 0.06 & 11.19 & $<.001$ & 0.46 & 0.41 & 0.64 & 0.67 & 2.00 \\ 
   & ctrl2 & $1.08\phantom{^+}$ & 0.12 & 8.91 & $<.001$ & 0.75 & 0.05 & 14.79 & $<.001$ & 0.56 &  &  &  &  \\ 
   & ctrl3 & $0.71\phantom{^+}$ & 0.09 & 7.47 & $<.001$ & 0.47 & 0.05 & 9.90 & $<.001$ & 0.22 &  &  &  &  \\ 
  aware & awa1 & $1.00^+$ &  &  &  & 0.70 & 0.05 & 15.13 & $<.001$ & 0.49 & 0.33 & 0.56 & 0.59 & 1.45 \\ 
   & awa2 & $0.97\phantom{^+}$ & 0.10 & 9.58 & $<.001$ & 0.75 & 0.05 & 13.73 & $<.001$ & 0.57 &  &  &  &  \\ 
   & awa3 & $0.79\phantom{^+}$ & 0.13 & 6.25 & $<.001$ & 0.39 & 0.05 & 7.34 & $<.001$ & 0.15 &  &  &  &  \\ 
  collect & coll1 & $1.00^+$ &  &  &  & 0.76 & 0.03 & 25.47 & $<.001$ & 0.58 & 0.63 & 0.86 & 0.87 & 6.55 \\ 
   & coll2 & $0.62\phantom{^+}$ & 0.06 & 11.00 & $<.001$ & 0.64 & 0.05 & 13.07 & $<.001$ & 0.41 &  &  &  &  \\ 
   & coll3 & $1.04\phantom{^+}$ & 0.06 & 17.97 & $<.001$ & 0.92 & 0.02 & 51.13 & $<.001$ & 0.85 &  &  &  &  \\ 
   & coll4 & $0.89\phantom{^+}$ & 0.05 & 16.70 & $<.001$ & 0.80 & 0.03 & 30.70 & $<.001$ & 0.64 &  &  &  &  \\ 
  iuipc & collect & $0.45\phantom{^+}$ & 0.07 & 6.42 & $<.001$ & 0.43 & 0.06 & 6.81 & $<.001$ & 0.18 &  &  &  &  \\ 
   & ctrl & $0.46\phantom{^+}$ & 0.06 & 7.69 & $<.001$ & 0.67 & 0.07 & 9.61 & $<.001$ & 0.45 &  &  &  &  \\ 
   & aware & $0.45\phantom{^+}$ & 0.06 & 7.97 & $<.001$ & 0.93 & 0.10 & 9.40 & $<.001$ & 0.87 &  &  &  &  \\ 
   \bottomrule
\multicolumn{14}{c}{\emph{Note:} $^+$ fixed parameter; the standardized solution is STDALL}\\
\end{tabular}
\endgroup
\end{table*}
}
\newcommand{\loadingsVredux}{
\begin{table*}[tbp]
\centering
\caption{Factor loadings and their standardized solution of the MLM CFA of IUIPC-8 on validation Sample~\textsf{V}} 
\label{tab:loadingsVredux}
\begingroup\footnotesize
\begin{tabular}{lllrrrrrrrrrrrr}
  \toprule
 \multirow{2}{*}{Factor} & \multirow{2}{*}{Indicator} & \multicolumn{4}{c}{Factor Loading} & \multicolumn{4}{c}{Standardized Solution} & \multicolumn{5}{c}{Reliability} \\ 
 \cmidrule(lr){3-6} \cmidrule(lr){7-10} \cmidrule(lr){11-15} 
  & & $\lambda$ & $\vari{SE}_{\lambda}$ & $Z_{\lambda}$ & $p_{\lambda}$ & $\beta$ & $\vari{SE}_{\beta}$ & $Z_{\beta}$ & $p_{\beta}$ & $R^2$ & $\mathit{AVE}$ & $\alpha$ & $\omega$ & $\mathit{S/\!N}_\omega$ \\
  \midrule
 ctrl & ctrl1 & $1.00^+$ &  &  &  & 0.67 & 0.06 & 10.51 & $<.001$ & 0.45 & 0.54 & 0.70 & 0.70 & 2.33 \\ 
   & ctrl2 & $1.16\phantom{^+}$ & 0.16 & 7.18 & $<.001$ & 0.79 & 0.06 & 12.49 & $<.001$ & 0.63 &  &  &  &  \\ 
  aware & awa1 & $1.00^+$ &  &  &  & 0.72 & 0.04 & 16.16 & $<.001$ & 0.52 & 0.55 & 0.71 & 0.71 & 2.41 \\ 
   & awa2 & $0.95\phantom{^+}$ & 0.11 & 8.68 & $<.001$ & 0.76 & 0.06 & 12.93 & $<.001$ & 0.58 &  &  &  &  \\ 
  collect & coll1 & $1.00^+$ &  &  &  & 0.76 & 0.03 & 25.41 & $<.001$ & 0.58 & 0.63 & 0.86 & 0.87 & 6.54 \\ 
   & coll2 & $0.62\phantom{^+}$ & 0.06 & 10.96 & $<.001$ & 0.64 & 0.05 & 12.93 & $<.001$ & 0.40 &  &  &  &  \\ 
   & coll3 & $1.04\phantom{^+}$ & 0.06 & 17.98 & $<.001$ & 0.93 & 0.02 & 50.74 & $<.001$ & 0.86 &  &  &  &  \\ 
   & coll4 & $0.89\phantom{^+}$ & 0.05 & 16.62 & $<.001$ & 0.80 & 0.03 & 30.48 & $<.001$ & 0.64 &  &  &  &  \\ 
  iuipc & collect & $0.39\phantom{^+}$ & 0.07 & 5.25 & $<.001$ & 0.37 & 0.07 & 5.39 & $<.001$ & 0.14 &  &  &  &  \\ 
   & ctrl & $0.40\phantom{^+}$ & 0.07 & 5.74 & $<.001$ & 0.59 & 0.09 & 6.90 & $<.001$ & 0.35 &  &  &  &  \\ 
   & aware & $0.46\phantom{^+}$ & 0.07 & 6.45 & $<.001$ & 0.93 & 0.13 & 7.43 & $<.001$ & 0.87 &  &  &  &  \\ 
   \bottomrule
\multicolumn{14}{c}{\emph{Note:} $^+$ fixed parameter; the standardized solution is STDALL}\\
\end{tabular}
\endgroup
\end{table*}
}
\newcommand{\residualsVreduxwls}{
\begin{table*}[p]
\centering\caption{Residuals of the WLSMVS validation CFA of IUIPC-8 on Sample \textsf{V}}
\label{tab:residualsVreduxwls}
\captionsetup{position=top}
\subfloat[Correlation residuals]{
\label{tab:residualsVreduxcorwls}
\centering
\begingroup\footnotesize
\begin{tabular}{rllllllll}
  \toprule
 & 1 & 2 & 3 & 4 & 5 & 6 & 7 & 8 \\ 
  \midrule
1. ctrl1 & --- &  &  &  &  &  &  &  \\ 
  2. ctrl2 & 0 & --- &  &  &  &  &  &  \\ 
  3. awa1 & 0.036 & 0.026 & --- &  &  &  &  &  \\ 
  4. awa2 & -0.015 & -0.027 & 0 & --- &  &  &  &  \\ 
  5. coll1 & -0.069 & -0.035 & -0.064 & -0.001 & --- &  &  &  \\ 
  6. coll2 & 0.083 & 0.086 & -0.036 & 0.057 & 0.023 & --- &  &  \\ 
  7. coll3 & -0.002 & -0.021 & -0.029 & 0.002 & 0.005 & -0.019 & --- &  \\ 
  8. coll4 & -0.045 & 0.012 & 0.043 & 0.045 & -0.012 & -0.014 & 0.007 & --- \\ 
   \bottomrule
\multicolumn{9}{c}{\emph{Note:} Correlation residuals in absolute $> 0.1$ are marked}\\
\end{tabular}
\endgroup
}

\subfloat[Covariance residuals]{%
\label{tab:residualsVreduxrawwls}
\centering
\begingroup\footnotesize
\begin{tabular}{rllllllll}
  \toprule
 & 1 & 2 & 3 & 4 & 5 & 6 & 7 & 8 \\ 
  \midrule
1. ctrl1 & --- &  &  &  &  &  &  &  \\ 
  2. ctrl2 & 0 & --- &  &  &  &  &  &  \\ 
  3. awa1 & 0.036 & 0.026 & --- &  &  &  &  &  \\ 
  4. awa2 & -0.015 & -0.027 & 0 & --- &  &  &  &  \\ 
  5. coll1 & -0.069 & -0.035 & -0.064 & -0.001 & --- &  &  &  \\ 
  6. coll2 & 0.083 & 0.086 & -0.036 & 0.057 & 0.023 & --- &  &  \\ 
  7. coll3 & -0.002 & -0.021 & -0.029 & 0.002 & 0.005 & -0.019 & --- &  \\ 
  8. coll4 & -0.045 & 0.012 & 0.043 & 0.045 & -0.012 & -0.014 & 0.007 & --- \\ 
   \bottomrule
\multicolumn{9}{c}{\emph{Note:} Standardized residuals are not available for estimator WLSMVS}\\
\end{tabular}
\endgroup
}
\end{table*}
}
\newcommand{\residualsVredux}{
\begin{table*}[p]
\centering\caption{Residuals of the MLM validation CFA of IUIPC-8 on Sample \textsf{V}}
\label{tab:residualsVredux}
\captionsetup{position=top}
\subfloat[Correlation residuals]{
\label{tab:residualsVreduxcor}
\centering
\begingroup\footnotesize
\begin{tabular}{rllllllll}
  \toprule
 & 1 & 2 & 3 & 4 & 5 & 6 & 7 & 8 \\ 
  \midrule
1. ctrl1 & --- &  &  &  &  &  &  &  \\ 
  2. ctrl2 & 0 & --- &  &  &  &  &  &  \\ 
  3. awa1 & 0.012 & 0.026 & --- &  &  &  &  &  \\ 
  4. awa2 & -0.009 & -0.022 & 0 & --- &  &  &  &  \\ 
  5. coll1 & -0.046 & -0.006 & -0.068 & 0.014 & --- &  &  &  \\ 
  6. coll2 & 0.082 & \textbf{0.108} & -0.031 & 0.051 & 0.037 & --- &  &  \\ 
  7. coll3 & 0.009 & -0.017 & -0.047 & 0.022 & 0.001 & -0.008 & --- &  \\ 
  8. coll4 & -0.042 & 0.011 & 0.028 & 0.048 & -0.016 & -0.014 & 0.005 & --- \\ 
   \bottomrule
\multicolumn{9}{c}{\emph{Note:} Correlation residuals in absolute $> 0.1$ are marked}\\
\end{tabular}
\endgroup
}

\subfloat[Standardized residuals]{%
\label{tab:residualsVreduxstd}
\centering
\begingroup\footnotesize
\begin{tabular}{rllllllll}
  \toprule
 & 1 & 2 & 3 & 4 & 5 & 6 & 7 & 8 \\ 
  \midrule
1. ctrl1 & --- &  &  &  &  &  &  &  \\ 
  2. ctrl2 & 0 & --- &  &  &  &  &  &  \\ 
  3. awa1 & 0.579 & 1.623 & --- &  &  &  &  &  \\ 
  4. awa2 & -0.524 & -1.648 & 0 & --- &  &  &  &  \\ 
  5. coll1 & -1.196 & -0.138 & -1.719 & 0.441 & --- &  &  &  \\ 
  6. coll2 & \textbf{2.041} & 1.895 & -0.681 & 1.371 & 1.003 & --- &  &  \\ 
  7. coll3 & 0.287 & -0.775 & -1.761 & 0.94 & 0.129 & -0.778 & --- &  \\ 
  8. coll4 & -1.179 & 0.332 & 0.745 & 1.634 & -0.887 & -0.552 & 0.928 & --- \\ 
   \bottomrule
\multicolumn{9}{c}{\emph{Note:} Statistically significant residuals ($\mathsf{abs} > 1.96$) are marked}\\
\end{tabular}
\endgroup
}
\end{table*}
}
\newcommand{\residualsVreduxml}{
\begin{table*}[p]
\centering\caption{Residuals of the ML validation CFA of IUIPC-8 on Sample \textsf{V}}
\label{tab:residualsVreduxml}
\captionsetup{position=top}
\subfloat[Correlation residuals]{
\label{tab:residualsVreduxmlcor}
\centering
\begingroup\footnotesize
\begin{tabular}{rllllllll}
  \toprule
 & 1 & 2 & 3 & 4 & 5 & 6 & 7 & 8 \\ 
  \midrule
1. ctrl1 & --- &  &  &  &  &  &  &  \\ 
  2. ctrl2 & 0 & --- &  &  &  &  &  &  \\ 
  3. awa1 & 0.012 & 0.026 & --- &  &  &  &  &  \\ 
  4. awa2 & -0.009 & -0.022 & 0 & --- &  &  &  &  \\ 
  5. coll1 & -0.046 & -0.006 & -0.068 & 0.014 & --- &  &  &  \\ 
  6. coll2 & 0.082 & \textbf{0.108} & -0.031 & 0.051 & 0.037 & --- &  &  \\ 
  7. coll3 & 0.009 & -0.017 & -0.047 & 0.022 & 0.001 & -0.008 & --- &  \\ 
  8. coll4 & -0.042 & 0.011 & 0.028 & 0.048 & -0.016 & -0.014 & 0.005 & --- \\ 
   \bottomrule
\multicolumn{9}{c}{\emph{Note:} Correlation residuals in absolute $> 0.1$ are marked}\\
\end{tabular}
\endgroup
}

\subfloat[Standardized residuals]{%
\label{tab:residualsVreduxmlstd}
\centering
\begingroup\footnotesize
\begin{tabular}{rllllllll}
  \toprule
 & 1 & 2 & 3 & 4 & 5 & 6 & 7 & 8 \\ 
  \midrule
1. ctrl1 & --- &  &  &  &  &  &  &  \\ 
  2. ctrl2 & 0 & --- &  &  &  &  &  &  \\ 
  3. awa1 & 0.635 & 1.745 & --- &  &  &  &  &  \\ 
  4. awa2 & -0.576 & -1.817 & 0 & --- &  &  &  &  \\ 
  5. coll1 & -1.277 & -0.19 & \textbf{-2.084} & 0.462 & --- &  &  &  \\ 
  6. coll2 & \textbf{2.039} & \textbf{2.871} & -0.834 & 1.415 & 1.762 & --- &  &  \\ 
  7. coll3 & 0.302 & -0.907 & \textbf{-2.085} & 1.13 & 0.194 & -1.251 & --- &  \\ 
  8. coll4 & -1.216 & 0.37 & 0.912 & 1.654 & -1.398 & -0.804 & 1.609 & --- \\ 
   \bottomrule
\multicolumn{9}{c}{\emph{Note:} Statistically significant residuals ($\mathsf{abs} > 1.96$) are marked}\\
\end{tabular}
\endgroup
}
\end{table*}
}
\newcommand{\residualsV}{
\begin{table*}[p]
\centering\caption{Residuals of the MLM validation CFA of IUIPC-10 on Sample \textsf{V}}
\label{tab:residualsV}
\captionsetup{position=top}
\subfloat[Correlation residuals]{
\label{tab:residualsVcor}
\centering
\begingroup\footnotesize
\begin{tabular}{rllllllllll}
  \toprule
 & 1 & 2 & 3 & 4 & 5 & 6 & 7 & 8 & 9 & 10 \\ 
  \midrule
1. ctrl1 & --- &  &  &  &  &  &  &  &  &  \\ 
  2. ctrl2 & 0.03 & --- &  &  &  &  &  &  &  &  \\ 
  3. ctrl3 & -0.029 & -0.041 & --- &  &  &  &  &  &  &  \\ 
  4. awa1 & -0.018 & 0.013 & \textbf{0.115} & --- &  &  &  &  &  &  \\ 
  5. awa2 & -0.046 & -0.041 & 0.071 & 0.023 & --- &  &  &  &  &  \\ 
  6. awa3 & -0.01 & 0.051 & 0.062 & -0.062 & -0.019 & --- &  &  &  &  \\ 
  7. coll1 & -0.082 & -0.036 & \textbf{0.153} & -0.092 & -0.015 & \textbf{0.177} & --- &  &  &  \\ 
  8. coll2 & 0.053 & 0.082 & \textbf{0.187} & -0.051 & 0.026 & \textbf{0.267} & 0.036 & --- &  &  \\ 
  9. coll3 & -0.034 & -0.054 & \textbf{0.173} & -0.075 & -0.013 & \textbf{0.196} & 0.001 & -0.009 & --- &  \\ 
  10. coll4 & -0.079 & -0.021 & \textbf{0.199} & 0.003 & 0.018 & \textbf{0.17} & -0.017 & -0.016 & 0.006 & --- \\ 
   \bottomrule
\multicolumn{11}{c}{\emph{Note:} Correlation residuals in absolute $> 0.1$ are marked}\\
\end{tabular}
\endgroup
}

\subfloat[Standardized residuals]{%
\label{tab:residualsVstd}
\centering
\begingroup\footnotesize
\begin{tabular}{rllllllllll}
  \toprule
 & 1 & 2 & 3 & 4 & 5 & 6 & 7 & 8 & 9 & 10 \\ 
  \midrule
1. ctrl1 & --- &  &  &  &  &  &  &  &  &  \\ 
  2. ctrl2 & \textbf{3.976} & --- &  &  &  &  &  &  &  &  \\ 
  3. ctrl3 & -1.571 & \textbf{-3.442} & --- &  &  &  &  &  &  &  \\ 
  4. awa1 & -0.69 & 0.514 & \textbf{3.117} & --- &  &  &  &  &  &  \\ 
  5. awa2 & -1.906 & -1.868 & 1.868 & \textbf{3.585} & --- &  &  &  &  &  \\ 
  6. awa3 & -0.262 & 0.925 & 1.449 & \textbf{-2.712} & -1.069 & --- &  &  &  &  \\ 
  7. coll1 & \textbf{-2.166} & -0.842 & \textbf{3.426} & \textbf{-2.297} & -0.467 & \textbf{2.963} & --- &  &  &  \\ 
  8. coll2 & 1.315 & 1.432 & \textbf{4.211} & -1.117 & 0.704 & \textbf{3.325} & 0.976 & --- &  &  \\ 
  9. coll3 & -1.217 & \textbf{-2.179} & \textbf{4.163} & \textbf{-2.835} & -0.574 & \textbf{3.148} & 0.24 & -0.839 & --- &  \\ 
  10. coll4 & \textbf{-2.303} & -0.633 & \textbf{4.52} & 0.089 & 0.62 & \textbf{2.992} & -0.901 & -0.624 & 0.956 & --- \\ 
   \bottomrule
\multicolumn{11}{c}{\emph{Note:} Statistically significant residuals ($\mathsf{abs} > 1.96$) are marked}\\
\end{tabular}
\endgroup
}
\end{table*}
}
\newcommand{\residualsVml}{
\begin{table*}[p]
\centering\caption{Residuals of the ML validation CFA of IUIPC-10 on Sample \textsf{V}}
\label{tab:residualsVml}
\captionsetup{position=top}
\subfloat[Correlation residuals]{
\label{tab:residualsVmlcor}
\centering
\begingroup\footnotesize
\begin{tabular}{rllllllllll}
  \toprule
 & 1 & 2 & 3 & 4 & 5 & 6 & 7 & 8 & 9 & 10 \\ 
  \midrule
1. ctrl1 & --- &  &  &  &  &  &  &  &  &  \\ 
  2. ctrl2 & 0.03 & --- &  &  &  &  &  &  &  &  \\ 
  3. ctrl3 & -0.029 & -0.041 & --- &  &  &  &  &  &  &  \\ 
  4. awa1 & -0.018 & 0.013 & \textbf{0.115} & --- &  &  &  &  &  &  \\ 
  5. awa2 & -0.046 & -0.041 & 0.071 & 0.023 & --- &  &  &  &  &  \\ 
  6. awa3 & -0.01 & 0.051 & 0.062 & -0.062 & -0.019 & --- &  &  &  &  \\ 
  7. coll1 & -0.082 & -0.036 & \textbf{0.153} & -0.092 & -0.015 & \textbf{0.177} & --- &  &  &  \\ 
  8. coll2 & 0.053 & 0.082 & \textbf{0.187} & -0.051 & 0.026 & \textbf{0.267} & 0.036 & --- &  &  \\ 
  9. coll3 & -0.034 & -0.054 & \textbf{0.173} & -0.075 & -0.013 & \textbf{0.196} & 0.001 & -0.009 & --- &  \\ 
  10. coll4 & -0.079 & -0.021 & \textbf{0.199} & 0.003 & 0.018 & \textbf{0.17} & -0.017 & -0.016 & 0.006 & --- \\ 
   \bottomrule
\multicolumn{11}{c}{\emph{Note:} Correlation residuals in absolute $> 0.1$ are marked}\\
\end{tabular}
\endgroup
}

\subfloat[Standardized residuals]{%
\label{tab:residualsVmlstd}
\centering
\begingroup\footnotesize
\begin{tabular}{rllllllllll}
  \toprule
 & 1 & 2 & 3 & 4 & 5 & 6 & 7 & 8 & 9 & 10 \\ 
  \midrule
1. ctrl1 & --- &  &  &  &  &  &  &  &  &  \\ 
  2. ctrl2 & \textbf{4.171} & --- &  &  &  &  &  &  &  &  \\ 
  3. ctrl3 & -1.483 & \textbf{-2.89} & --- &  &  &  &  &  &  &  \\ 
  4. awa1 & -0.699 & 0.582 & \textbf{3.193} & --- &  &  &  &  &  &  \\ 
  5. awa2 & \textbf{-2.001} & \textbf{-2.142} & \textbf{1.991} & \textbf{3.945} & --- &  &  &  &  &  \\ 
  6. awa3 & -0.264 & 1.344 & 1.492 & \textbf{-2.926} & -1.091 & --- &  &  &  &  \\ 
  7. coll1 & \textbf{-2.308} & -1.122 & \textbf{3.492} & \textbf{-2.778} & -0.488 & \textbf{4.002} & --- &  &  &  \\ 
  8. coll2 & 1.326 & \textbf{2.154} & \textbf{4.256} & -1.369 & 0.727 & \textbf{5.944} & 1.697 & --- &  &  \\ 
  9. coll3 & -1.282 & \textbf{-2.428} & \textbf{4.139} & \textbf{-3.224} & -0.67 & \textbf{4.605} & 0.36 & -1.352 & --- &  \\ 
  10. coll4 & \textbf{-2.347} & -0.701 & \textbf{4.589} & 0.108 & 0.613 & \textbf{3.971} & -1.424 & -0.908 & 1.68 & --- \\ 
   \bottomrule
\multicolumn{11}{c}{\emph{Note:} Statistically significant residuals ($\mathsf{abs} > 1.96$) are marked}\\
\end{tabular}
\endgroup
}
\end{table*}
}
\newcommand{\residualsVwls}{
\begin{table*}[p]
\centering\caption{Residuals of the WLSMVS validation CFA of IUIPC-10 on Sample \textsf{V}}
\label{tab:residualsVwls}
\captionsetup{position=top}
\subfloat[Correlation residuals]{
\label{tab:residualsVwlscor}
\centering
\begingroup\footnotesize
\begin{tabular}{rllllllllll}
  \toprule
 & 1 & 2 & 3 & 4 & 5 & 6 & 7 & 8 & 9 & 10 \\ 
  \midrule
1. ctrl1 & --- &  &  &  &  &  &  &  &  &  \\ 
  2. ctrl2 & 0.066 & --- &  &  &  &  &  &  &  &  \\ 
  3. ctrl3 & \textbf{-0.108} & \textbf{-0.155} & --- &  &  &  &  &  &  &  \\ 
  4. awa1 & 0.024 & 0.023 & 0.086 & --- &  &  &  &  &  &  \\ 
  5. awa2 & -0.021 & -0.022 & 0.021 & 0.055 & --- &  &  &  &  &  \\ 
  6. awa3 & -0.067 & -0.027 & -0.028 & \textbf{-0.104} & -0.068 & --- &  &  &  &  \\ 
  7. coll1 & \textbf{-0.137} & \textbf{-0.105} & 0.092 & \textbf{-0.123} & -0.06 & 0.098 & --- &  &  &  \\ 
  8. coll2 & 0.02 & 0.021 & \textbf{0.142} & -0.092 & -0.001 & \textbf{0.178} & 0.017 & --- &  &  \\ 
  9. coll3 & -0.082 & \textbf{-0.104} & 0.097 & -0.097 & -0.068 & \textbf{0.103} & 0.011 & -0.024 & --- &  \\ 
  10. coll4 & \textbf{-0.116} & -0.062 & \textbf{0.139} & -0.02 & -0.019 & 0.084 & -0.012 & -0.023 & 0.01 & --- \\ 
   \bottomrule
\multicolumn{11}{c}{\emph{Note:} Correlation residuals in absolute $> 0.1$ are marked}\\
\end{tabular}
\endgroup
}

\subfloat[Covariance residuals]{%
\label{tab:residualsVwlsraw}
\centering
\begingroup\footnotesize
\begin{tabular}{rllllllllll}
  \toprule
 & 1 & 2 & 3 & 4 & 5 & 6 & 7 & 8 & 9 & 10 \\ 
  \midrule
1. ctrl1 & --- &  &  &  &  &  &  &  &  &  \\ 
  2. ctrl2 & 0.066 & --- &  &  &  &  &  &  &  &  \\ 
  3. ctrl3 & -0.108 & -0.155 & --- &  &  &  &  &  &  &  \\ 
  4. awa1 & 0.024 & 0.023 & 0.086 & --- &  &  &  &  &  &  \\ 
  5. awa2 & -0.021 & -0.022 & 0.021 & 0.055 & --- &  &  &  &  &  \\ 
  6. awa3 & -0.067 & -0.027 & -0.028 & -0.104 & -0.068 & --- &  &  &  &  \\ 
  7. coll1 & -0.137 & -0.105 & 0.092 & -0.123 & -0.06 & 0.098 & --- &  &  &  \\ 
  8. coll2 & 0.02 & 0.021 & 0.142 & -0.092 & -0.001 & 0.178 & 0.017 & --- &  &  \\ 
  9. coll3 & -0.082 & -0.104 & 0.097 & -0.097 & -0.068 & 0.103 & 0.011 & -0.024 & --- &  \\ 
  10. coll4 & -0.116 & -0.062 & 0.139 & -0.02 & -0.019 & 0.084 & -0.012 & -0.023 & 0.01 & --- \\ 
   \bottomrule
\multicolumn{11}{c}{\emph{Note:} Standardized residuals are not available for estimator WLSMVS}\\
\end{tabular}
\endgroup
}
\end{table*}
}
\newcommand{\corfitV}{
\begin{table}[ht]
\centering
\caption{Implied correlations of the CFA model on Sample \textsf{V}} 
\label{tab:corfitV}
\begingroup\footnotesize
\begin{tabular}{rllll}
  \toprule
 & 1 & 2 & 3 & 4 \\ 
  \midrule
1. ctrl & 0.638 &  &  &  \\ 
  2. aware & 0.625 & 0.572 &  &  \\ 
  3. collect & 0.285 & 0.399 & 0.793 &  \\ 
  4. iuipc & 0.668 & 0.935 & 0.427 & 0.717 \\ 
   \bottomrule
\multicolumn{5}{c}{\emph{Note:} The diagonal contains the $\sqrt{\mathit{AVE}}$}\\
\end{tabular}
\endgroup
\end{table}
}
\newcommand{\corfitVredux}{
\begin{table}[ht]
\centering
\caption{Implied correlations of the improved CFA model on Sample \textsf{V}} 
\label{tab:corfitVredux}
\begingroup\footnotesize
\begin{tabular}{rllll}
  \toprule
 & 1 & 2 & 3 & 4 \\ 
  \midrule
1. ctrl & 0.735 &  &  &  \\ 
  2. aware & 0.547 & 0.74 &  &  \\ 
  3. collect & 0.218 & 0.344 & 0.793 &  \\ 
  4. iuipc & 0.588 & 0.93 & 0.37 & 0.774 \\ 
   \bottomrule
\multicolumn{5}{c}{\emph{Note:} The diagonal contains the $\sqrt{\mathit{AVE}}$}\\
\end{tabular}
\endgroup
\end{table}
}

\newcommand{\instumentReliabilityTable}{
\begin{table*}[ht]
\centering
\caption{Item/construct reliability metrics of IUIPC-10 on Sample \textsf{B}} 
\label{tab:instruments.reliability}
\begingroup\footnotesize
\begin{tabular}{lrrrrrrlrr}
  \toprule
Construct & $\alpha$ & $\mathsf{LL}(\alpha)$ & $\mathsf{UL}(\alpha)$ & $\lambda_{6}$ & $\mathsf{avg}(r)$ & $r_{\mathsf{i/w}}$ & Bartlett $p$ & \textsf{KMO} & \textsf{MSA} \\ 
  \midrule
Control & 0.63 & 0.57 & 0.68 & 0.56 & 0.36 &  & $<.001$ & 0.59 &  \\ 
  \qquad ctrl1 &  &  &  &  &  & 0.79 &  &  & 0.56 \\ 
  \qquad ctrl2 &  &  &  &  &  & 0.80 &  &  & 0.56 \\ 
  \qquad ctrl3 &  &  &  &  &  & 0.69 &  &  & 0.76 \\ 
  Awareness & 0.68 & 0.64 & 0.73 & 0.62 & 0.42 &  & $<.001$ & 0.60 &  \\ 
  \qquad awa1 &  &  &  &  &  & 0.77 &  &  & 0.57 \\ 
  \qquad awa2 &  &  &  &  &  & 0.77 &  &  & 0.57 \\ 
  \qquad awa3 &  &  &  &  &  & 0.78 &  &  & 0.79 \\ 
  Collection & 0.91 & 0.89 & 0.92 & 0.89 & 0.71 &  & $<.001$ & 0.83 &  \\ 
  \qquad coll1 &  &  &  &  &  & 0.89 &  &  & 0.88 \\ 
  \qquad coll2 &  &  &  &  &  & 0.83 &  &  & 0.88 \\ 
  \qquad coll3 &  &  &  &  &  & 0.93 &  &  & 0.77 \\ 
  \qquad coll4 &  &  &  &  &  & 0.89 &  &  & 0.81 \\ 
   \bottomrule
\end{tabular}
\endgroup
\end{table*}
}

\newcommand{\pathPlotCFAB}{
\begin{figure*}[p]
\centering
\vspace{0cm}

\includegraphics[width=\maxwidth]{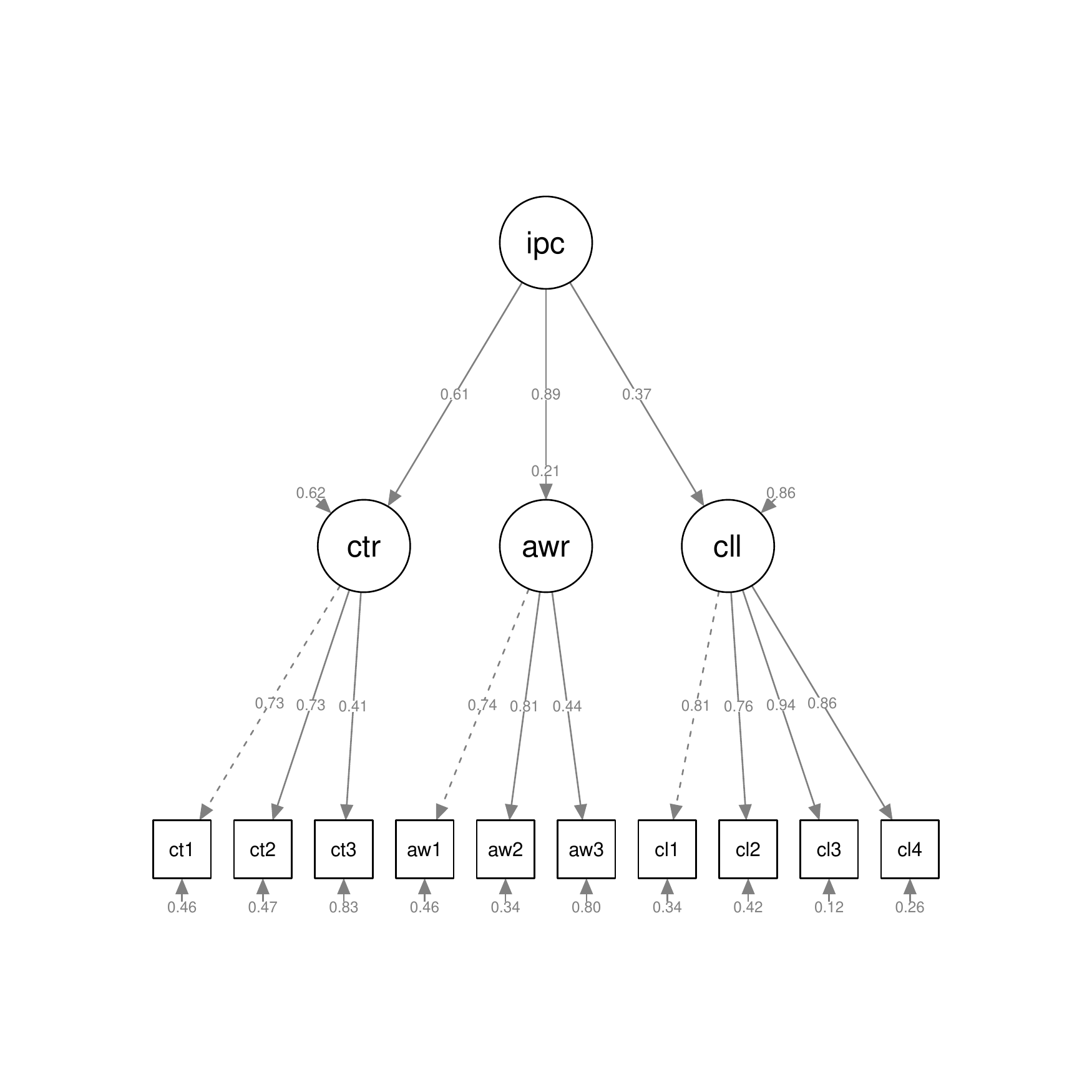} 
\vspace{0cm}
\caption{CFA paths plot with standardized estimates of IUIPC-10 on Sample \textsf{B}}
\label{fig:pathPlotCFAB}
\end{figure*}
}
\newcommand{\pathPlotCFABredux}{
\begin{figure*}[p]
\centering
\vspace{0cm}

\includegraphics[width=\maxwidth]{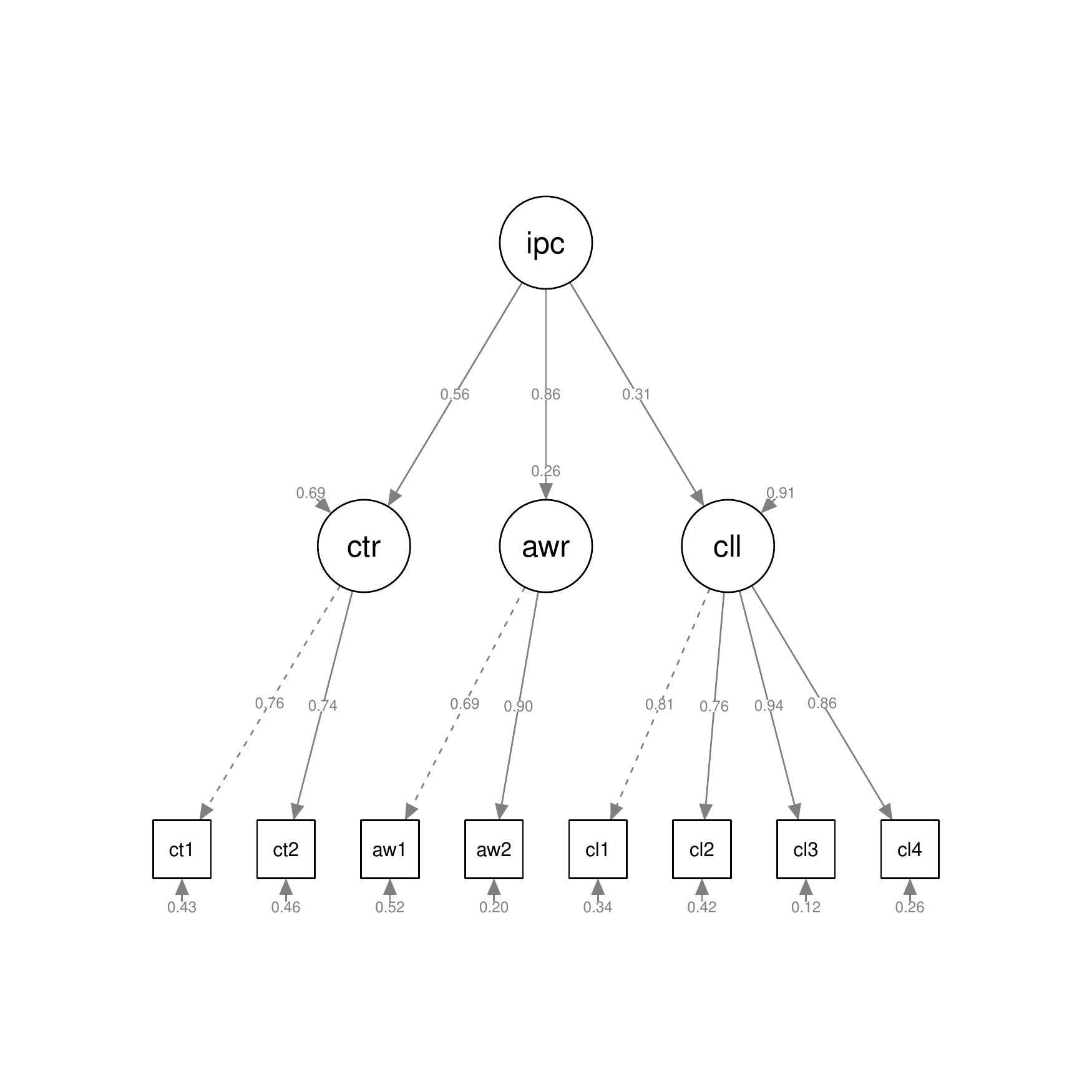} 
\vspace{0cm}
\caption{CFA paths plot with standardized estimates of the respecified IUIPC-8 on Sample \textsf{B}}
\label{fig:pathPlotCFABredux}
\end{figure*}
}

\newcommand{\screePlotB}{
\begin{figure}[htb]
\centering\includegraphics[keepaspectratio,width=\maxwidth]{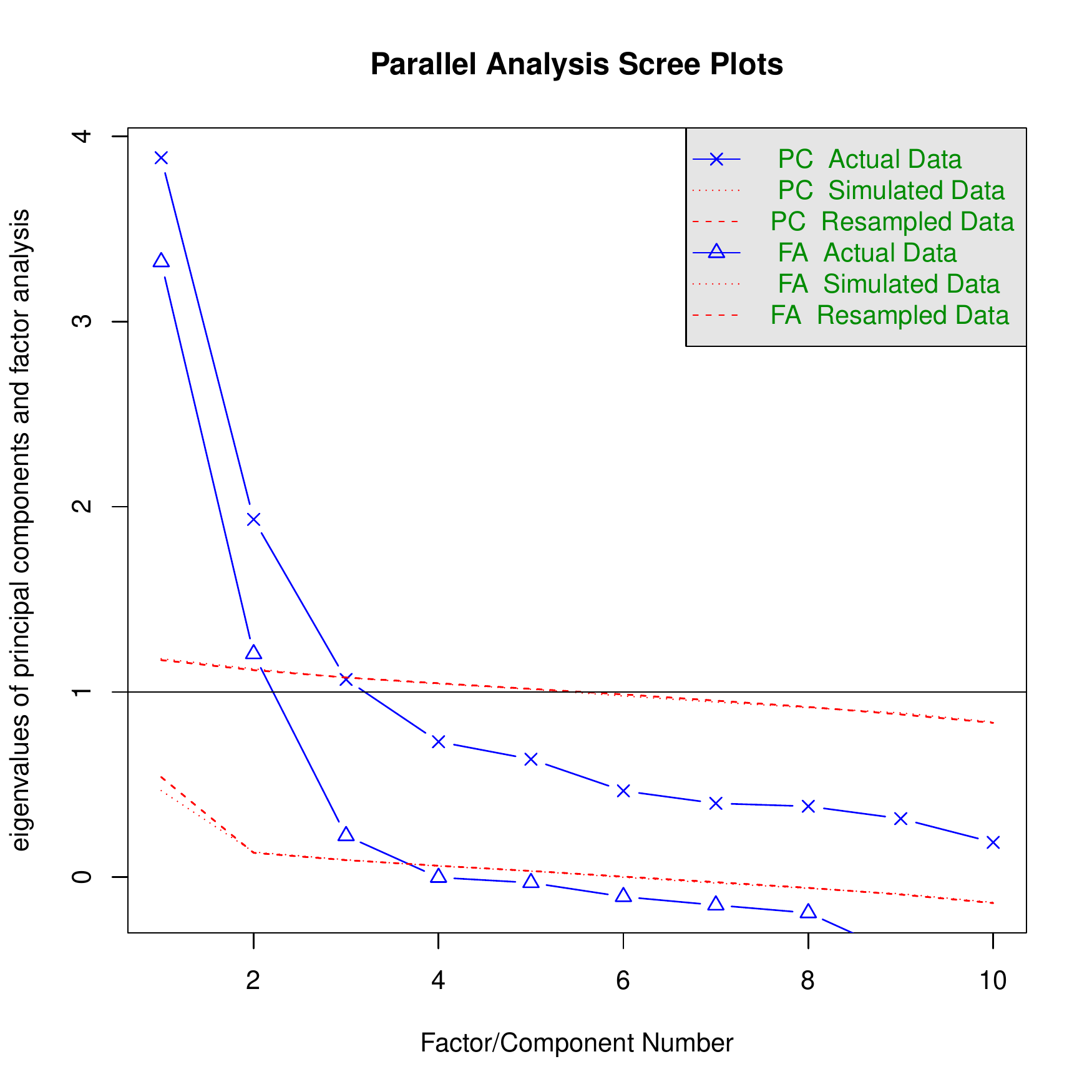}\caption{Screeplot for factor number of Sample \textsf{B}}
\label{fig:screePlotB}
\end{figure}
}

%

\newcommand{\descSubScalesOrig}{
\begin{table}[htb]
\centering\caption{Means (SDs) of the parceled sub-scales of IUIPC-10}
\label{tab:descSubScalesOrig}
\begin{adjustbox}{max width=\columnwidth}
\begingroup\footnotesize
\begin{tabular}{rllll}
  \toprule
 & Sample \textsf{A} & Sample \textsf{B} & Sample \textsf{V} & Malhotra et al. \\ 
  \midrule
\textsf{ctrl} & $5.82$ $(0.99)$ & $5.93$ $(0.78)$ & $5.86$ $(0.84)$ & $5.67$ $(1.06)$ \\ 
  \textsf{awa} & $6.22$ $(0.78)$ & $6.51$ $(0.52)$ & $6.43$ $(0.66)$ & $6.21$ $(0.87)$ \\ 
  \textsf{coll} & $5.48$ $(1.12)$ & $5.58$ $(1.12)$ & $5.60$ $(1.04)$ & $5.63$ $(1.09)$ \\ 
  \textsf{iuipc} & $5.84$ $(0.75)$ & $6.00$ $(0.61)$ & $5.96$ $(0.64)$ & $5.84$ $(1.01)$ \\ 
   \bottomrule
\end{tabular}
\endgroup
\end{adjustbox}\end{table}
}

\newcommand{\bootBSorig}{
\begin{table}[ht]
\centering
\caption{Bootstrapped CFA for IUIPC-10} 
\label{tab:bootBSorig}
\begingroup\footnotesize
\begin{tabular}{rlllllllll}
  \toprule
 & $\chi^2$ & $p$ & $\mathsf{cn}_{05}$ & \textsf{CFI} & \textsf{TLI} & \textsf{RMSEA} & \textsf{LL} & \textsf{UL} & \textsf{SRMR} \\ 
  \midrule
mean &     45.900 &    0.178 &    1112.251 &    0.996 &    0.994 &    0.018 &    0.006 &    0.032 &    0.020 \\ 
  std.dev &     13.029 &    0.231 &     331.202 &    0.004 &    0.005 &    0.010 &    0.008 &    0.008 &    0.003 \\ 
  min &     14.430 &    0.000 &     400.236 &    0.973 &    0.962 &    0.000 &    0.000 &    0.000 &    0.011 \\ 
  max &    117.789 &    0.997 &    3259.877 &    1.000 &    1.007 &    0.051 &    0.042 &    0.061 &    0.039 \\ 
   \bottomrule
\end{tabular}
\endgroup
\end{table}
}
\newcommand{\bootBSredux}{
\begin{table}[ht]
\centering
\caption{Bootstrapped CFA for IUIPC-8} 
\label{tab:bootBSredux}
\begingroup\footnotesize
\begin{tabular}{rlllllllll}
  \toprule
 & $\chi^2$ & $p$ & $\mathsf{cn}_{05}$ & \textsf{CFI} & \textsf{TLI} & \textsf{RMSEA} & \textsf{LL} & \textsf{UL} & \textsf{SRMR} \\ 
  \midrule
mean &     26.939 &    0.239 &    1291.895 &    0.997 &    0.996 &    0.017 &    0.005 &    0.035 &    0.021 \\ 
  std.dev &      9.381 &    0.258 &     496.468 &    0.003 &    0.004 &    0.012 &    0.008 &    0.010 &    0.005 \\ 
  min &      4.988 &    0.000 &     393.554 &    0.981 &    0.972 &    0.000 &    0.000 &    0.000 &    0.009 \\ 
  max &     78.170 &    0.999 &    6153.324 &    1.000 &    1.006 &    0.055 &    0.043 &    0.068 &    0.052 \\ 
   \bottomrule
\end{tabular}
\endgroup
\end{table}
}

\newcommand{\discriminantB}{
\begin{table*}[p]
\centering\caption{Evidence for discriminant validity of IUIPC-10 (Fornell-Larcker Criterion and Heterotrait-Monotrait Ratio) on Sample \textsf{B}\explain{.\newline{}The Fornell-Larcker criterion of $\sqrt{\vari{AVE}}$ greater than any correlation with any other factor and the $\vari{HTMT} < .85$ are fulfilled.}}
\captionsetup{position=top}
\label{tab:discriminantB}
\subfloat[Fornell-Larcker]{
\label{tab:corfitB}
\centering
\begingroup\footnotesize
\begin{tabular}{rllll}
  \toprule
 & 1 & 2 & 3 & 4 \\ 
  \midrule
1. ctrl & 0.634 &  &  &  \\ 
  2. aware & 0.546 & 0.627 &  &  \\ 
  3. collect & 0.227 & 0.329 & 0.848 &  \\ 
  4. iuipc & 0.613 & 0.891 & 0.37 & 0.769 \\ 
   \bottomrule
\multicolumn{5}{c}{\emph{Note:} The diagonal contains the $\sqrt{\mathit{AVE}}$}\\
\end{tabular}
\endgroup
}~\subfloat[HTMT Ratio]{%
\label{tab:htmtB}
\centering
\begingroup\footnotesize
\begin{tabular}{rlll}
  \toprule
 & 1 & 2 & 3 \\ 
  \midrule
1. ctrl & --- &  &  \\ 
  2. aware & 0.67 & --- &  \\ 
  3. collect & 0.32 & 0.45 & --- \\ 
   \bottomrule
\end{tabular}
\endgroup
}
\end{table*}
}
\newcommand{\discriminantV}{
\begin{table*}[p]
\centering\caption{Evidence for discriminant validity of IUIPC-10 (Fornell-Larcker Criterion and Heterotrait-Monotrait Ratio) on Sample \textsf{V}}
\captionsetup{position=top}
\label{tab:discriminantV}
\subfloat[Fornell-Larcker]{
\label{tab:corfitV}
\centering
\begingroup\footnotesize
\begin{tabular}{rllll}
  \toprule
 & 1 & 2 & 3 & 4 \\ 
  \midrule
1. ctrl & 0.638 &  &  &  \\ 
  2. aware & 0.625 & 0.572 &  &  \\ 
  3. collect & 0.285 & 0.399 & 0.793 &  \\ 
  4. iuipc & 0.668 & 0.935 & 0.427 & 0.717 \\ 
   \bottomrule
\multicolumn{5}{c}{\emph{Note:} The diagonal contains the $\sqrt{\mathit{AVE}}$}\\
\end{tabular}
\endgroup
}~\subfloat[HTMT Ratio]{%
\label{tab:htmtV}
\centering
\begingroup\footnotesize
\begin{tabular}{rlll}
  \toprule
 & 1 & 2 & 3 \\ 
  \midrule
1. ctrl & --- &  &  \\ 
  2. aware & 0.73 & --- &  \\ 
  3. collect & 0.39 & 0.53 & --- \\ 
   \bottomrule
\end{tabular}
\endgroup
}
\end{table*}
}
\newcommand{\discriminantBredux}{
\begin{table*}[p]
\centering\caption{Evidence for discriminant validity of IUIPC-8 (Fornell-Larcker Criterion and Heterotrait-Monotrait Ratio) on Sample \textsf{B}}
\captionsetup{position=top}
\label{tab:discriminantBredux}
\subfloat[Fornell-Larcker]{
\label{tab:corfitBredux}
\centering
\begingroup\footnotesize
\begin{tabular}{rllll}
  \toprule
 & 1 & 2 & 3 & 4 \\ 
  \midrule
1. ctrl & 0.747 &  &  &  \\ 
  2. aware & 0.477 & 0.803 &  &  \\ 
  3. collect & 0.171 & 0.265 & 0.848 &  \\ 
  4. iuipc & 0.556 & 0.859 & 0.308 & 0.825 \\ 
   \bottomrule
\multicolumn{5}{c}{\emph{Note:} The diagonal contains the $\sqrt{\mathit{AVE}}$}\\
\end{tabular}
\endgroup
}~\subfloat[HTMT Ratio]{%
\label{tab:htmtBredux}
\centering
\begingroup\footnotesize
\begin{tabular}{rlll}
  \toprule
 & 1 & 2 & 3 \\ 
  \midrule
1. ctrl & --- &  &  \\ 
  2. aware & 0.48 & --- &  \\ 
  3. collect & 0.16 & 0.28 & --- \\ 
   \bottomrule
\end{tabular}
\endgroup
}
\end{table*}
}
\newcommand{\discriminantVredux}{
\begin{table*}[p]
\centering\caption{Evidence for discriminant validity of IUIPC-8 (Fornell-Larcker Criterion and Heterotrait-Monotrait Ratio) on Sample \textsf{V}}
\captionsetup{position=top}
\label{tab:discriminantVredux}
\subfloat[Fornell-Larcker]{
\label{tab:corfitVredux}
\centering
\begingroup\footnotesize
\begin{tabular}{rllll}
  \toprule
 & 1 & 2 & 3 & 4 \\ 
  \midrule
1. ctrl & 0.735 &  &  &  \\ 
  2. aware & 0.547 & 0.74 &  &  \\ 
  3. collect & 0.218 & 0.344 & 0.793 &  \\ 
  4. iuipc & 0.588 & 0.93 & 0.37 & 0.774 \\ 
   \bottomrule
\multicolumn{5}{c}{\emph{Note:} The diagonal contains the $\sqrt{\mathit{AVE}}$}\\
\end{tabular}
\endgroup
}~\subfloat[HTMT Ratio]{%
\label{tab:htmtVredux}
\centering
\begingroup\footnotesize
\begin{tabular}{rlll}
  \toprule
 & 1 & 2 & 3 \\ 
  \midrule
1. ctrl & --- &  &  \\ 
  2. aware & 0.55 & --- &  \\ 
  3. collect & 0.24 & 0.35 & --- \\ 
   \bottomrule
\end{tabular}
\endgroup
}
\end{table*}
}

\newcommand{\corrgramScores}{
\begin{figure}[tb]
\centering
\includegraphics[width=\maxwidth]{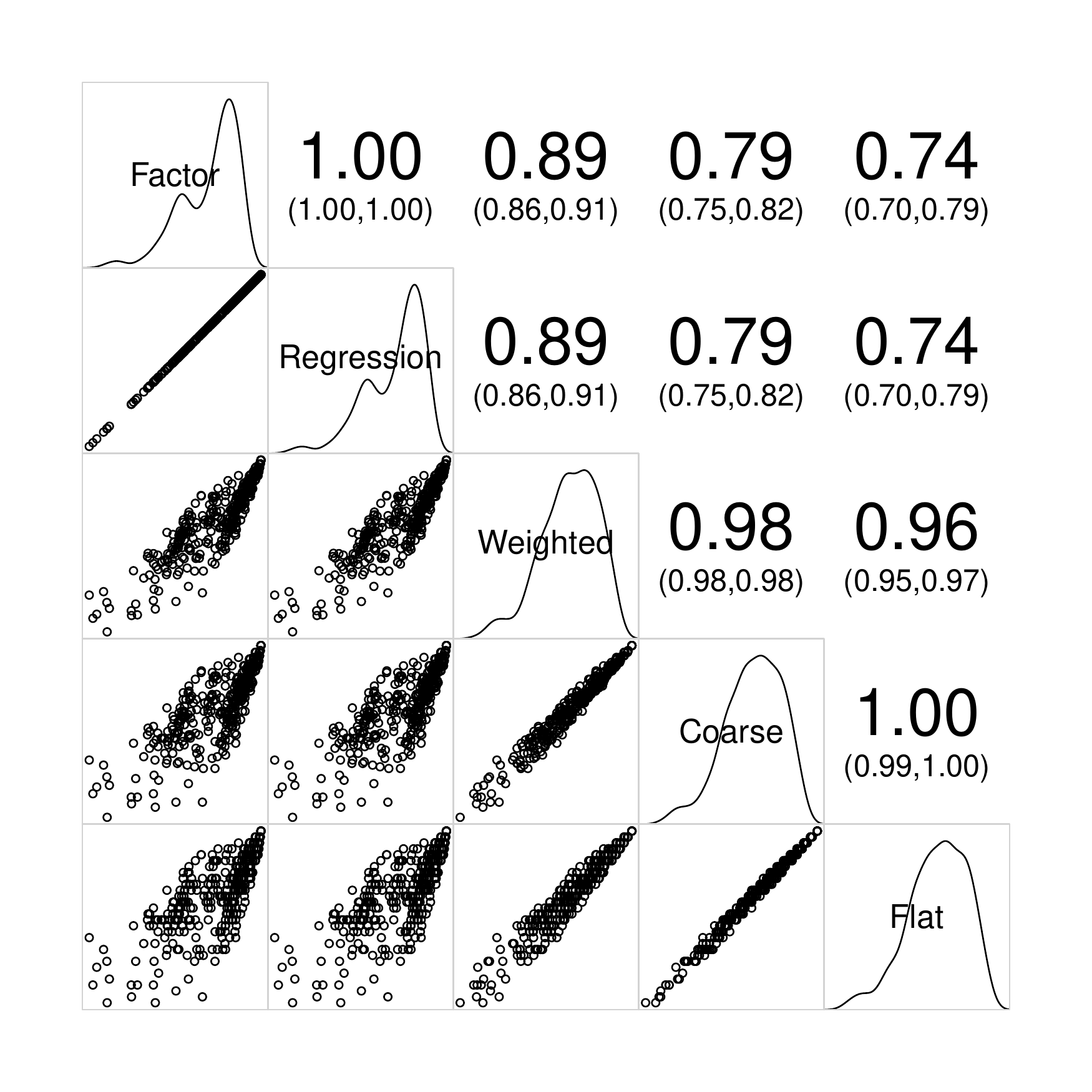} 
\caption{Relation of the IUIPC-10 factors to approximate scores\explain{.\newline{}While allowing for factor-analysis models also yielding estimates, overall, na{\"{i}}ve score computations, such as flat averaging, lose information over more sophisticated factor scores.}}
\label{fig:corrgramScores}
\end{figure}
}


\newcommand{\ACSlong}{Attribute-Based Credential System (ACS)\xspace}
\newcommand{\ACSname}{Attribute-Based Credential System\xspace}
\newcommand{\ACSpllong}{Attribute-Based Credential Systems (ACS)\xspace}
\newcommand{\ACSplname}{Attribute-Based Credential Systems\xspace}
\newcommand{\ACS}{ACS\xspace}

\begin{DocumentVersionTR}
\title{Validity and Reliability of the Scale Internet Users' Information Privacy Concern (IUIPC) [Extended Version]\thanks{Open Science Framework: \protect\url{https://osf.io/5pywm}. This is the author's version of this work.}}

\begin{NamedAuthors}
  \author{Thomas Gro{\ss}\\
   Newcastle University\\
   United Kingdom\\
   \textit{thomas.gross@newcastle.ac.uk}}
\end{NamedAuthors}
\end{DocumentVersionTR}

\maketitle

\begin{DocumentVersionTR}
\begin{abstract}
{
Internet Users' Information Privacy Concerns (IUIPC-10) is one of the most endorsed privacy concern scales. It is widely used in the evaluation of human factors of PETs and the investigation of the privacy paradox.  Even though its predecessor Concern For Information Privacy (CFIP) has been evaluated independently and the instrument itself seen some scrutiny, we are still missing a dedicated confirmation of IUIPC-10, itself.
%
We aim at closing this gap by systematically analyzing IUIPC's construct validity and reliability.
%
We obtained three mutually independent samples \processifversion{DocumentVersionTR}{with sample sizes $N_\mathsf{A} = 205$, $N_\mathsf{B} = 379$, and $N_\mathsf{V} = 433$.}\processifversion{DocumentVersionConference}{with a total of $N = 1031$ participants.}
We conducted a confirmatory factor analysis (CFA) on our main sample. Having found weaknesses, we established further factor analyses to assert the dimensionality of IUIPC-10. We proposed a respecified instrument IUIPC-8 with improved psychometric properties.
Finally, we validated our findings on a validation sample.
%
While we could confirm the overall three-dimensionality of IUIPC-10, we found that IUIPC-10 consistently failed construct validity and reliability evaluations, calling into question the unidimensionality of its sub-scales Awareness and Control.
Our respecified scale IUIPC-8 offers a statistically significantly better model and outperforms IUIPC-10's construct validity and reliability.
%
The disconfirming evidence on the construct validity raises doubts how well IUIPC-10 measures the latent variable information privacy concern. The sub-par reliability could yield spurious and erratic results as well as attenuate relations with other latent variables, such as behavior.
Thereby, the instrument could confound studies of human factors of PETs or the privacy paradox, in general.}
\end{abstract}

\end{DocumentVersionTR}

\section{Introduction}

Sound measurement instruments are a key ingredient in the investigation of privacy concerns and their impact on human behavior. They act as a measuring stick for privacy concerns themselves as well as a foundational component for substantive composite models. Thereby, they are a crucial keystone in evaluating human factors of PETs and studying the privacy paradox.

While there has been a diversification of instruments on privacy concern and behaviors~\cite{smith1996information,malhotra2004internet,dinev2004internet,xu2008examining,buchanan2007development,braunstein2011indirect,preibusch2013guide} also documented in systematic reviews on the privacy paradox~\cite{kokolakis2017privacy,gerber2018explaining}, {Internet users' information privacy concerns} (IUIPC)~\cite{malhotra2004internet} stands out as a widely adopted scale diligently created in an evolutionary fashion, with a sound theoretical underpinning and with systematic factor analyses.

IUIPC is based on {Concerns for information privacy} (CFIP)~\cite{smith1996information}, itself a popular scale measuring organizational information privacy concern, which has been validated in independent empirical studies~\cite{stewart2002empirical,harborth2018german}. Both CFIP and IUIPC scales have been endorsed by Preibusch~\cite{preibusch2013guide} as a sound instrument.

We are interested in IUIPC-10, a 10-item privacy concern scale with the three dimensions Control, Awareness, and Collection, because of its strong pedigree for the use in the investigation of human factors of PETs and of the privacy paradox.
While it has seen some scrutiny as part of other studies~\cite{sipior2013empirically,morton2013measuring} and questions of its validity have become apparent, there has not yet been a dedicated adequately sized factor analysis to confirm its validity.

We aim at two complementary research questions:
\begin{inparaenum}[(i)]
  \item First, we investigate to what extent the the validity and reliability of IUIPC-10 can be confirmed, focusing especially on validating its factor structure.
  \item Second, we consider under which circumstances IUIPC-10 can be employed most reliably, considering especially the estimation method used.
\end{inparaenum}
The latter line of inquiry is motivated by design decisions made by Malhotra et al.~\cite{malhotra2004internet}\processifversion{DocumentVersionTR}{on the handling measurement level, distribution characteristics, outliers and choice of estimation method}, which are at odds with contemporary recommendations for a sound CFA methodology~\cite{kline2012assumptions,bovaird2012measurement,finney2006non}. Hence, we thereby aim at ruling out possible confounders of our direct replication and at offering empirically grounded recommendations derived from our conceptual replications on \emph{how} to best use IUIPC.

\paragraph*{Our Contributions.}
To the best of our knowledge, we established the first dedicated adequately-sized independent confirmatory factor analysis of IUIPC-10. We, thereby, offer the first comprehensive disconfirming evidence on the construct validity and reliability of this scale, with wide implications for studies measuring information privacy concern in their endeavor to evaluate the users attitude to PETs or to study the privacy paradox overall.
While we could confirm the overall three-dimensionality of the scale, we found weaknesses in factorial and convergent validity as well as reliability, especially rooted the sub-scales Control and Awareness.
Those weaknesses appeared consistently across our independent samples and irrespective of CFA estimators used. 
We propose a respecified scale IUIPC-8 that consistently offers a statistically significantly better fit, stronger construct validity and reliability.
In terms of analysis methodology, we build a bridge between replicating the exact design decisions of Malhotra et al.~\cite{malhotra2004internet} to factor analyses especially adept on non-normal, ordinal following contemporary recommendations.

\section{Background}
\label{sec:background}

\begin{DocumentVersionTR}
We first introduce the role of privacy concern scales and IUIPC in the research of PETs and the privacy paradox, give an overview of the genesis of IUIPC and discuss earlier analyses vis-{\`a}-vis of the contributions of this study in Section~\ref{sec:priv_concern}.
Second, we consider requirements for sound measurement instruments in Section~\ref{sec:reqs}.
Finally, we explain factor analysis as tool to establish a scale's construct validity in Section~\ref{sec:fa}. 
\end{DocumentVersionTR}

%
%
%

\subsection{Information Privacy Concern}
\label{sec:priv_concern}

In defining \defterm{ipc}{information privacy concern} we focus on the conceptual framework of IUIPC.
Malhotra et al.~\cite[p. 337]{malhotra2004internet} refer to Westin's definition of information privacy as a foundation of their understanding of privacy concern: ``the claim of individuals, groups, or institutions to determine for them selves when, how, and to what extent information about them is communicated to others.'' Information privacy concern is then defined as ``an individual's subjective views of fairness within the context of information privacy.'' 

This framing of information privacy concern is well aligned with the interdisciplinary review of privacy studies by Smith et al.~\cite{smith2011information}, which considered privacy concern as the central antecedent of related behavior in their privacy macro model.
Of course, the causal impact of privacy concern on behavior has been under considerable scrutiny with the observation of the privacy attitude-behavior dichotomy---the \defterm{privacy_paradox}{privacy paradox}~\cite{gerber2018explaining}.
The intense inquiry of the privacy community into the paradox calls for measuring information privacy concern accurately and reliably. This conviction is rooted in the fact that measurement errors could confound the assessment of users' privacy concerns and, thereby, alternative explanation for the privacy paradox: If one has not actually measured privacy concern, it can hardly be expected to align with exhibited behavior.

There has been a proliferation of related and distinct instruments for measuring information privacy concern\processifversion{DocumentVersionTR}{~\cite{smith1996information,malhotra2004internet,dinev2004internet,xu2008examining,buchanan2007development,braunstein2011indirect}}. As a comprehensive comparison would be beyond the scope of this study, we refer to Preibusch's excellent guide to measuring privacy concern~\cite{preibusch2013guide} for an overview of the field and shall focus on specific comparisons to IUIPC itself.
First, we mention  \defterm{CFIP}{Concern for information privacy} (CFIP)~\cite{smith1996information} a major influence  on IUIPC. It consists of four dimensions---Collection, Unauthorized Secondary Use, Improper Access and Errors. While both questionnaires share questions, CFIP focuses on individuals' concerns about organizational privacy practices and the organization's responsibilities, IUIPC shifts this focus to Internet users framed as consumers and their perception of fairness and justice in the context of information privacy and online companies. 

\defterm{IPC}{Internet Privacy Concerns} (IPC)~\cite{dinev2004internet} considered internet privacy concerns with antecedents of perceived vulnerability and control, antecedents familiar from the Protection Motivation Theory (PMT). In terms of the core scale of privacy concern, Dinev and Hart identified two factors 
\begin{inparaenum}[(i)] 
\item Abuse (concern about misuse of information submitted on the Internet) and
\item Finding (concern about being observed and specific private information being found out).
\end{inparaenum}
IPC differs from IUIPC in its focus on misuse rather than just collection of information and of concerns of surveillance.

Buchanan et al.'s \defterm{OPC}{Online Privacy Concern and Protection for Use on the Internet} (OPC)~\cite{buchanan2007development} measure considered three sub-scales---General Caution, Technical Protection (both on behaviors), and Privacy Attitude. Compared to IUIPC, OPC sports a strong focus on item stems eliciting being concerned and on measures through a range of concrete privacy risks. The authors considered concurrent validity with IUIPC, observing a correlation of $r= .246$ between OPC's privacy concern and the total IUIPC score.

CFIP, IPC and OPC have in common that---unlike IUIPC---they do not explicitly mention the \refterm{loaded_word}{loaded word} ``privacy.'' \processifversion{DocumentVersionTR}{This is even more so a distinctive feature of the \defterm{ICP}{Indirect Content Privacy} (ICP) survey developed by Braunstein et al.~\cite{braunstein2011indirect}. Compared to IUIPC it focuses more on sharing specific kinds of sensitive information, carefully avoiding to mention ``privacy'' or other \refterm{loaded_word}{loaded words}. ICP does not offer a dedicated factor analysis confirming its construct validity and reliability.}

\subsubsection{Genesis of IUIPC}
\label{sec:genesis}
The scale \defterm{IUIPC}{Internet users' information privacy concern} (IUIPC) was developed by Malhotra et al.~\cite{malhotra2004internet}, by predominately adapting questions of the earlier 15-item scale Concern for information privacy (CFIP) by Smith et al.~\cite{smith1996information} and by framing the questionnaire for Internet users.

CFIP received independent empirical confirmations of its factor structure, first by Stewart and Segars~\cite{stewart2002empirical}, but also by Harborth and Pape~\cite{harborth2018german} on its German translation.
 \processifversion{DocumentVersionTR}{In a somewhat related yet largely unnoticed analysis of general privacy concern, Casta{\~n}da et al.~\cite{castaneda2007dimensionality} considered the factor structure on \emph{collection} and \emph{use} sub-factors, bearing distinct similarity to the first two dimensions of CFIP.}

Malhotra et al.~\cite[pp. 338]{malhotra2004internet} conceived IUIPC-10 as a second-order reflective scale of \defterm{dimensions_IUIPC}{information privacy concern}, with the dimensions Control, Awareness, and Collection. The authors considered the ``act of collection, whether it is legal or illegal,'' as the starting point of information privacy concerns. The sub-scale control is founded on the conviction that ``individuals view procedures as fair when they are vested with control of the procedures.'' Finally, they considered being ``informed about data collection and other issues'' as central concept to the sub-scale awareness.
The authors developed IUIPC in exploratory and confirmatory factor analysis, which we shall review systematically in Section~\ref{sec:review}.

\subsubsection{The Role of Privacy Concern Scales in the Investigation of PETs}
\label{sec:relevance}

The role of privacy concern scales in the investigation of the privacy paradox was well documented in the Systematic Literature Review by Gerber et al.~\cite{gerber2018explaining}: More than a dozen studies used privacy concern as variable. For instance, Schwaig et al.~\cite{schwaig2013model} used IUIPC as instrument in their privacy paradox study.

On human factors of PETs, we have the, e.g., technology acceptance of Tor/JonDonym~\cite{harborth2020explaining} and anonymous credentials~\cite{benenson2015user}.  While these two studies used ``perceived anonymity'' as a three-item scale, subsequent work by Harborth and Pape used IUIPC as privacy concern scale to evaluate JonDonym\cite{harborth2018jondonym} and Tor~\cite{harborth2019privacy}. 

Furthermore, IUIPC has not only been used in the narrow sense of evaluating the privacy paradox or PETs. Let us highlight a few examples in PoPETS: Pu and Grossklags~\cite{pu2016towards} used a scale adopted from IUIPC Collection (or one of its predecessors) to measure own and friends' privacy concern in their study social app users' valuation of interdependent privacy.  Gerber et al.~\cite{gerber2019investigating} used IUIPC to contextualize their investigation on privacy risk perception.  Barbosa et al.~\cite{barbosa2019if} used IUIPC as main privacy measure to predict changes in smart home device use.

Given that Smith et al.'s privacy macro model~\cite{smith2011information} centered on privacy concern as antecedent of behavior and that Preibusch~\cite{preibusch2013guide} recommended IUIPC as a ``safe bet,'' it is not surprising that IUIPC is high on the list of instruments to analyze privacy paradox or PETs. Thereby, we would expect prolific use of IUIPC in future privacy research and perceive a strong need to substantiate the evidence of its \refterm{construct_validity}{construct validity} and \refterm{reliability}{reliability}.

\subsection{Validity and Reliability}
\label{sec:reqs}
When evaluating privacy concern instruments, the dual key questions for privacy researchers interested in the investigation of the human factors of PETs and the privacy paradox are:
\begin{inparaenum}[(i)]
  \item Are we measuring the hidden latent construct ``privacy concern'' accurately? (validity)
  \item Are we measuring privacy concern consistently and with an adequate signal-to-noise ratio? (reliability)
\end{inparaenum}

\defterm{validity}{Validity} refers to whether an instrument measures what it purports to measure. Messick offered an early well-regarded definition of validity as the ``integrated evaluative judgment of the degree to which empirical evidence and theoretical rationales support the adequacy and appropriateness of inferences and actions based on test scores''~\cite{messick1987validity}. Validity is inferred---judged in degrees---not measured. \processifversion{DocumentVersionTR}{This classic authoritative definition is at odds with a newer ontological approach proposed by Borsboom et al.~\cite{borsboom2004concept} stating that ``a test is valid for measuring an attribute if 
\begin{inparaenum}[(a)] 
  \item the attribute exists and
  \item variations in the attribute causally produce variation in the measurement outcomes.''
\end{inparaenum}}
In our work, we take a pragmatic empiricist's approach focusing on the validation procedure and evidence.

\subsubsection{Content Validity}
\defterm{content_validity}{Content validity} refers to the relevance and representativeness of the content of the instrument, typically assessed by expert judgment. 
We shall evaluate content validity together in keeping with evidence on the craft of the questionnaire design, incl. question format, language used, question order, and, hence, assess \defterm{barriers}{psychometric barriers} in the form of biases rooted in the questionnaire wording~\cite[p. 128]{Oppenheim1992}.
\begin{compactdesc}
  \item[Priming] \defterm{priming}{Priming}\processifversion{DocumentVersionTR}{~\cite{Priming2008}} means that mentioning a concept activates it in respondents minds and makes it more easily accessible in subsequent questions. Priming is, for instance, created by the use of a \defterm{loaded_word}{loaded word}\processifversion{DocumentVersionTR}{~\cite[pp. 137]{Oppenheim1992}} such as ``security.'' It can invoke the respondents' \defterm{social_desirability}{social desirability bias} \processifversion{DocumentVersionTR}{~\cite{SocialDesirability2008}}.
  \item[Leading questions] \defterm{leading_question}{Leading questions} elicit agreement or specific instances of a general term, introduce a bias towards that lead.\processifversion{DocumentVersionTR}{~\cite[pp. 137]{Oppenheim1992}}
  \item[Double-barreled questions] \defterm{double-barreled_question}{Double-barreled questions}\processifversion{DocumentVersionTR}{~\cite[p. 128]{Oppenheim1992}} consist of two or more questions or clauses, making it difficult for the respondent to decide what to answer to, causing nondifferentiated responses.
  \item[Positively-oriented questions] Exclusively using positively framed wording for questions leads to \defterm{nondifferentiation}{nondifferentiation}/straightlining\processifversion{DocumentVersionTR}{~\cite{Nondifferentiation2008}} and, thereby, to accommodating the \defterm{acquiescent_response}{acquiescent response bias}\processifversion{DocumentVersionTR}{~\cite{AcquiescenceResponseBias2008}}.
\end{compactdesc}

\subsubsection{Construct Validity}
\label{sec:construct_validity}
While Messick~\cite{messick1980test} considers \defterm{construct_validity}{construct validity}~\cite{cronbach1955construct}\fyi{Check out how Cook and Campbell 1979 defines construct validity, cf. CFIP p. 182/17}, the interpretive meaningfulness, the extent to which an instrument accurately represents a construct, as foremost evaluation of validity, there is a range of confirmatory and disconfirmatory evaluations of validity, often referred to as kinds of validity, even if they only constitute a lens on a whole. For the purposes of this paper, we summarize the terms we will use in our own investigation. 
\processifversion{DocumentVersionConference}{Appendix~\ref{sec:fa} details used factor-analysis concepts.}

First, we seek evidence of \defterm{factorial_validity}{factorial validity}, that is, evidence that that factor composition and dimensionality are sound. While IUIPC is a \defterm{multidimensional}{multidimensional} scale with three correlated designated dimensions, we require \defterm{unidimensionality}{unidimensionality} of each sub-scale, a requirement discussed at length by Gerbing and Anderson~\cite{gerbing1988updated}. 
\begin{DocumentVersionTR}
Unidimensional measurement models for sub-scales often correspond to expecting \defterm{congeneric}{congeneric} measures, that is, the scores on an item are the expression of a true score weighted by the item's loading plus some measurement error, where in the congeneric case neither the loadings nor error variances across items are required to be equal.
This property entails that the items of each sub-scale must be conceptually homogeneous. 
\end{DocumentVersionTR}
The empirical evidence for factorial validity is the found in the adequacy of the hypothesized model's fit and passing the corresponding fit hypotheses of a confirmatory factor analysis for the designated factor structure~\cite{anderson1987assessment,gerbing1988updated,kline2015principles}. Specifically, we prioritize the following fit metrics and hypotheses, refering the interested reader to Appendix~\ref{sec:fit.tests} for further explanations:
\begin{compactdesc}
  \item[Goodness-of-fit $\chi^2$:] Measures the exact fit of a model and gives rise to the \refterm{accept-support}{accept-support} exact-fit test against null hypothesis $H_{\chi^2, 0}$.
  \item[\textsf{RMSEA}:] \refterm{RMSEA}{Root Mean Square Estimate of Approximation}, an absolute badness-of-fit measure estimated as $\hat{\varepsilon}$ with its $90\%$ confidence interval, yielding a range of fit-tests: \refterm{close_fit}{close fit}, \refterm{not_close_fit}{not-close fit}, and \refterm{poor_fit}{poor fit} with decreasing tightness requirements.
\end{compactdesc}
\processifversion{DocumentVersionConference}{Further common criteria, such as \textsf{CFI} or \textsf{SRMR}, are defined in Appendix~\ref{sec:fit.tests}.}

\defterm{convergent_validity}{Convergent validity}~\cite[pp. 675]{hair2019multivariate} (convergent coherence) on an item-construct level means that items belonging together, that is, to the same construct, should be observed as related to each other.
Similarly, \defterm{discriminant_validity}{discriminant validity}~\cite[pp. 676]{hair2019multivariate} (discriminant distinctiveness) means that items not belonging together, that is, not belonging to the same construct, should be observed as not related to each other. Similarly, on a sub-scale level, we expect factors of the same higher-order construct to relate to each other and, on hierarchical factor level, we expect all 1\textsuperscript{st}-order factors to load strongly on the 2\textsuperscript{nd}-order factor.

While a poor local fit and tell-tale residual patterns yield disconfirming evidence for convergent and discriminant validity, further evidence is found evidence inter-item correlation matrices. We expect items belonging to the same sub-scale to be highly correlated (converge on the same construct).  At the same time correlation to items of other sub-scales should be low, especially lower than the in-construct correlations~\cite[pp. 196]{kline2015principles}.

These judgments are substantiated with empirical criteria, where we highlight the following metrics:
\begin{compactdesc}
  \item[Standardized Factor Loadings $\beta$:] $Z$-transformed factor scores, typically reported in factor analysis.
  \item[Variance Extracted $R^2$:] The factor variance accounted for, giving rise to the Average Variance Extracted ($\vari{AVE}$) defined subsequently in reliability Section~\ref{sec:reliability}.
  \item[Heterotrait-Monotrait Ratio (HTMT)] The \defterm{heterotrait_monotrait_ratio}{Heterotrait-Monotrait Ratio} is the ratio of the the avg. correlations of indicators across constructs measuring different phenomena to the avg. correlations of indicators within the same construct~\cite{henseler2015new}.
\end{compactdesc}
We gain empirical evidence in favor of convergent validity~\cite[pp. 675]{hair2019multivariate}
\begin{inparaenum}[(i)]
  \item if the variance extracted by an item $R^2 > .50$ entailing that the standardized factor loading are significant and $\beta > .70$, and
   \item if the \refterm{internal_consistency}{internal consistency} (defined in Section~\ref{sec:reliability}) is sufficient ($\vari{AVE} > .50$, $\omega > \vari{AVE}$, and $\omega > .70$).
\end{inparaenum}
The analysis yields empirical evidence of discriminant validity~\cite[pp. 676]{hair2019multivariate}
\begin{inparaenum}[(i)]
   \item if the square root of \vari{AVE} of a latent variable is greater than the max correlation with any other latent variable (Fornell-Larcker criterion~\cite{fornell1981evaluating}),
   \item if the \refterm{heterotrait_monotrait_ratio}{Heterotrait-Monotrait Ratio} (HTMT) is less than $.85$~\cite{henseler2015new,ab2017discriminant}\processifversion{DocumentVersionTR}{\cite{henseler2015new,ab2017discriminant}}.
\end{inparaenum}

\subsubsection{Reliability}
\label{sec:reliability}
\defterm{reliability}{Reliability} is the extent to which a variable is consistent in what is being measured~\cite[p. 123]{hair2019multivariate}. It can further be understood as the capacity of ``separating signal from noise''~\cite[p. 709]{RevCon2018}, quantified by the ratio of true score to observed score variance.~\cite[pp. 90]{kline2015principles}
We evaluate \defterm{internal_consistency}{internal consistency} as a means to estimate reliability from a single test application.
Internal consistency entails that items that purport to measure the same construct produce similar scores~\cite[p. 91]{kline2015principles}.
\processifversion{DocumentVersionTR}{We note that the validity of indices, such as Cronbach's $\alpha$, is contingent on the scale being congeneric.}
We will use the following internal consistency measures:
\begin{compactdesc}
  \item[Cronbach's $\alpha$:] Is based on the average inter-item correlations. \processifversion{DocumentVersionTR}{We report it as comparison to IUIPC, even though it is limited to assess internal consistency reliably only if a scale is truly unidimensional\processifversion{DocumentVersionTR}{~\cite[pp. 733]{RevCon2018}\cite[p. 91]{kline2015principles}}.}
  \processifversion{DocumentVersionTR}{\item[Guttman's $\lambda_6$:] Measures internal consistency with squared multiple correlations of a set of items.}
  \item[Congeneric Reliability $\omega$:] The amount of general factor saturation (also called \defterm{CR}{composite reliability}~\cite[pp. 313]{kline2015principles} or construct reliability (CR)~\cite[p. 676]{hair2019multivariate} depending on the source).
  \item[AVE:] \defterm{AVE}{Average Variance Extracted} (AVE)~\cite[pp. 313]{kline2015principles} is the average of the squared standardized loadings of indicators belonging to the same factor.
\end{compactdesc}
Thresholds for reliability estimates like Cronbach's $\alpha$ or Composite Reliability $\omega$ are debated in the field, where many recommendations are based on Nunnally's original treatment of the subject, but equally often misstated~\cite{lance2006sources}. The often quoted $\alpha \geq .70$ was described by Nunnally only to ``save time and energy,'' whereas a greater threshold of $.80$ was endorsed for basic research~\cite{lance2006sources}. While that would be beneficial for privacy research as well, we shall adopt reliability metrics $\alpha, \omega \geq .70$ as suggested by Hair et al.~\cite[p. 676]{hair2019multivariate}. We further require $\vari{AVE} > .50$.

\begin{DocumentVersionConference}
\subsection{Factor Analysis as Validation Tool}
\refterm{FA}{Factor analysis} is an excellent tool to establish \refterm{construct_validity}{construct validity} and \refterm{reliability}{reliability} of an instrument. We introduce factor analysis concepts in the Appendix and assume knowledge of the standard factor analysis with Maximum Likelihood estimation. Here we shall only mention properties of estimators MLM and robust WLS, developed to handle non-normal, ordinal data.

\processifversion{DocumentVersionTR}{\paragraph*{Alternative Estimators}}
A maximum-likelihood estimation with  robust standard errors and Satorra-Bentler scaled test statistic (MLM) is robust against some deviations from normality~\cite[p. 122]{kline2012assumptions}.
Lei and Wu~\cite[p. 172]{LeiWu2012} submit that Likert items with more than five points approximate continuous measurement enough to use MLM, this stance is not universally endorsed~\processifversion{DocumentVersionConference}{\cite{finney2006non,bovaird2012measurement}}\processifversion{DocumentVersionTR}{\cite{finney2006non,bovaird2012measurement,ChengHsien2016}}.

Kline recommends to use of estimators specializing on ordinal data~\cite[p. 122]{kline2012assumptions}, such as robust weighted least squares (\defterm{wlsmvs}{WLSMVS}\footnote{weighted least squares with robust standard errors and a Satterthwaite mean- and variance-adjusted test statistic}). The following explanation summarizes Kline~\cite[pp. 324]{kline2015principles}. in general, robust WLS methods associate each indicator with a latent response variable for which the method estimates thresholds that relate the ordinal responses on the indicator to a continuous distribution on the indicator's latent response variable. Then, the measurement model is evaluated with the latent response variables as indicators, while optimizing to approximate the observed polychoric correlations between the original indicator variables. In addition, the method will compute robust standard errors and corrected test statistics. Because the method also needs to estimate the thresholds, the number of free parameters will be greater than in a comparable ML estimation. Reported estimates and $R^2$ relate to the latent response variable, not the original indicators themselves. 

Because of the level of indirection and the estimation of non-linearly related thresholds, robust WLS is a quite different kettle of fish than Maximum Likelihood estimation on assumedly continuous variables: they are not easily compared by trivially examining blunt fit indices. The estimates of robust WLS need to be interpreted differently than in ML: they are \emph{probit} of individuals' response to an indicator instead of a linear change in an indicator~\cite{bovaird2012measurement}. The thresholds show what factor score is necessary for the respective option of an indicator or higher to be selected with $50\%$ probability.

The advantages and disadvantages of robust WLS have been carefully evaluated in simulation studies with known ground truth\processifversion{DocumentVersionConference}{~\cite{distefano2002impact,ChengHsien2016}}\processifversion{DocumentVersionTR}{~\cite{distefano2002impact,finney2006non,ChengHsien2016}}:
\begin{inparaenum}[(i)]
   \item Robust WLS models show little bias in the parameter estimation even es the level of skewness and kurtosis increased. Unlike ML models, do not suffer from out-of-bounds estimations of indicator variables.
   \item There is contradictory evidence on standard errors, where robust WLS may be subject to greater amounts of bias.
   \item $\chi^2$ fit indices of robust WLS are inflated for larger models or smaller samples.
   \item Robust WLS may inflate correlation estimates.
\end{inparaenum}
\end{DocumentVersionConference}

\begin{DocumentVersionTR}
\processifversion{DocumentVersionTR}{\subsection{Factor Analysis}}
\label{sec:fa}
Factor analysis is a powerful tool for evaluating the \refterm{construct_validity}{construct validity} and \refterm{reliability}{reliability} of privacy concern instruments.
\defterm{FA}{Factor analysis} refers to a set of statistical methods that are meant to determine the number and nature of \defterm{LV}{latent variables} (LVs) or \defterm{factor}{factors} that account for the variation and covariation among a set of observed measures commonly referred to as \defterm{IV}{indicators}\processifversion{DocumentVersionTR}{~\cite{brown2015confirmatory}}.

In general, we distinguish \defterm{EFA}{exploratory factor analysis} (EFA) and  \defterm{CFA}{confirmatory factor analysis}  (CFA), both of which are used to establish and evaluate psychometric instruments, respectively. They are both based on the \defterm{CFM}{common factor model}, which holds that each indicator variable contributes to the variance of one or more common factors and one unique factor. Thereby, \defterm{communality}{common variance} (communality) of related observed measures is attributed to the corresponding latent factor, and \defterm{uniqueness}{unique variance} (uniqueness) seen either as variance associated with the indicator or as error variance. \processifversion{DocumentVersionTR}{While in EFA the association of indicators to factors is unconstrained, in CFA this association is governed by a specified measurement model.} IUIPC is based on a \defterm{reflective}{reflective measurement}, that is, the observed measure of an indicator variable is seen as \emph{caused} by some latent factor. 
Indicators are thereby \defterm{endogenous}{endogenous} variables, latent variables \defterm{exogenous}{exogenous} variables.
Reflective measurement requires that all items of the sub-scale are interchangeable~\cite[pp. 196]{kline2015principles}\processifversion{DocumentVersionTR}{ and is closely related to the \refterm{congeneric}{congeneric} measure introduced above}.

In this paper, we are largely concerned with \defterm{CB-CFA}{covariance-based confirmatory factor analysis} (CB-CFA). Therein, the statistical tools aim at estimating coefficients for parameters of the measurement model that best fit the covariance matrix of the observed data.
The difference between an observed covariance of the sample and an implied covariance of the model is called a \defterm{residual}{residual}.

\processifversion{DocumentVersionTR}{\subsubsection{Estimators and their Assumptions}}
\processifversion{DocumentVersionConference}{\subsection{Estimators and their Assumptions}}
\label{sec:est_assumptions}
The purpose of a \refterm{FA}{factor analysis} is to estimate free parameters of the model (such as loadings or error variance), which is facilitated by \defterm{estimator}{estimators}.
The choice of estimator matters, because each comes with different strengths and weaknesses, requirements and assumptions that need to be fulfilled for the validity of their use. 
\processifversion{RedundantContent}{While practitioners sometimes endorse estimators as robust in face of violations, there is an ongoing debate what uses are appropriate or ill-advised.}
\processifversion{DocumentVersionTR}{\paragraph*{ML Assumptions}}
The most commonly used method for confirmatory factor analysis is \defterm{ML}{maximum likelihood} (ML) estimation. Among other \defterm{ml_assumptions}{assumptions}, this estimator requires according to Kline~\cite[pp. 71]{kline2015principles}:
\begin{inparaenum}[(i)]
  \item a \defterm{continuous}{continuous measurement level},
  \item \defterm{normal_distribution}{multi-variate normal distribution} (entailing the absence of extreme \defterm{skewness}{skewness})~\cite[pp. 74]{kline2015principles}, and
  \item treatment of influential cases and \defterm{outlier}{outliers}.
\end{inparaenum}
The distribution requirements are placed on the \refterm{endogenous}{endogenous} variables: the indicators.

\processifversion{DocumentVersionTR}{\paragraph*{Violations of Assumptions}}
These requirements are not always fulfilled in samples researchers are interested in. For instance, a common case at odds with ML-based CFA is the use of Likert items as indicator variables. Likert items are \defterm{ordinal}{ordinal}~\cite[p. 11]{hair2019multivariate} in nature, that is, ordered categories in which the distance between categories is not constant; they thereby require special treatment~\cite[pp. 323]{kline2015principles}. 

Lei and Wu~\cite{LeiWu2012} held based on a number of empirical studies that the fit indices of approximately normal \refterm{ordinal}{ordinal} variables with at least five categories are not greatly misleading. However, when ordinal and non-normal is treated as continuous and normal, the fit is underestimated and there is a more pronounced negative bias in estimates and standard errors. While Bovaird and Kozoil~\cite{bovaird2012measurement} acknowledge robustness of the ML estimator with \refterm{normal_distribution}{\emph{normally distributed}} \refterm{ordinal}{ordinal} data, they stress that increasingly skewed and kurtotic ordinal data inflate the Type I error rate. In the same vein, Kline~\cite[p. 122]{kline2012assumptions} holds the normality assumption for \refterm{endogenous}{endogenous} variables---the indicators---to be critical.

\begin{DocumentVersionTR}

\end{DocumentVersionTR}

\processifversion{DocumentVersionTR}{\subsubsection{Global and Local Fit}}
\processifversion{DocumentVersionConference}{\subsection{Global and Local Fit}}
\label{sec:fit.tests}

The \defterm{fit}{closeness of fit} of a factor model to an observed sample is evaluated globally with fit indices as well as locally by inspecting the residuals.
We shall focus on the ones Kline~\cite[p. 269]{kline2015principles} required as minimal reporting.
\processifversion{DocumentVersionTR}{\paragraph*{Fit Indices}}
\begin{compactdesc}
  \item[$\chi^2(\vari{df})$:] The $\chi^2$ for given degrees of freedom is the likelihood ratio chi-square, as a measure of exact fit. 
  \item[\textsf{CFI}:] The \defterm{CFI}{Bentler Comparative Fit} Index is an incremental fit index based on the non-centrality measure comparing selected against the null model.
  \item[\textsf{RMSEA}:] \defterm{RMSEA}{Root Mean Square Estimate of Approximation} ($\hat{\varepsilon}$) is an absolute index of bad fit, reported with its 90\% Confidence Interval $[\hat{\varepsilon}_{\mathsf{UL}}, \hat{\varepsilon}_{\mathsf{LL}}]$.
  \item[\textsf{SRMR}:] \defterm{SRMR}{Standardized Root Mean Square Residual} is a standardized version of the mean absolute covariance residual, where zero indicates excellent fit.
\end{compactdesc}
We mention the \defterm{GFI}{Goodness-of-Fit index} (GFI) reported by IUIPC, which approximates the proportion of variance explained in relation to the estimated population covariance. It is not recommended as it is substantially impacted by sample size and number of indicators\processifversion{DocumentVersionTR}{~\cite{Kenny2015fit,sharma2005simulation}}. 

Malhotra et al.~\cite{malhotra2004internet} adopted a \defterm{combination_rule}{combination rule} referring to Hu and Bentler\processifversion{DocumentVersionTR}{~\cite{hu1999cutoff}}, which we will report as HBR, staying comparable with their analysis:
``A model is considered to be satisfactory if
\begin{inparaenum}[(i)]
  \item $\mathsf{CFI} > .95$, 
  \item $\mathsf{GFI} > .90$, and
  \item $\mathsf{RMSEA} < .06$.''
\end{inparaenum}
\processifversion{DocumentVersionTR}{Kline~\cite{kline2015principles} spoke against the reliability of such combination rules.}

\processifversion{DocumentVersionTR}{\paragraph*{Model Comparison}}
\defterm{nested_model}{Nested models}~\cite[p. 280]{kline2015principles}, that is, models with can be derived from each other by restricting free parameters, can be well-compared with a \defterm{LRT}{Likelihood Ratio $\chi^2$ Difference Test} (LRT)~\cite[p. 270]{kline2015principles}. 
However, the models we are interested in are \defterm{non-nested}{non-nested}~\cite[p. 287]{kline2015principles}, because they differ in their observed variables. On ML-estimations, we have the \defterm{vuong}{Vuong Likelihood Ratio Test}~\cite{vuong1989likelihood} at our disposal to establish statistical inferences on such models.

In addition, we introduce the \textsf{AIC}/\textsf{BIC} family of metrics with formulas proposed by Kline~\cite[p. 287]{kline2015principles}:
\[ \mathsf{AIC} := \chi^2 + 2q \qquad\qquad \mathsf{BIC} := \chi^2 + q\ln(N), \]
where $q$ is the number of free parameters and $N$ the sample size.
We compute \textsf{CAIC} as the even-weighted mean between \textsf{AIC} and \textsf{BIC}.
These criteria can be used to compare different models estimated on the \emph{same} samples, on the same variables, but theoretically also on different subsets of variables. Smaller is better. 

\paragraph*{Statistical Inference} 
The $\chi^2$ and \textsf{RMSEA} indices offer us \defterm{fit_tests}{statistical inferences of global fit}. Such tests can either be \defterm{accept-support}{accept-support}, that is, accepting the null hypothesis supports the selected model, or \defterm{reject-support}{reject-support}, that is, rejecting the null hypothesis supports the selected model. We present them in the order of decreasing demand for close approximation.
\begin{compactdesc}
  \item[\defterm{exact_fit}{Exact Fit}:] An \refterm{accept-support}{accept-support} test, in which rejecting the null hypothesis on the model $\chi^2$ test implies the model is not an exact approximation of the data. The test is sensitive to the sample size and may reject well-fitting models at greater $N$.
    \begin{compactdesc}
      \item[$H_{\chi^2, 0}$:] The model is an exact fit in terms of residuals of the covariance structure.
      \item[$H_{\chi^2, 1}$:] The residuals of the model are considered too large for an exact fit.
    \end{compactdesc}
  \item[\defterm{close_fit}{Close Fit}:] An \refterm{accept-support}{accept-support} test, evaluated on the \refterm{RMSEA}{RMSEA} $\hat{\varepsilon}$ with zero as best result indicating approximate fit.
  \begin{compactdesc}
    \item[$H_{\hat{\varepsilon} \leq .05, 0}$:] The model has an approximate fit with RMSEA being less or equal $.05$, $\hat{\varepsilon} \leq .05$.
    \item[$H_{\hat{\varepsilon} \leq .05, 1}$:] The model does not evidence a close fit.
  \end{compactdesc}
  \item[\defterm{not_close_fit}{Not-close Fit}:] A \refterm{reject-support}{reject-support} hypothesis operating on the upper limit if the RMSEA 90\% CI, $\hat{\varepsilon}_\mathsf{UL}$, a significant $p$-value rejecting the not-close fit.
    \begin{compactdesc}
      \item[$H_{\hat{\varepsilon} \geq .05, 0}$:] The model is not a close fit, $\hat{\varepsilon}_\mathsf{UL} \geq .05$.
      \item[$H_{\hat{\varepsilon} \geq .05, 1}$:] Model approximate fit, $\hat{\varepsilon}_\mathsf{UL} < .05$.
    \end{compactdesc}
  \item[\defterm{poor_fit}{Poor Fit}:] A \refterm{reject-support}{reject-support} test on the upper limit if the RMSEA 90\% CI, $\hat{\varepsilon}_\mathsf{UL}$ checking for a poor fit.
    \begin{compactdesc}
      \item[$H_{\hat{\varepsilon} \geq .10, 0}$:] The model is a poor fit with $\hat{\varepsilon}_\mathsf{UL} \geq .10$.
      \item[$H_{\hat{\varepsilon} \geq .10, 1}$:] Model not a poor fit, $\hat{\varepsilon}_\mathsf{UL} < .10$.
    \end{compactdesc}
 \end{compactdesc}
\processifversion{DocumentVersionTR}{We note that the results for these different hypotheses will not necessarily be consistent. Hence, researchers are encouraged to consider the point estimate $\hat{\varepsilon}$ and its entire 90\% confidence interval $[\hat{\varepsilon_\mathsf{LL}}, \hat{\varepsilon_\mathsf{UL}}]$.
We recommend to interested readers Kline's excellent summary of these \textsf{RMSEA} fit tests~\cite[Topic Box 12.1]{kline2015principles}.}

\processifversion{DocumentVersionTR}{\paragraph*{Local Fit}}
Even with excellent global fit indices, the inspection of the local fit---evidenced by the residuals---must not be neglected. In fact, Kline~\cite[p. 269]{kline2015principles} drives home ``Any report of the results without information about the residuals is incomplete.'' \processifversion{DocumentVersionTR}{For that reason, we also stress the importance of evaluating the local fit, which we will do on correlation and (standardized) covariance residuals.}
Simply put, a large absolute \refterm{residual}{residual} indicates covariation that the model does not approximate well and that may thereby lead to spurious results.
\end{DocumentVersionTR}

\section{Related Work}
\label{sec:related}

\processifversion{DocumentVersionTR}{While IUIPC-10 has been one of the most-adopted privacy concern scales in the field, its independent confirmation was limited to analyses in the context of other research questions.}

Sipior et al.~\cite{sipior2013empirically} observed that IUIPC was under-studied at the time, offered a considerate review of the related literature, and focused their lens on the role of trusting and risk beliefs in IUIPC's causal model. With the caveat of being executed on a small sample of $N = 63$ students, their research could ``not confirm the use of the IUIPC construct to measure information privacy concerns.'' Even though they also excluded one \textsf{control} item, their concerns, however, were mostly focused on the causal structure of IUIPC and not on the soundness of the underlying measurement model of IUIPC-10.
Our study goes beyond Sipior et al.'s by evaluating the heart of IUIPC, that is, its measurement model, and by doing so in adequately sized confirmatory factor analyses.

Morton~\cite[p. 472]{morton2013measuring} conducted an adequately sized exploratory factor analysis of IUIPC-10 ($N = 353$) as part of his pilot study for the development of the scale on Dispositional Privacy Concern (DPC). He observed a misloading of the item we call \textsf{awa3} between the dimensions Awareness and Control in that EFA. He chose to exclude the two items we call \textsf{ctrl3} and \textsf{awa3} from IUIPC-10. Even though not highlighted as a main point of the paper, Morton's EFA raised concerns on the validity and factor structure of IUIPC-10.
Our analysis differs from Morton's by employing confirmatory factor analysis poised to systematically establish construct validity, by offering a wide range of diagnostics beyond the factor loadings and by re-confirming our analyses on an independent validation sample. While Morton's pilot EFA largely yields statements on his sample, our analysis is a pre-registered confirmatory study yields at results holding generally for the instrument itself and irrespective of estimation methods used.

To the best of our knowledge, the factor structure of IUIPC-10 has not undergone an adequately sized dedicated independent analysis, to date. As a starting point for that inquiry, we offer a rigorous evaluation of the original IUIPC-10 scale in Section~\ref{sec:review}.

\section{Review of IUIPC-10}
\label{sec:review}

\refterm{IUIPC}{Internet users' information privacy concerns} (IUIPC-10)~\cite{malhotra2004internet}  
was created in two studies, determining a preliminary factor structure in an EFA on Study 1 ($N_{\mathsf{IUIPC, 1}} = 293$) and confirming it in a LISREL covariance-based ML CFA on Study 2 ($N_{\mathsf{IUIPC, 2}} = 449$). 
\processifversion{DocumentVersionTR}{This diligent approach---to first establish a factor structure and, then, confirming it on an independent sample---supported its wide-spread endorsement in the community.}
We have asked the authors of the original IUIPC scale for a dataset or covariance matrix to directly compare against their results. Sung S. Kim~\cite{Kim2020} was so kind to respond promptly and stated that they could locate these data.

\subsection{Content Validity}
The authors~\cite[pp. 338]{malhotra2004internet}  make a compelling and well-argued case for the content relevance of the \refterm{dimensions_IUIPC}{information privacy concern dimensions} of collection, control and awareness (cf. in Section~\ref{sec:genesis}). Being rooted in Social Contract (SC) theory, the authors focus on one of the key SC principles they quoted as ``norm-generating microsocial contracts must be founded in informed consent, buttressed by rights of exit and voice,''\processifversion{DocumentVersionTR}{~\cite{dunfee1999social} }which, in turn, underpins the respondents perception of fairness of information collection contingent on their granted control and awareness of intended use.

The questionnaire consisted of ten Likert 7-point items anchored on 1=``Strongly disagree'' to 7=``Strongly Agree.'' We included the questionnaire in the Materials Appendix~\ref{app:materials} Table~\ref{tab:iuipc}.
In terms of question format, we observe two types of questions present
\begin{inparaenum}[(i)]
   \item statements of belief or conviction, e.g., ``Consumer control of personal information lies at the heart of consumer privacy.'' (\textsf{ctrl2}) and
   \item statements of concern, e.g., ``It usually bothers me when online companies ask me for personal information.''
\end{inparaenum}
We would classify \textsf{ctrl1}, \textsf{ctrl2}, \textsf{awa1} and \textsf{awa2} as belief statements, the remainder as concern statements.

Considering the temporal reference point of the questions, we observe that the questionnaire aims at long-term \defterm{trait}{trait} statements, evoked for instance by keywords like ``usually.'' Consequently, we believe IUIPC not to respond strongly to respondents short-term changes in \defterm{state}{state}.

When it comes to content relating to \refterm{barriers}{psychometric barriers} and biases, we find that the questionnaire mentions the \refterm{loaded_word}{loaded word} ``privacy'' four times. It further uses \refterm{loaded_word}{loaded words}, such as ``autonomy.'' Hence, we expect the questionnaire to exhibit a systematic \refterm{priming}{priming bias} and a \refterm{social_desirability}{social desirability bias}, leading to a negative skew of measured scores.
The priming is aggravated by the question order, in which the loaded words are most predominant in the first three items.

We find \refterm{leading_question}{leading questions} too: \textsf{ctrl3} ``I believe that online privacy is invaded when control is lost or unwillingly reduced as a result of a marketing transaction,'' is a leading question towards thinking about the more specific theme of ``marketing transactions.'' Other questions, such as \textsf{awa1} (``Companies seeking information online should disclose the way the data are collected, processed, and used.'' induce agreement---why would respondent disagree with such a statement?

Two items exhibit a \refterm{double-barreled_question}{double-barreled structure}: \textsf{ctrl3}---``I believe that online privacy is invaded when control 
\begin{inparaenum}[(i)]
  \item is lost 
  \emph{or}
  \item unwillingly reduced\dots''
\end{inparaenum}
We find for \textsf{awa3}---``It is very important to me that I am 
\begin{inparaenum}[(i)]
  \item aware \emph{and} 
  \item knowledgeable\dots
\end{inparaenum}
In both cases, we can ask how a participant will answer if only one of the two clauses is fulfilled, or both.

Finally, we find that the questionnaire only contains positively-oriented items. The absence of reverse-coding may lead to \refterm{nondifferentiation}{nondifferentiation}, a risk also observed by Preibusch~\cite{preibusch2013guide}. This can set up the respondents' \refterm{acquiescent_response}{acquiescent response bias}.

\subsection{Sample}
The samples were obtained by ``students in a marketing research class at a large southeastern university 
in the United States [\dots] collecting the survey data''~\cite[p. 343]{malhotra2004internet} from households in the catchment area of the university in one-to-one interviews. 
No explicit survey population or sampling frame was reported, placing the sampling process in the realm of judgment sampling.
While there was no information given how the sample size was determined, the total sample for Study 2 ($N_{\mathsf{IUIPC, 2}} = 449$) was not unreasonable.
\begin{DocumentVersionTR}
We noticed that the sample of Study 2 used for the IUIPC-10 confirmatory factor analysis was composed of two sub-samples differing in scenarios presented: Study 2A ($N_{\mathsf{IUIPC, 2A}} = 217$) contained a scenario with less sensitive information; Study 2B ($N_{\mathsf{IUIPC, 2B}} = 232$) contained a scenario with more sensitive information. 
While these two sub-samples did not seem to differ significantly in their demographics, we could not discern from the paper, how the variance from the difference in the scenarios was taken into account in the factor analysis of IUIPC-10.
\end{DocumentVersionTR}

\subsection{Construct Validity}
\subsubsection{Assumptions}
In terms of \refterm{ml_assumptions}{assumptions and requirements} of Maximum Likelihood estimation, the paper did not mention how distribution, univariate and multivariate outliers were handled. Kim~\cite{Kim2020} clarified that they ``did not check the distribution or outliers'' and that they ``relied on the robustness of maximum likelihood estimation'', a case Bovaird and Koziol~\cite[p. 497]{bovaird2012measurement} called ``ignoring ordinality.''
In the field, there are practitioners considering the ML estimator robust enough to handle ordinal data with more than five levels as well as empirical analyses cautioning against this practice~\cite{finney2006non,bovaird2012measurement}.
We found IUIPC-10 surveys to yield non-normal data with a negative skew throughout, rendering the questionnaire scores less suitable to be covered by ML-robustness results\processifversion{DocumentVersionTR}{~\cite[p. 258]{kline2015principles}}.

\subsubsection{Factorial Validity}
\processifversion{DocumentVersionTR}{\paragraph{Global Fit}}
In terms of global fit as evidence for \refterm{factorial_validity}{factorial validity}, we outlined the fit measures reported for the original IUIPC instrument~\cite{malhotra2004internet} in Table~\ref{tab:iuipc_original_fit}. Though not reported, we estimated the $\chi^2$ $p$-value, the \textsf{RMSEA} 90\% confidence interval, and $p_{\hat{\epsilon}_0 \leq .05}$ from the $\chi^2$ test statistic. \processifversion{DocumentVersionTR}{The confidence interval was estimated with the \textsf{R} package \textsf{MBESS}, the close-fit $p$-value estimated directly from the $\chi^2$ test statistic, using the method introduced by Browne and Cudeck~\cite[p. 146]{BroCud1993alternative}.}

In terms of statistical inferences, we observed that the original IUIPC model failed the \refterm{exact_fit}{exact-fit} test and
the \refterm{not_close_fit}{not-close fit} test ($\hat{\epsilon}_\mathsf{UL} < .05$).
It passed the the \refterm{close_fit}{close-fit} test, \refterm{poor_fit}{poor-fit} test and the \refterm{combination_rule}{combination rule (HBR)}. IUIPC did not report the model's \textsf{SRMR}.

\begin{table}[tb]
\centering
\footnotesize
\caption{Fit of IUIPC-10's second-order model~\cite[p. 346]{malhotra2004internet}%
\explain{.\newline{}The tests of the exact-fit hypothesis on the $\chi^2$ and the not-close fit hypothesis on the \textsf{RMSEA} failed. The tests of close-fit and poor-fit hypotheses passed. The combination of \textsf{CFI}, \textsf{GFI}, and \textsf{RMSEA} supports a satisfactory fit.}}
\label{tab:iuipc_original_fit}
\begin{tabular}{r@{ }r@{ }rrrr@{ }r}
  \toprule
$\chi^{2}$ & $\vari{df}$ & \textit{p} & \textsf{CFI} & \textsf{GFI} & \textsf{RMSEA} & $p_{\hat{\epsilon}_0 \leq .05}$\\ 
  \midrule
73.19 & 32 & $< .001$ & $.98$ & $.97$ & .054 [.037, .070]  &  $.338$ \\
   \bottomrule
\end{tabular}
\\\emph{Note:} $N_{\mathsf{IUIPC}} = 449$. \textsf{RMSEA} with inferred 90\% CI. 
\end{table}

\processifversion{DocumentVersionTR}{\paragraph{Local Fit}}
The original IUIPC paper did not contain an analysis of residuals. We did not have the data at our disposal to compute it ourselves~\cite{Kim2020} and could, thereby, not evaluate the local fit. Hence, the evidence for \refterm{factorial_validity}{factorial validity} was incomplete.

\subsubsection{Convergent and Discriminant Validity}
In the item-level evaluation of \refterm{convergent_validity}{convergent validity}, 
we first examined the factor loadings reported for IUIPC.
The EFA of IUIPC is stated by the authors to have retained items that loaded greater than $.70$ on their designated factors, and less than $.40$ on other factors. The CFA of the measurement model was reported to have had a minimal factor loading of $.61$ for awareness. A detailed factor loading table was not reported; neither standardized loadings or $R^2$ for individual items.
Criteria for $\vari{AVE}$ and composite reliability were fulfilled, just so for $\vari{AVE}$.
In terms of discriminant validity, we find that the Fornell-Larcker criterion is violated for awareness (\textsf{awa}) and control (\textsf{ctrl}), where the their correlation is greater than the square root of their respective $\vari{AVE}$.

\begin{table}[tb]
\centering
\footnotesize
\caption{Validity and reliability evidence on IUIPC~\cite[Tab. 2]{malhotra2004internet}\explain{. Low $\vari{AVE}$ with \textsf{awa} shy of the $\vari{AVE} > .50$ criterion cautions against low internal consistency, even if construct reliability $\omega > .70$ is sufficient, entailing a moderate signal-to-noise ratio $\vari{S/\!N}_\omega$. That the $\sqrt{\vari{AVE}}$ of \textsf{awa} and \textsf{ctrl} on the diagonal of the correlation table is less than the correlation with the respective other factor violates the Fornell-Larcker criterion for discriminant validity.}}
\label{tab:iuipc_reliability}
\begin{adjustbox}{max width=\columnwidth}
\begin{tabular}{llrrrrrrrr}
\toprule
  & & & & & & \multicolumn{3}{c}{Correlations}\\
\cmidrule(lr){7-9}
  &                     &  $M$ & $\vari{SD}$ &  $\vari{AVE}$ & $\omega$ & $\vari{S/\!N}_\omega$ & 1 & 5 & 6 \\
\midrule
1. & \textsf{coll} &  5.63 & 1.09 &         .55 & .83 & 4.88  &  .74 & & \\ 
5. & \textsf{awa} & 6.21 & 0.87 &         .50 & .74 & 2.85  & .66 & .71 \\
6. & \textsf{ctrl}   & 5.67 & 1.06 &         .54 & .78 & 3.55 & .53 & .75 & .73 \\
\bottomrule
\end{tabular}
\end{adjustbox}
\\\emph{Note:} Value on the diagonal is the square root of AVE.
\end{table}

\subsection{Reliability}
Considering the internal consistency criteria vis-{\`a}-vis of Table~\ref{tab:iuipc_reliability}, we find that the Average Variance Extracted criterion $\vari{AVE} > .50$ is just so fulfilled, with awareness on the boundary of acceptable. The composite reliability $\omega$ is greater than $.70$ throughout, with \textsf{coll} achieving the best value ($.83$). The reported reliability is low for \vari{AVE} but decent for composite reliability.

\subsection{Summary}
Content validity is a strong point of IUIPC as the argument on relevance seems compelling. We observed problems in the questionnaire wording that could induce systematic biases into the scale, at the same time.
Considering the two-step process with a preliminary EFA and a subsequent CFA, there is an impression that IUIPC has been diligently done.
In terms of construct validity, we found that IUIPC reported a satisfactory global fit, while unchecked assumptions, an estimation at odds with non-normal, ordinal data, and the missing information on local fit weakened the case. While evidence of convergent validity was scarce without a factor loading table or standardized loadings to work with, discriminant validity was counter-indicated. Finally, the low---if not disqualifying---\vari{AVE} in the reliability inquiry will caution privacy researchers to expect an only moderate signal-to-noise ratio ($\vari{S/\!R}_\omega$ between $2.85$ and $4.88$) and attenuation of effects on other variables.

\section{Aims}
\label{sec:aims}

\subsection{Construct Validity and Reliability}
\label{sec:aims.fc}

Our main goal is an independent confirmation of the IUIPC-10 instrument by Malhotra et al.~\cite{malhotra2004internet}, where we focus on \refterm{construct_validity}{construct validity} and \refterm{reliability}{reliability}.
\begin{researchquestion}[Confirmation of IUIPC]
\label{rq:confirm}
To what extent can we confirm IUIPC-10's construct validity and reliability?
\end{researchquestion}
This aim largely entails confirming the \refterm{factorial-validity}{factorial validity}, that is, the three-dimensional factor structure, of IUIPC. The first inquiry there is to compare alternative models of the IUIPC with different factor solutions\processifversion{DocumentVersionTR}{ based on the following statistical hypotheses:
\begin{compactdesc}
  \item[$H_{\const{LRT}, \chi^2, 0}$:] The models' approximations do not differ beyond measurement error.
  \item[$H_{\const{LRT}, \chi^2, 1}$:] The models' approximations differ
\end{compactdesc}}.

Second, we will gather further evidence for \refterm{factorial-validity}{factorial validity} by seeking to fit IUIPC-10 hypothesized second-order model, where the \refterm{unidimensionality}{unidimensionality} of its sub-scales, that is, the absence of cross-loadings is a key consideration. This will be tested with \refterm{fit_tests}{statistical inferences on the models global fit} based on the \refterm{fit_tests}{statistical hypotheses} introduced in \processifversion{DocumentVersionTR}{Section~\ref{sec:fit.tests}}\processifversion{DocumentVersionConference}{Appendix~\ref{sec:fit.tests}} indicating increasingly worse approximations:
\begin{inparaenum}[(i)]
  \item \refterm{exact_fit}{Exact Fit} ($H_{\chi^2, 0}$),
  \item \refterm{close_fit}{Close Fit} ($H_{\hat{\varepsilon} \leq .05, 0}$),
  \item \refterm{not_close_fit}{Not Close Fit} ($H_{\hat{\varepsilon} \geq .05, 0}$),
  \item \refterm{poor_fit}{Poor Fit} ($H_{\hat{\varepsilon} \geq .10, 0}$).
\end{inparaenum}
We further evaluated the \refterm{combination_rule}{combination rule} used by Malhotra et al.~\cite{malhotra2004internet}:
\begin{inparaenum}[(i)]
  \item $\mathsf{CFI} > .95$,
  \item $\mathsf{GFI} > .90$,
  \item $\mathsf{RMSEA} < .06$.
\end{inparaenum}
This global fit analysis will be complemented by an assessment of local fit on residuals.

This inquiry is complemented by analyses of \refterm{convergent_validity}{convergent} and \refterm{discriminant_validity}{discriminant} validity on the criteria established in Section~\ref{sec:construct_validity} similarly to analyses of \refterm{internal_consistency}{internal-consistency} \refterm{reliability}{reliability} on criteria from Section~\ref{sec:reliability}

Overall, the aim of evaluating the construct validity and reliability of IUIPC-10 is not just about a binary judgment, but a fine-grained diagnosis of possible problems and viable improvements. This wealth of evidence to enable privacy researchers to form their own opinions.

\subsection{Estimator Appraisal}
\label{sec:aims.estimator}

\processifversion{DocumentVersionTR}{While the creators of IUIPC-10 have employed a particular data preparation and estimation approach, given the current state of the field, we would advocate other design choices. However, the differences in our and their designs could act as confounders on a direct replication. For that reason, a}\processifversion{DocumentVersionConference}{A}s second line of inquiry, we considered multiple estimation methods in conceptual replications shown in Figure~\ref{fig:aims} on the horizontal axis.

\begin{researchquestion}[Estimator Invariance]
\label{rq:invariance}
To what extent do the confirmation results in regards to RQ~\ref{rq:confirm} hold irrespective of the estimator used?
\end{researchquestion}
For that, we expect the \refterm{fit_tests}{statistical hypotheses} of RQ~\ref{rq:confirm} to yield the same outcome across estimation methods. For respecifications, we expect the fit indices (especially \textsf{CFI} and \textsf{CAIC}) to show appreciable improvements comparing the models on their respective estimators shown in Figure~\ref{fig:aims} on the vertical axis.

In addition, we aim at gaining an empirical underpinning to design decisions made in the field with respect to the methodological setup of CFAs and SEMs with IUIPC and similar scales.
\begin{researchquestion}
\label{rq:estimator}
Which estimator is most viable to create models with IUIPC, measured in ordinal 7-point Likert items?
\end{researchquestion}
We aim at investigating the viability of alternatives to the maximum likelihood (ML) estimation:
\begin{inparaenum}[(i)] 
  \item scaled estimation (MLM) and
  \item estimation specializing on ordinal variables (\refterm{wlsmvs}{robust WLS}).
\end{inparaenum}
As discussed in Section~\ref{sec:est_assumptions}, \refterm{wlsmvs}{robust WLS} is a far cry from ML/MLM estimation. Hence, a plain comparison on their fit indices, such as on the consistent Akaike Information Criteria (\textsf{CAIC}), may lead us astray: their fit measures are not directly comparable in a fair manner. 

Thereby, the question becomes: Are the respective estimations viable in their own right, everything else being equal?
To what extent do the estimators offer us a plausible approximation of IUIPC?
For these questions we aim at estimating mean structures throughout such that we can assess their first-order predictions on indicators. Without knowing the ground truth of the true IUIPC scores of our samples, the assessment on what estimator is most viable will be largely qualitative.

\begin{DocumentVersionTR}
\subsection{Recommendations for the Privacy Community}
We intended to gather empirically grounded recommendations for the privacy community on how to treat questionnaires, such as IUIPC-10. We focused especially on non-normal, Likert-type/ordinal instruments. Our recommendations included the use of data preparation as well as estimators.
\end{DocumentVersionTR}

\section{Method}

The project was registered on the Open Science Framework\footnote{OSF: \url{https://osf.io/5pywm}}. The OSF project contains the registration document as well as technical supplementary materials, incl. 
\processifversion{DocumentVersionTR}{
\begin{inparaenum}[(i)]
  \item correlation-SD tables for all samples,
  \item SEM and residual tables for all estimated models,
  \item \textsf{R} importable covariance matrices of all samples and related data needed to reproduce the ML- and MLM-estimated CFAs in \textsf{lavaan}.
\end{inparaenum}
}\processifversion{DocumentVersionConference}{\textsf{R} covariance matrices and related data needed to reproduce the ML- and MLM-estimated CFAs.}
The statistics were computed in \textsf{R} \processifversion{DocumentVersionTR}{(v. 3.6.2)} largely using the package \textsf{lavaan}\processifversion{DocumentVersionTR}{ (v. 0.6-5)}, where graphs and tables were largely produced with \textsf{knitr}.
The significance level was set to $\alpha = .05$.

\subsection{Ethics}
The ethical requirements of the host institution were followed and ethics cases registered.
Participants were recruited under informed consent.
They could withdraw from the study at any point.
They were enabled to ask questions about the study to the principal investigator.
They agreed to offer their demographics (age, gender, mother tongue) as well as the results of the questionnaires for the study.
Participants were paid standard rates for Prolific Academic, \pounds{12}/hour, which is greater than the UK minimum wage of \pounds{8.21} during the study's timeframe.
The data of participants was stored on encrypted hard disks, their Prolific ID only used to ensure independence of observations and to arrange payment.

\subsection{Sample}
We used three independent samples, \textsf{A}, \textsf{B}, \textsf{V} in different stages of the analysis.
Auxiliary sample~\textsf{A} was collected in a prior study and aimed at a sample size of $200$ cases.

Base sample~\textsf{B} and validation sample~\textsf{V} were collected for a current investigation. They had a designated sample size of $420$ each, based on an \textit{a priori} power analysis for structural equation modelling with RMSEA-based significance tests. 

While all three samples were recruited on Prolific Academic, \textsf{B} and \textsf{V} were recruited to be representative of the UK census by age and gender. The sampling frame was Prolific users who were registered to be residents of the UK, consisting of $48,454$ users at sampling time. 
The sampling process was as follows:
\begin{inparaenum}[1.]
  \item Prolific presented our studies to all users with matching demographics,
  \item the users could choose themselves whether they would participate or not.
\end{inparaenum}
We prepared to enforce sample independence by uniqueness of the participants' Prolific ID.

We planned for excluding observations from the sample, without replacement, because
\begin{inparaenum}[(i)]
  \item observations were incomplete,
  \item observations were duplicates by Prolific ID,
  \item participants failed more than one attention check,
  \item observations constituted multi-variate outliers determined with a Mahalanobis distance of $12$ or greater.
\end{inparaenum}

\subsection{Analysis Approach}
\begin{figure}
\includegraphics[keepaspectratio,width=\columnwidth]{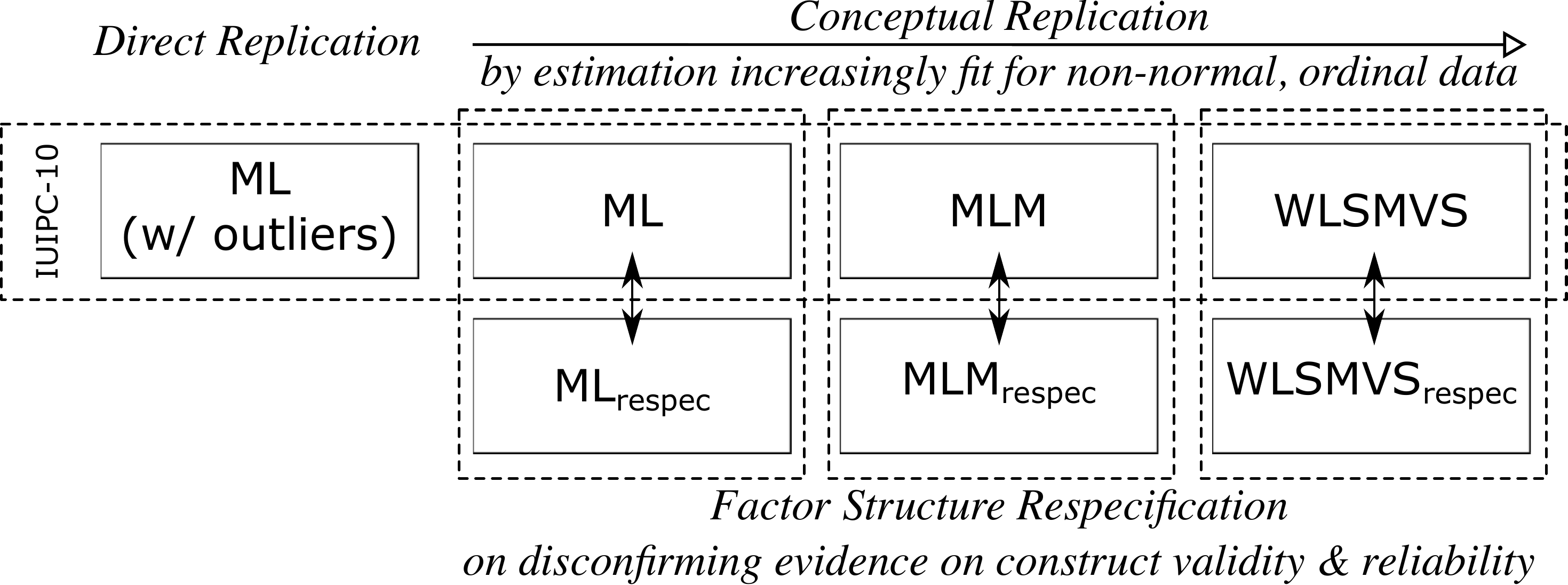}
\caption{Analysis approach}
\label{fig:aims}
\end{figure}

We bridged between a direct replication of the IUIPC-10 analysis approach and conceptual replications adopting to non-normal, ordinal data. We illustrate the dimensions of our approach in Figure~\ref{fig:aims}. As a direct replication with ML estimation and without distribution or outlier consideration would have exposed this study to unpredictable confounders, we computed our analysis with three estimators \textit{ceteris paribus}. We computed all models including their mean structures.

We faced the didactic challenge that even though WLSMVS would be best suited for the task at hand~\cite{bovaird2012measurement}, it is least used in the privacy community. And its probit estimation is interpreted differently to other methods. Hence, we chose to make the MLM model the primary touch stone for our analysis. It carries the advantages of being robust to moderately skewed non-normal data and of yielding interpretations natural to the community.

\begin{figure}
\centering
\includegraphics[keepaspectratio, width=\columnwidth]{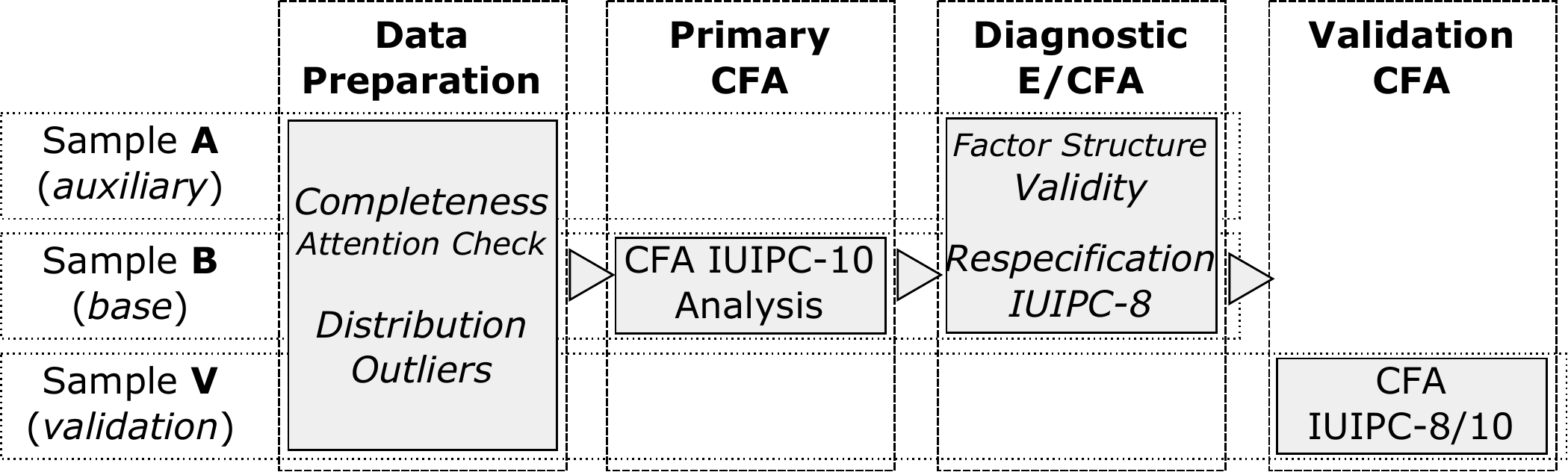}
\caption{Which steps were taken on what sample}
\label{fig:process}
\end{figure}

In our analysis process depicted in Figure~\ref{fig:process}, we used our three samples deliberately. For the factor analysis of IUIPC-10, we used base Sample \textsf{B} as main dataset to work with.
We retained Sample \textsf{A} as an auxiliary sample to conduct exploratory factor analyses and to have respecification proposals informed by more than one dataset, thereby warding against the impact of chance.
Sample~\textsf{V} was reserved for validation after a final model was chosen.
Figure~\ref{fig:process} illustrate this relationship of the different samples to analysis stages.

First, we established a sound data preparation, including consideration for measurement level and distribution as well as outliers.
Second, we computed a covariance-based confirmatory factor analysis on the IUIPC-10 second-order model~\cite{malhotra2004internet}, complemented with alternative one-factor and two-factor models. This comparison served to confirm the three-dimensionality of IUIPC.
We evaluated the hypothesized IUIPC-10 model on Sample~\textsf{B}, gathering evidence for construct validity in the form of factorial validity evident in global and local fit, convergent and discriminant validity, as well as reliability.

Having found inconsistencies, we then engaged in a diagnosis and respecification stage. Therein, we also computed a parallel polychoric factor analysis and EFAs on samples~\textsf{A} and \textsf{B} to re-assess the three-dimensional factor structure itself and to hunt down patterns of weaknesses. From this evaluation, we prepared a respecified IUIPC-8 which was first evaluated on Sample \textsf{B}. We compared the non-nested models of IUIPC-10 and IUIPC-8 with the Vuong Likelihood Ratio Test on the ML estimation. Otherwise, we compared between fit indices, focusing on an evenly weighted CAIC for non-nested comparisons.

Finally, once respecification and design decisions were settled, we entered the CFA validation stage. Therein, we compared the performance of the original IUIPC-10 and the respecified IUIPC-8 on the independent validation Sample~\textsf{V}.

\begin{RedundantContent}
\subsection{Estimator Choice}

For the line of inquiry on which estimator to use, we computed three confirmatory factor analyses on the respective samples and IUIPC variants, differing in their estimators.
Everything else being equal, these CFAs compared between
\begin{compactdesc}
  \item[ML:] Standard Maximum Likelihood estimation,
  \item[MLM:] Maximum Likelihood estimation with robust standard errors and a Satorra-Bentler scaled test statistic, tailored for non-normal continuous data,
  \item[WLSMVS:] Weighted Least Square estimation with Satterthwaite means/variance adjusted test statistics, a preferred option for ordinal data.
\end{compactdesc}
These three estimators are increasingly better prepared to handle non-normal, ordinal data.

The comparison across estimators bears the note of caution that the estimation methods are quite different in their underlying approach and their fit measures not directly comparable. Hence, while the fit measures can show that an estimation performed well in its own right, it would stretch the interpretation to compare them across estimators.
\end{RedundantContent}


\section{Results}
\label{sec:results}

\subsection{Sample}
\label{sec:sample}

\begin{table}
\centering
\caption{Sample Refinement}
\label{tab:sample}
\begin{adjustbox}{max width=\columnwidth}
\begin{tabular}{lc@{ }cc@{ }cc@{ }c}
\toprule
\multirow{2}{*}{Phase} &  \multicolumn{2}{c}{\textsf{A}} & \multicolumn{2}{c}{\textsf{B}} & \multicolumn{2}{c}{\textsf{V}}\\
\cmidrule(lr){2-3} \cmidrule(lr){4-5} \cmidrule(lr){6-7}
  & Excl. & Size & Excl. & Size & Excl. & Size\\ 
\midrule
Starting Sample &   & $226$ & & $473$ & & $467$\\
Incomplete & $0$ & $226$
   & $58$ & $415$ 
   & $34$ & $433$\\
Duplicate & $7$ & $219$
   & $25$ & $390$ 
   & $0$ & $433$\\
\textsf{FailedAC}$ > 1$ & $14$ & $205$
   & $11$ & $379$ 
   & $0$ & $433$\\
MV Outlier & $4$ & $201$
   & $9$ & $370$ 
   & $14$ & $419$\\
\midrule
Final Sample & \multicolumn{2}{r}{$N_{\mathsf{A}}^\prime = 201$} &
               \multicolumn{2}{r}{$N_{\mathsf{B}}^\prime = 370$} &  
               \multicolumn{2}{r}{$N_{\mathsf{V}}^\prime = 419$}\\
\bottomrule
\end{tabular}
\end{adjustbox}
\\\emph{Note:} $N_{\mathsf{A}} = 205$, $N_{\mathsf{B}} = 379$, $N_{\mathsf{V}} = 433$ are after attention checks.
\end{table}

%
%
\newcommand{\demoCombinedAll}{
\begin{table}[tb]
\centering\caption{Demographics of Samples \textsf{B} and \textsf{V}}
\captionsetup{position=top}
\label{tab:demoCombined}
\subfloat[Sample \textsf{B}]{
\label{tab:demoCombinedB}
\centering
\begin{tabular}{ll}
\toprule
  & Overall\\
\midrule
$N_{\mathsf{B}}$ & 379\\
Gender (\%) & \\
Female & 197 (52.0)\\
Male & 179 (47.2)\\
Rather not say & 3 ( 0.8)\\
\addlinespace
Age (\%) & \\
18-24 & 41 (10.9)\\
25-34 & 72 (19.0)\\
35-44 & 84 (22.2)\\
\addlinespace
45-54 & 57 (15.0)\\
55-64 & 97 (25.6)\\
65+ & 28 ( 7.4)\\
Mothertongue (\%) & \\
English & 374 (98.7)\\
\addlinespace
Greek & 1 ( 0.3)\\
Portuguese & 1 ( 0.3)\\
Tagalog & 1 ( 0.3)\\
Other & 2 ( 0.5)\\
\bottomrule
\end{tabular}
}~\subfloat[Sample \textsf{V}]{%
\label{tab:demoCombinedV}
\centering
\begin{tabular}{ll}
\toprule
  & Overall\\
\midrule
$N_{\mathsf{V}}$ & 433\\
Gender (\%) & \\
Female & 217 (50.1)\\
Male & 212 (49.0)\\
Rather not say & 4 ( 0.9)\\
\addlinespace
Age (\%) & \\
18-24 & 92 (21.2)\\
25-34 & 143 (33.0)\\
35-44 & 83 (19.2)\\
\addlinespace
45-54 & 58 (13.4)\\
55-64 & 44 (10.2)\\
65+ & 13 ( 3.0)\\
Mothertongue (\%) & \\
English & 422 (97.5)\\
Arabic & 1 ( 0.2)\\
\addlinespace
Greek & 2 ( 0.5)\\
Hindi & 1 ( 0.2)\\
Italian & 2 ( 0.5)\\
Polish & 1 ( 0.2)\\
Russian & 1 ( 0.2)\\
\addlinespace
Spanish & 1 ( 0.2)\\
Urdu & 1 ( 0.2)\\
Other & 1 ( 0.2)\\
\bottomrule
\end{tabular}
}
\end{table}
}

\newcommand{\demoCombined}{
\begin{table}[tb]
\centering\caption{Demographics of Samples \textsf{B} and \textsf{V}}
\captionsetup{position=top}
\label{tab:demoCombined}
\subfloat[Sample \textsf{B}]{
\label{tab:demoCombinedB}
\centering
\begin{tabular}{ll}
\toprule
  & Overall\\
\midrule
$N_{\mathsf{B}}$ & 379\\
Gender (\%) & \\
Female & 197 (52.0)\\
Male & 179 (47.2)\\
Rather not say & 3 ( 0.8)\\
\addlinespace
Age (\%) & \\
18-24 & 41 (10.9)\\
25-34 & 72 (19.0)\\
35-44 & 84 (22.2)\\
\addlinespace
45-54 & 57 (15.0)\\
55-64 & 97 (25.6)\\
65+ & 28 ( 7.4)\\
\bottomrule
\end{tabular}
}~\subfloat[Sample \textsf{V}]{%
\label{tab:demoCombinedV}
\centering
\begin{tabular}{ll}
\toprule
  & Overall\\
\midrule
$N_{\mathsf{V}}$ & 433\\
Gender (\%) & \\
Female & 217 (50.1)\\
Male & 212 (49.0)\\
Rather not say & 4 ( 0.9)\\
\addlinespace
Age (\%) & \\
18-24 & 92 (21.2)\\
25-34 & 143 (33.0)\\
35-44 & 83 (19.2)\\
\addlinespace
45-54 & 58 (13.4)\\
55-64 & 44 (10.2)\\
65+ & 13 ( 3.0)\\
\bottomrule
\end{tabular}
}
\end{table}
}

\newcommand{\demoCombinedWithA}{
\begin{table*}[tb]
\centering\caption{Demographics}
\captionsetup{position=top}
\label{tab:demoCombinedWithA}
\subfloat[Sample \textsf{A}]{
\label{tab:demoCombinedA}
\centering
\begin{tabular}{ll}
\toprule
   & Overall\\
   \midrule
   $N_{\mathsf{A}}$ & 205\\
Gender (\%) & \\
Female & 80 (39.0) \\
Male & 125 (61.0)\\
Rather not say & 0 ( 0.0)\\
\addlinespace
Age (\%) & \\
18-24 & 109 (53.2)\\
25-34 & 71 (34.6)\\
35-44 & 18 ( 8.8)\\
\addlinespace
45-54 & 4 ( 2.0)\\
55-64 & 3 ( 1.5)\\
65+ &    0 ( 0.0)\\
\bottomrule
\end{tabular}
}
\subfloat[Sample \textsf{B}]{
\centering
\label{tab:demoCombinedB}
\centering
\begin{tabular}{ll}
\toprule
  & Overall\\
\midrule
$N_{\mathsf{B}}$ & 379\\
Gender (\%) & \\
Female & 197 (52.0)\\
Male & 179 (47.2)\\
Rather not say & 3 ( 0.8)\\
\addlinespace
Age (\%) & \\
18-24 & 41 (10.9)\\
25-34 & 72 (19.0)\\
35-44 & 84 (22.2)\\
\addlinespace
45-54 & 57 (15.0)\\
55-64 & 97 (25.6)\\
65+ & 28 ( 7.4)\\
\bottomrule
\end{tabular}
}~\subfloat[Sample \textsf{V}]{%
\label{tab:demoCombinedV}
\centering
\begin{tabular}{ll}
\toprule
  & Overall\\
\midrule
$N_{\mathsf{V}}$ & 433\\
Gender (\%) & \\
Female & 217 (50.1)\\
Male & 212 (49.0)\\
Rather not say & 4 ( 0.9)\\
\addlinespace
Age (\%) & \\
18-24 & 92 (21.2)\\
25-34 & 143 (33.0)\\
35-44 & 83 (19.2)\\
\addlinespace
45-54 & 58 (13.4)\\
55-64 & 44 (10.2)\\
65+ & 13 ( 3.0)\\
\bottomrule
\end{tabular}
}
\\\footnotesize{\emph{Note:} Samples \textsf{B} and \textsf{V} were drawn to be representative of the UK census by age and gender; Sample \textsf{A} was not.}
\end{table*}
}
\demoCombinedWithA

We refined the three samples~\textsf{A}, \textsf{B} and \textsf{V} in stages, where Table~\ref{tab:sample} accounts for the refinement process. First, we removed incomplete cases without replacement. Second, we removed duplicates across samples by the participants' Prolific ID. Third, we removed cases in which participants failed more than one attention check ($\mathsf{FailedAC} > 1$). Overall, of the $N_{\mathsf{C}} = 1074$ complete cases, only $5.3 \%$ were removed due to duplicates or failed attention checks.

The demographics the samples are outlined in Table~\ref{tab:demoCombinedWithA}. In samples \textsf{B} and \textsf{V} meant to be UK representative, we found a slight under-representation of elderly participants compared to the UK census age distribution.
Interested readers can reproduce the ML and MLM CFA from the correlation matrices and the standard deviations of the samples included in Appendix~\ref{app:sample}, \processifversion{DocumentVersionTR}{Table~\ref{tab:inputCorSDBV}}\processifversion{DocumentVersionConference}{Table~\ref{tab:inputCorSDB}}. 

\begin{DocumentVersionTR}
\subsection{Data Preparation}\label{sec:dataprep}
\processifversion{DocumentVersionTR}{\subsubsection{Extreme Multi-Collinearity}}
While we found considerable correlation between indicator variables (cf. Appendix Table~\ref{tab:inputCorSDBV}),
there was no extreme case of multi-collinearity.

\processifversion{DocumentVersionTR}{\subsubsection{Non-Normality}}
We checked the input variables in all samples for indications of uni-variate \refterm{normal_distribution}{non-normality}. 
\processifversion{RedundantContent}{The distribution histograms and density plots are available in Figure~\ref{fig:histograms} in Appendix~\ref{app:sample}.}
All input variables (of all samples) were moderately \refterm{skewness}{negatively skewed} with the most extreme skew being $-2.09$.
In general, all input variables apart from \textsf{coll1} showed a \refterm{normal_distribution}{positive kurtosis}, less than $2.4$.

While already \refterm{ordinal}{ordinal} in measurement level (7-point Likert scales), all input items need to be considered \refterm{normal_distribution}{non-normal}, yet not extremely so.

\processifversion{DocumentVersionTR}{\subsubsection{Outliers}}
We checked for uni-variate and multivariate outliers. These checks were computed on parcels of indicator variables of 1\textsuperscript{st}-order sub-scales \textsf{Control}, \textsf{Awareness} and \textsf{Collection}.
We checked for univariate outliers with the robust outlier labeling rule and multi-variate outliers on all three variables with the Mahalanobis distance (at a fixed threshold of $12$). We checked the sample distributions and marked outliers with 3-D scatter plots on the three variables.

All our IUIPC-10 datasets yielded $6\%$ univariate outliers by the robust outlier labeling rule and $3\%$ multi-variate outliers with a Mahalanobis distance of $12$ or greater.
We removed multi-variate outliers from the samples without replacement, yielding sample sizes $N_{\mathsf{A}}^\prime$, $N_{\mathsf{B}}^\prime$, and $N_{\mathsf{V}}^\prime$, respectively, as indicated in Table~\ref{tab:sample}.
\end{DocumentVersionTR}

\subsection{Descriptive Statistics}
\label{sec:desc}
We found all indicator variables of all samples to be substantively \refterm{skewness}{negatively skewed}, meaning that there are relatively few small values and that the distribution tails off to the left~\cite[p. 48]{hair2019multivariate}, with the most extreme skew being $-2.09$. In general, all indicators apart from \textsf{coll1} showed a substantive positive kurtosis~\cite[pp. 74]{kline2015principles}, that is, peakedness, less than $2.4$. While this pattern of substantive \refterm{normal_distribution}{non-normality} was present in the indicator distributions, we also found it in the IUIPC sub-scales and illustrate these distributions in Table~\ref{tab:descSubScalesOrig} and Figure~\ref{fig:densitySubScales}. 
We observed that the three samples had approximately equal distributions by sub-scales. 
\begin{DocumentVersionTR}
Explicitly controlling for the difference between Samples \textsf{B} and \textsf{V}, we found that none of their sub-scale means were statistically significantly different\processifversion{DocumentVersionTR}{:}\processifversion{DocumentVersionConference}{, the maximal absolute standardized mean difference being 0.13---a small magnitude.}%
\begin{inparaenum}[(i)]
   \item \textsf{ctrl}, $t(807.35) = 1.08, p = .279$, $g = 0.08$, 95\% CI $[-0.06, 0.21]$;
  \item \textsf{awa}, $t(801.11) = 1.9, p = .058$, $g = 0.13$, 95\% CI $[-0.01, 0.27]$;
  \item \textsf{coll}, $t(776.13) = -0.25, p = .802$, $g = -0.02$, 95\% CI $[-0.16, 0.12]$;
  \item \textsf{iuipc}, $t(804.51) = 0.92, p = .356$, $g = 0.06$, 95\% CI $[-0.07, 0.2]$.
\end{inparaenum}
%
We are, thereby, confident that the obtained descriptive statistics in Table~\ref{tab:descSubScalesOrig} generalize well as benchmarks for IUIPC-10 in the UK population.
We observed that the standardized mean differences between IUIPC-10 and our Sample \textsf{B} by sub-scales were small\processifversion{DocumentVersionTR}{:
\begin{inparaenum}[(i)]
  \item \textsf{ctrl}: $p < .001$, $g = -0.27$, 95\% CI $[-0.41, -0.13]$;
  \item \textsf{awa}: $p < .001$, $g = -0.4$, 95\% CI $[-0.54, -0.27]$;
  \item \textsf{coll}: $p = .538$, $g = 0.04$, 95\% CI $[-0.09, 0.18]$;
  \item \textsf{iuipc}: $p = .005$, $g = -0.2$, 95\% CI $[-0.33, -0.06]$.
\end{inparaenum}}\processifversion{DocumentVersionConference}{, with a maximum absolute standardized mean difference of 0.4. Sample \textsf{B}'s mean estimates of \textsf{ctrl} and \textsf{awa} were statistically significantly greater than IUIPC's, both $p < .001$.}
\end{DocumentVersionTR}

Our IUIPC-10 samples yielded $6\%$ \refterm{outlier}{univariate outliers} by the robust outlier labeling rule and $3\%$ multi-variate outliers with a Mahalanobis distance of $12$ or greater~\cite[pp. 72]{kline2015principles}.

Of the requirements for Maximum Likelihood estimation, we find the \refterm{normal_distribution}{multi-variate normality} violated in a case of \refterm{continuous}{non-continuous measurement}. Our data preparation handled the \refterm{outlier}{outliers} as recommended.

\descSubScalesOrig
\densitySubScales

\subsection{Construct Validity}
\label{sec:construct_val}

\subsubsection{Factorial Validity}
\label{sec:factor_val}
To confirm the hypothesized factor structure of IUIPC-10, we computed confirmatory factor analyses on one-factor, two-factor and the hypothesized three-dimensional second-order model. We present the fit of the respective estimations in Table~\ref{tab:comparison.models}. By a \refterm{LRT}{likelihood-ratio $\chi^2$ difference test}, we concluded that the two-factor solution was statistically significantly better than the one-factor solution, $\chi^2(1) = 138.761, p < .001$. In turn, the three-factor solutions were statistically significantly better than the two-factor solution, $\chi^2(2) = 49.957, p < .001$. We \processifversion{DocumentVersionTR}{rejected the respective hypotheses $H_{\const{LRT}, \chi^2, 0}$ and }accepted the hypothesized three-factor second-order model, offering confirming evidence for its \refterm{factorial_validity}{factorial validity}.

\begin{DocumentVersionTR}

\begin{table*}[tbp]
\centering
\caption{Comparison of different model structures of IUIPC-10 on Sample \textsf{B} with MLM estimation\explain{.\newline{}The models show increasingly better fits in scaled $\chi^2$, \textsf{CFI}, and \textsf{CAIC}, supporting the predicted hierarchical three-factor model.}}
\label{tab:comparison.models}
\begin{adjustbox}{max width=\textwidth}
\begin{tabular}{@{}rrrrrrrrrrrrr@{}}
\toprule
& \multicolumn{3}{c}{One Factor}& \multicolumn{3}{c}{Two Factors}& \multicolumn{3}{c}{Three Factors (1\textsuperscript{st} Order)}& \multicolumn{3}{c}{Three Factors (2\textsuperscript{nd} Order)}\tabularnewline 
\midrule
$\chi^{2} (\mathit{df})$& 481.87 (35)& & & 239.28 (34)& & & 163.69 (32)& &  & 163.69 (32)& & \tabularnewline
$\chi^2/\mathit{df}$ & 13.77 &&& 7.04 &&& 5.12 &&& 5.12 \tabularnewline
\textsf{CFI} & .73& & & .87& & & .92& & & .92& & \tabularnewline
\textsf{GFI}& 1.00& & & 1.00& & & 1.00& & & 1.00& & \tabularnewline
\textsf{RMSEA} & .19& \multicolumn{2}{c}{[.17, .20]} & .13& \multicolumn{2}{c}{[.11, .14]} & .11& \multicolumn{2}{c}{[.09, .12]} & .10& \multicolumn{2}{c}{[.09, .12]} \tabularnewline
\textsf{SRMR}& .14& & & .09& & & .10& & & .10& & \tabularnewline
Scaled $\chi^{2} (\mathit{df})$& 377.87 (35)& & & 189.71 (34)& & & 131.42 (32)& & & 131.42 (32)& & \tabularnewline
\textsf{CAIC} & 600.572 &&& 361.941 &&& 294.264 &&& 294.264 \tabularnewline
Scaled \textsf{CAIC} & 496.570 &&& 312.366 &&& 261.990 &&& 261.990 \tabularnewline
\bottomrule
\end{tabular}
\end{adjustbox}
\end{table*}
\end{DocumentVersionTR}

\processifversion{DocumentVersionTR}{\paragraph*{Model Overview}}
To further test the \refterm{construct_validity}{construct validity} of the three-factor second-order model, we conducted a confirmatory factor analysis of the IUIPC-10 measurement model on Sample~\textsf{B}\processifversion{DocumentVersionTR}{ ($N_{\mathsf{B}^\prime} = 370$)}. 

\processifversion{DocumentVersionTR}{The second-order model considers the indicator variables as first-level reflective measurement of the latent variables (\textsf{Control}, \textsf{Awareness}, and \textsf{Collection}). It further models that these first-level latent variables are caused by the second-order factor: \textsf{IUIPC}. We included the model's path plot in Figure~\ref{fig:pathPlotCFABref}.}
\processifversion{DocumentVersionConference}{We included the model's path plot in Figure~\ref{fig:pathPlotCFABref} in Appendix~\ref{app:validity}.}

\newcommand{\pathPlotCFABref}{
\begin{figure*}[tbp]
\centering
\vspace{-1.3cm}
\includegraphics[keepaspectratio,width=0.8\maxwidth]{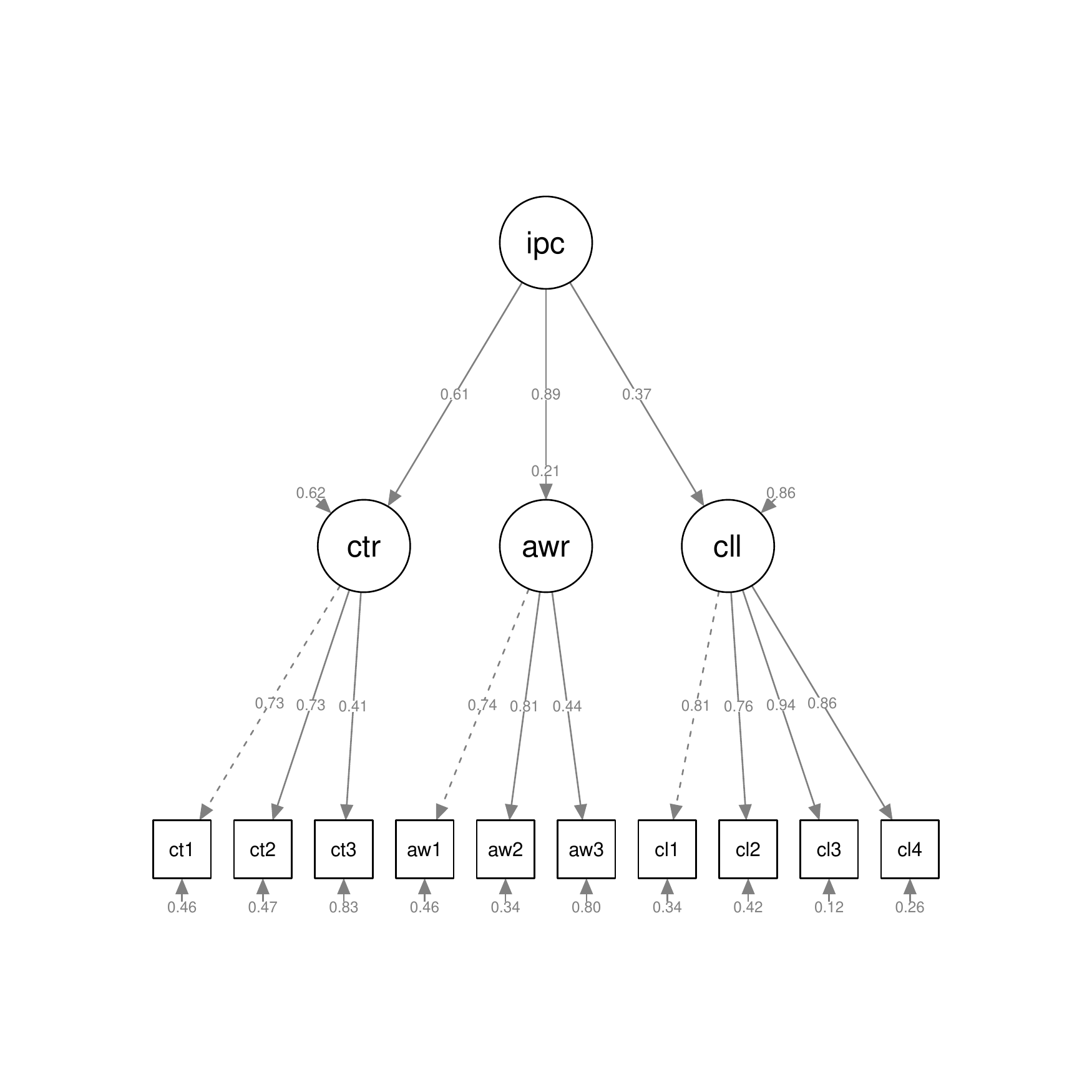} 
\vspace{-1.5cm}
\caption{CFA paths plot with standardized estimates of IUIPC-10 on Sample \textsf{B}. \emph{Note:} The coefficients are standardized}
\label{fig:pathPlotCFABref}
\end{figure*}
}
\newcommand{\pathPlotCFABreduxref}{
\begin{figure*}[tbp]
\centering
\vspace{-1.3cm}
\includegraphics[keepaspectratio,width=0.8\maxwidth]{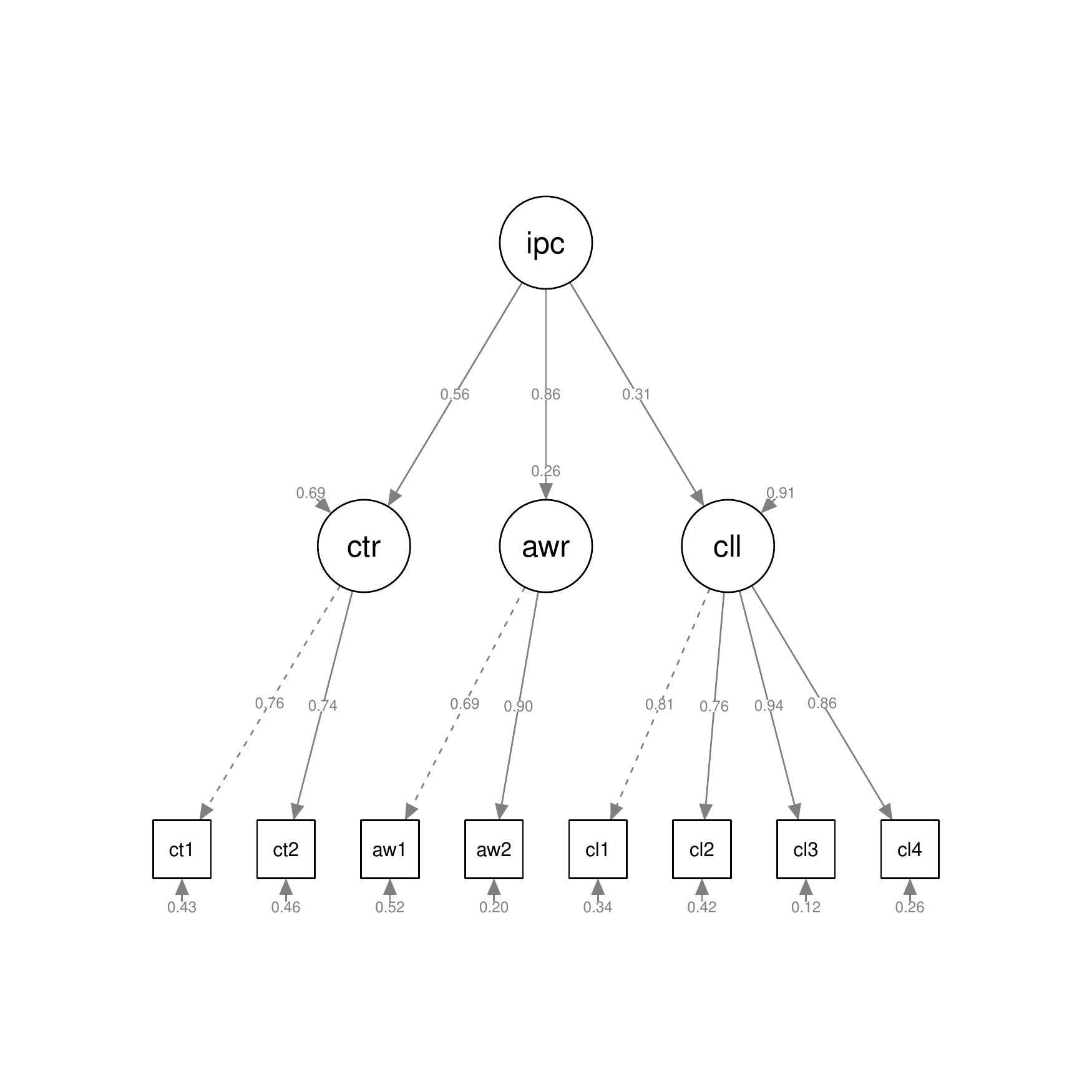} 
\vspace{-1.5cm}
\caption{CFA paths plot with standardized estimates of the respecified IUIPC-8 on Sample \textsf{B}. \emph{Note:} The coefficients are standardized}
\label{fig:pathPlotCFABreduxref}
\end{figure*}
}

\processifversion{DocumentVersionTR}{\pathPlotCFABref}

\paragraph*{Global Fit}
Our first point of call for further evaluating the \refterm{factorial_validity}{factorial validity} of IUIPC-10 is is global fit.
We included an overview of the fit measures on different samples in Table~\ref{tab:respec}, drawing attention to the top row.

First, we observed that the \refterm{exact_fit}{exact-fit test} failed for IUIPC-10 irrespective of estimator, that is, the exact-fit null hypotheses $H_{\chi^2, 0}$ were rejected with the $\chi^2$-tests being statistically signficant.
For the \textsf{RMSEA}-based hypotheses we have:
\begin{inparaenum}[(i)]
  \item The \refterm{close_fit}{close-fit test} evaluating whether \textsf{RMSEA} is likely less or equal $.05$ failed irrespective of estimator. 
 \item The \refterm{not_close_fit}{not-close-fit test} could not be rejected for either estimator, withholding support for the models.
  \item Finally, the \refterm{poor_fit}{poor-fit test} could not be rejected either for any models, with the upper bound of the \textsf{RMSEA} CI being greater than or equal as $.10$, indicating a poor fit.
\end{inparaenum}

The \refterm{fit}{fit indices} \textsf{CFI} and \textsf{SRMR} yielded $.92$ and $.10$ for the ML based models, respectively, not supporting the models. None of the models passed the \vari{HBR} \refterm{combination_rule}{combination rule} used by Malhotra et al.
The direct replication of IUIPC with ML estimation and outliers present fared more poorly than the corresponding ML models implementing the stated assumption: $\chi^2(32) = 202.789, p < .001$; \textsf{CFI}=$.90$; \textsf{RMSEA}=$.12$ $[.10, .13]$; \textsf{SRMR}=$.11$; \textsf{CAIC}=333.8.

Overall, we conclude that the global fit of the model was poor and that we found disconfirming evidence for IUIPC-10's factorial validity.
This disconfirmation of the CFA held irrespective of the data preparation and estimator employed.
Our further examination of construct validity will be on the touch-stone MLM model.

\paragraph*{Local Fit}
\label{sec:local.fit}
\begin{DocumentVersionTR}
We analyzed the correlation and standardized covariance residuals of the MLM model presented in Table~\ref{tab:residualsB} on page~\pageref{tab:residualsB}.

First, we observed that the correlation residuals in Table~\ref{tab:residualsBcor} showed a number of positive correlations greater than $.10$. We noticed especially that there were patterns of residual correlations with \textsf{ctrl3} and \textsf{awa3}. 
Second, we saw in Table~\ref{tab:residualsBstd} that there were a majority of 
statistically significant residual standardized covariances, indicating a wide-spread poor local fit and a repetition of the pattern seen in the correlations. These statistically significant residuals were too frequent to be attributed to chance alone.
Positive covariance residuals were large and prevalent, which showed that---in many cases---the CFA model underestimated the association between the variables present in the sample.

From from the positive covariance residuals with \textsf{awa} and \textsf{coll} indicators, we concluded that \textsf{ctrl3} misloaded both on factors \textsf{Awareness} and \textsf{Collection}. From the positive covariance residuals between \textsf{awa3} and the \textsf{coll} variables, we inferred that this indicator loaded on the factor \textsf{Collection}. These observations yielded further disconfirming evidence for the \refterm{factorial_validity}{factorial validity} of IUIPC-10, especially regarding the \refterm{unidimensionality}{unidimensionality} of its sub-scales.
\end{DocumentVersionTR}
\begin{DocumentVersionConference}
The correlation residuals showed patterns of positive correlations greater than $.10$ with \textsf{ctrl3} and \textsf{awa3}, matched with statistically significant standardized covariances. This indicated considerable misloading on these indicators and disconfirmed the \refterm{unidimensionality}{unidimensionality} of the corresponding sub-scales.
\end{DocumentVersionConference}

\subsubsection{Convergent and Discriminant Validity}
We first analyzed the standardized loadings and the variance explained in Table~\ref{tab:loadingsB}. Therein, we found that \textsf{ctrl3} and \textsf{awa3} only explained 
$.17$ and 
$.20$ 
of the variance, respectively. Those values were unacceptably below par ($R^2 > .50$), yielding a poor convergent validity.

\loadingsB

\processifversion{DocumentVersionTR}{While we assessed the intra- and inter-factor correlations~\cite{bollen1989measurement}, let focus here on the criteria specified in Section~\ref{sec:construct_validity}.}
In terms of \refterm{convergent_validity}{convergent validity}, we evaluated the Average Variance Extracted (AVE) and Composite Reliability (CR) $\omega$ in Table~\ref{tab:loadingsB}. While the CR $\omega$ being greater than the AVE for all three dimensions indicated support, we observed that the AVE being less than $.50$ for both \textsf{Control} and \textsf{Awareness}, showing that there is more error variance extracted than factor variance. Similarly, $\omega < .70$ implied sub-par convergent validity.

For \refterm{discriminant_validity}{discriminant validity}, we found the Fornell-Larcker and \refterm{heterotrait_monotrait_ratio}{HTMT}-criteria fulfilled, offering support for the specified models.

\subsection{Reliability: Internal Consistency}
\processifversion{DocumentVersionTR}{We summarized a range of item/construct reliability metrics of main Sample \textsf{B} in Table~\ref{tab:instruments.reliability}, observing that Guttman's $\lambda_6$ dropped to less than $.6$ and that the average correlation for \textsf{Control} and \textsf{Awareness} were low.}

Let us consider the reliability criteria derived from the MLM CFA model in Table~\ref{tab:loadingsB}.
Considering Cronbach's $\alpha$, we observed estimates for \textsf{Control} and \textsf{Awareness} less than the $.70$, what Nunnally classified only acceptable to ``save time and energy.'' The Composite Reliability estimate $\omega$ was equally sub-par.

\processifversion{DocumentVersionTR}{\instumentReliabilityTable}

\begin{DocumentVersionConference}
\subsection{Respecification}
\label{sec:respec}
In face of the disconfirming evidence discovered on construct validity, we decided to remove the items \textsf{ctrl3} and \textsf{awa3} from the scale, at the risk of losing identification.
We compared the non-nesteds models IUIPC-10 and IUIPC-8 with the Vuong test on the ML estimation.
The variance test indicated the two models as distinguishable, $\omega^2 = 1.926$, $p < .001$.
The Vuong non-nested likelihood-ratio test rejected the null hypothesis that both models were equal. The IUIPC-8 model fitted statistically significantly better than the IUIPC-10 model, LRT $z = -34.541$, $p < .001$. Table~\ref{tab:respec} illustrates the comparison of the two models. This consitutes evidence of the \refterm{factorial_validity}{factorial validity} of the revised scale, including a confirmation of the \refterm{unidimensionality}{unidimensionality} of its sub-scales.
\end{DocumentVersionConference}

\begin{DocumentVersionTR}
\subsection{Respecification}
One possibility to handle the situation is to specify explicitly in the measurement model of IUIPC-10 that \textsf{ctrl3} and \textsf{awa3} are indicators to multiple factors.
While this approach could offer a better overall fit, the IUIPC-10 scale would still suffer from poor convergent and discriminant validity.

We can advocate to remove \textsf{ctrl3} and \textsf{awa3} from the scale and
create a reduced IUIPC-8 scale instead. 
Clearly, such a step comes at a price.
In CFA, the possible dire price lies in the loss of identification.
Kelly's rule of thumb on number of indicators states: ``Two \emph{might} be fine, three is better, four is best, and anything more is gravy.''

Ideally, to be identified in a nonstandard CFA model, for each factor there should be at least three indicators with mutually uncorrelated errors. For \textsf{Control} and \textsf{Awareness}, we are running the risk of violating this rule of identification. However, with a two-indicator situation as presented here, we could reach identification if their errors are uncorrelated within them and they are not correlated with the error of another factor.

We compared the \emph{non-nested} pair IUIPC-10 and IUIPC-8 with the Vuong test on the ML estimation.
The variance test indicated the two models as distinguishable, $\omega^2 = 1.926$, $p < .001$.
The Vuong non-nested likelihood-ratio test rejected the null hypothesis that both models are equal for the focal population. The IUIPC-8 model fits statistically significantly better than the IUIPC-10 model, LRT $z = -34.541$, $p < .001$. We display the comparison of IUIPC-10 and the trimmed model in Table~\ref{tab:respec}. This is strong evidence of the \refterm{factorial_validity}{factorial validity} of the revised scale, including a confirmation of the \refterm{unidimensionality}{unidimensionality} of its sub-scales.

While the IUIPC-8 model yields a significantly better fit, there are still two significant standardized residuals with appreciable negative correlations of concern: between \textsf{coll1} and \textsf{awa1}/\textsf{awa2}, respectively. The negative covariance residual means that the model overestimates the covariances between the first collection item and the awareness items.

\begin{RedundantContent}
\begin{remark}[Correlated errors]
We inspected the questionnaire itself for evidence of method correlation (cf. Appendix~\ref{app:materials}). We found specifically that \textsf{coll1} and \textsf{coll3} follow a parallel construction with the stem ``It bothers me\dots,'' which could lead to correlated error variances on substantive grounds. Testing the this variant with an LRT difference test on nested models, we found that the difference is statistically significant in the ML estimation, but not under Satorra-Bentler correction, $\chi^2(1) = 3.975, p = .046$ and $\chi^2(1) = 2.548, p = .110$, respectively. The difference between the variants is even less pronounced under WLSMVS estimation. Hence, we choose to keep the model without correlated errors.
\end{remark}
\end{RedundantContent}
\end{DocumentVersionTR}

The respecification is still highly correlated with IUIPC-10:
\begin{inparaenum}[(i)]
  \item \textsf{ctrl}, $r = .91$, 95\% CI $[.89, .93]$;
  \item \textsf{awa}, $r = .86$, 95\% CI $[.83, .88]$;
  \item \textsf{iuipc}, $r = .96$, 95\% CI $[.96, .97]$,
\end{inparaenum}
all statistically significant at $p < .001$, giving us evidence for IUIPC-8's concurrent validity.

\begin{DocumentVersionConference}
\begin{table*}
\centering
\caption{Respecification of IUIPC-10 to IUIPC-8 on Sample \textsf{B}%
\explain{.\\All IUIPC-10 models failed the poor-fit tests irrespective of estimator. The respecified IUIPC-8 yielded a statistically significantly better fit by the Vuong test on the ML estimation, with better \textsf{CAIC} throughout. The $\mathsf{RMSEA}$ on IUIPC-8 still asks us to mind the residuals.}}
\label{tab:respec}
\begin{tabular}{llccc}
\toprule
\multirow{2}{*}{Instrument} & \multirow{2}{*}{Respecification} & \multicolumn{3}{c}{Estimator}\\
\cmidrule(lr){3-5}
&& ML & MLM$^\ddagger$ & WLSMVS$^\ddagger$ \\
\midrule
\multirow{4}{*}{IUIPC-10} &            & $\chi^2(32) = 163.691, p < .001$ 
                                       & $\chi^2(32) = 131.417, p < .001$ 
                                       & $\chi^2(15) = 181.149, p < .001$\\ 
 && \textsf{CFI}=$.92$; 
    \textsf{GFI}=$1.00$ &
    \textsf{CFI}=$.92$;
    \textsf{GFI}=$1.00$&
    \textsf{CFI}=$.95$;
    \textsf{GFI}=$.99$\\
 && \textsf{RMSEA}=$.11$ $[.09, .12]$ &
    \textsf{RMSEA}=$.10$ $[.08, .12]$ &
    \textsf{RMSEA}=$.17$ $[.15, .19]$\\
 && \textsf{SRMR}=$.10$; 
    \textsf{CAIC}=294.3 &
    \textsf{SRMR}=$.10$;
    \textsf{CAIC}=262&
    \textsf{SRMR}=$.10$;
    \textsf{CAIC}=414.6\\
\cmidrule(lr){3-3}\cmidrule(lr){4-4}\cmidrule(lr){5-5}
                          & & $\uparrow$ & $\uparrow$ & $\uparrow$ \\
                          & \multirow{2}{*}{Trim \textsf{ctrl3} \& \textsf{awa3}} & 
                          $\Delta{\mathsf{CFI}} = 0.05$; $\Delta{\mathsf{CAIC}} = -132.57$  & 
                          \multirow{2}{*}{$\Delta{\mathsf{CFI}} = 0.05$; $\Delta{\mathsf{CAIC}} = -112.32$} &
                          \multirow{2}{*}{$\Delta{\mathsf{CFI}} = 0.04$; $\Delta{\mathsf{CAIC}} = -191.39$}\\
                          & & Vuong LRT $z = -34.541$, $p < .001$\\
                          & & $\downarrow$ & $\downarrow$ & $\downarrow$ \\
\cmidrule(lr){3-3}\cmidrule(lr){4-4}\cmidrule(lr){5-5}
IUIPC-8                  & & $\chi^2(17) = 54.863, p < .001$ 
                                       & $\chi^2(17) = 42.836, p < .001$ 
                                       & $\chi^2(10) = 33.282, p < .001$\\
 && \textsf{CFI}=$.97$;
    \textsf{GFI}=$1.00$&
    \textsf{CFI}=$.98$;
    \textsf{GFI}=$1.00$ &
    \textsf{CFI}=$.99$;
    \textsf{GFI}=$1.00$\\
 && \textsf{RMSEA}=$.08$ $[.06, .10]$ &
    \textsf{RMSEA}=$.07$ $[.05, .10]$ &
    \textsf{RMSEA}=$.08$ $[.04, .12]$\\
 && \textsf{SRMR}=$.03$; 
    \textsf{CAIC}=161.7 &
    \textsf{SRMR}=$.03$;
    \textsf{CAIC}=149.7&
    \textsf{SRMR}=$.04$;
    \textsf{CAIC}=223.2\\
\bottomrule
\end{tabular}
\\\emph{Note:} $^\ddagger$ Robust estimation with scaled test statistic. $\const{RMSEA}$ reported wih 90\% CI.
\end{table*}
\end{DocumentVersionConference}

\begin{DocumentVersionTR}
\begin{table*}
\centering
\caption{Respecification of IUIPC-10 to IUIPC-8 on Sample \textsf{B}}
\label{tab:respec}
\begin{adjustbox}{max width=\textwidth}
\begin{tabular}{lccc}
\toprule
Instrument \& & \multicolumn{3}{c}{Estimator}\\
\cmidrule(lr){2-4}
Respecification & ML & MLM$^\ddagger$ & WLSMVS$^\ddagger$ \\
\midrule
\multirow{4}{*}{IUIPC-10}              & $\chi^2(32) = 163.691, p < .001$ 
                                       & $\chi^2(32) = 131.417, p < .001$ 
                                       & $\chi^2(15) = 181.149, p < .001$\\ 
 & \textsf{CFI}=$.92$; 
    \textsf{GFI}=$1.00$ &
    \textsf{CFI}=$.92$;
    \textsf{GFI}=$1.00$&
    \textsf{CFI}=$.95$;
    \textsf{GFI}=$.99$\\
 & \textsf{RMSEA}=$.11$ $[.09, .12]$ &
    \textsf{RMSEA}=$.10$ $[.08, .12]$ &
    \textsf{RMSEA}=$.17$ $[.15, .19]$\\
 & \textsf{SRMR}=$.10$; 
    \textsf{CAIC}=294.3 &
    \textsf{SRMR}=$.10$;
    \textsf{CAIC}=262&
    \textsf{SRMR}=$.10$;
    \textsf{CAIC}=414.6\\
\cmidrule(lr){2-2}\cmidrule(lr){3-3}\cmidrule(lr){4-4}
                          & $\uparrow$ & $\uparrow$ & $\uparrow$ \\
\multirow{2}{*}{\emph{Trim} \textsf{ctrl3} \& \textsf{awa3}} & 
                          $\Delta{\mathsf{CFI}} = 0.05$; $\Delta{\mathsf{CAIC}} = -132.57$  & 
                          \multirow{2}{*}{$\Delta{\mathsf{CFI}} = 0.05$; $\Delta{\mathsf{CAIC}} = -112.32$} &
                          \multirow{2}{*}{$\Delta{\mathsf{CFI}} = 0.04$; $\Delta{\mathsf{CAIC}} = -191.39$}\\
                          & Vuong LRT $z = -34.541$, $p < .001$\\
                          & $\downarrow$ & $\downarrow$ & $\downarrow$ \\
\cmidrule(lr){2-2}\cmidrule(lr){3-3}\cmidrule(lr){4-4}
IUIPC-8                   & $\chi^2(17) = 54.863, p < .001$ 
                                       & $\chi^2(17) = 42.836, p < .001$ 
                                       & $\chi^2(10) = 33.282, p < .001$\\
 & \textsf{CFI}=$.97$;
    \textsf{GFI}=$1.00$&
    \textsf{CFI}=$.98$;
    \textsf{GFI}=$1.00$ &
    \textsf{CFI}=$.99$;
    \textsf{GFI}=$1.00$\\
 & \textsf{RMSEA}=$.08$ $[.06, .10]$ &
    \textsf{RMSEA}=$.07$ $[.05, .10]$ &
    \textsf{RMSEA}=$.08$ $[.04, .12]$\\
 & \textsf{SRMR}=$.03$; 
    \textsf{CAIC}=161.7 &
    \textsf{SRMR}=$.03$;
    \textsf{CAIC}=149.7&
    \textsf{SRMR}=$.04$;
    \textsf{CAIC}=223.2\\
\bottomrule
\end{tabular}
\end{adjustbox}
\\\emph{Note:} $^\ddagger$ Robust estimation with scaled test statistic.
\end{table*}
\end{DocumentVersionTR}

\processifversion{DocumentVersionTR}{We included the path model for the MLM estimation on Sample \textsf{B} in Figure~\ref{fig:pathPlotCFABreduxref}. We put the corresponding residuals table in Table~\ref{tab:residualsBredux} on page~\pageref{tab:residualsBredux}.}
\processifversion{DocumentVersionTR}{\pathPlotCFABreduxref}

\processifversion{DocumentVersionTR}{\loadingsBredux}

\processifversion{DocumentVersionTR}{\discriminantB}

\subsection{Validation}
We validated the respecified IUIPC-8 model on the independent validation sample \textsf{V} and compared against he performance of IUIPC-10. We offered this comparison under consideration of all three estimators in Table~\ref{tab:respec.V}.

First, we observed under ML estimation that the two models are indeed statistically significantly distinguishable with the variance test, $\omega^2 = 2.489$, $p < .001$. Furthermore, applied to the validation sample \textsf{V}, IUIPC-8 was the statistically significantly better model according to the Vuong test, LRT $z = -34.541$, $p < .001$.

For IUIPC-10 on \textsf{V}, we noticed that the ML estimation failed all test criteria, incl. the poor-fit hypothesis with a $\hat{\epsilon} = .08$ and $\hat{\epsilon}_{\mathsf{UL}} \geq .10$. The residuals showed a similar pattern as on Sample~\textsf{B}. Again, the direct replication of IUIPC with outliers showed a poorer fit than the other ML estimations: $\chi^2(32) = 137.368, p < .001$; \textsf{CFI}=$.93$; \textsf{RMSEA}=$.09$ $[.07, .10]$; \textsf{SRMR}=$.09$; \textsf{CAIC}=270.5.

The respecified IUIPC-8 fared better. Here, the ML estimator already offered a good fit. The estimators, MLM and WLSMVS, performed equally well, if not better.

Overall, we concluded that IUIPC-8 could be validated on an independent dataset as a well fitting model, whereas IUIPC-10 was disconfirmed once more. 

\begin{DocumentVersionConference}
\newcommand{\comparisonV}{
\begin{table*}
\centering
\caption{Comparison of IUIPC-10 and IUIPC-8 on Validation Sample \textsf{V}}
\label{tab:respec.V}
\begin{tabular}{llccc}
\toprule
\multirow{2}{*}{Instrument} & \multirow{2}{*}{Respecification} & \multicolumn{3}{c}{Estimator}\\
\cmidrule(lr){3-5}
&& ML & MLM$^\ddagger$ & WLSMVS$^\ddagger$ \\
\midrule
\multirow{4}{*}{IUIPC-10} &            & $\chi^2(32) = 122.015, p < .001$ 
                                       & $\chi^2(32) = 80.69, p < .001$ 
                                       & $\chi^2(16) = 151.118, p < .001$\\ 
 && \textsf{CFI}=$.94$; 
    \textsf{GFI}=$1.00$ &
    \textsf{CFI}=$.95$;
    \textsf{GFI}=$1.00$&
    \textsf{CFI}=$.96$;
    \textsf{GFI}=$.99$\\
 && \textsf{RMSEA}=$.08$ $[.07, .10]$ &
    \textsf{RMSEA}=$.07$ $[.05, .09]$ &
    \textsf{RMSEA}=$.14$ $[.12, .16]$\\
 && \textsf{SRMR}=$.08$; 
    \textsf{CAIC}=254.6 &
    \textsf{SRMR}=$.08$;
    \textsf{CAIC}=213.3&
    \textsf{SRMR}=$.07$;
    \textsf{CAIC}=404.3\\
\cmidrule(lr){3-3}\cmidrule(lr){4-4}\cmidrule(lr){5-5}
                          & & $\uparrow$ & $\uparrow$ & $\uparrow$ \\
                          & \multirow{2}{*}{Trim \textsf{ctrl3} \& \textsf{awa3}} & 
                          $\Delta{\mathsf{CFI}} = 0.05$; $\Delta{\mathsf{CAIC}} = -111.6$  & 
                          \multirow{2}{*}{$\Delta{\mathsf{CFI}} = 0.05$; $\Delta{\mathsf{CAIC}} = -82.76$} &
                          \multirow{2}{*}{$\Delta{\mathsf{CFI}} = 0.04$; $\Delta{\mathsf{CAIC}} = -174.5$}\\
                          & & Vuong LRT $z = -35.146$, $p < .001$\\
                          & & $\downarrow$ & $\downarrow$ & $\downarrow$ \\
\cmidrule(lr){3-3}\cmidrule(lr){4-4}\cmidrule(lr){5-5}
IUIPC-8                  & & $\chi^2(17) = 34.532, p = .007$ 
                                       & $\chi^2(17) = 22.04, p = .183$ 
                                       & $\chi^2(10) = 24.844, p = .005$\\
 && \textsf{CFI}=$.99$;
    \textsf{GFI}=$1.00$&
    \textsf{CFI}=$.99$;
    \textsf{GFI}=$1.00$ &
    \textsf{CFI}=$.99$;
    \textsf{GFI}=$1.00$\\
 && \textsf{RMSEA}=$.05$ $[.03, .07]$ &
    \textsf{RMSEA}=$.03$ $[.00, .07]$ &
    \textsf{RMSEA}=$.06$ $[.03, .09]$\\
 && \textsf{SRMR}=$.03$; 
    \textsf{CAIC}=143 &
    \textsf{SRMR}=$.03$;
    \textsf{CAIC}=130.6&
    \textsf{SRMR}=$.03$;
    \textsf{CAIC}=229.8\\
\bottomrule
\end{tabular}
\\\emph{Note:} $^\ddagger$ Robust estimation with scaled test statistic.
\end{table*}
}
\end{DocumentVersionConference}

\begin{DocumentVersionTR}
\newcommand{\comparisonVext}{
\begin{table*}
\centering
\caption{Comparison of IUIPC-10 and IUIPC-8 on Validation Sample \textsf{V}}
\label{tab:respec.V}
\begin{adjustbox}{max width=\textwidth}
\begin{tabular}{lccc}
\toprule
Instrument \& & \multicolumn{3}{c}{Estimator}\\
\cmidrule(lr){2-4}
Respecification & ML & MLM$^\ddagger$ & WLSMVS$^\ddagger$ \\
\midrule
\multirow{4}{*}{IUIPC-10}              & $\chi^2(32) = 122.015, p < .001$ 
                                       & $\chi^2(32) = 80.69, p < .001$ 
                                       & $\chi^2(16) = 151.118, p < .001$\\ 
 & \textsf{CFI}=$.94$; 
    \textsf{GFI}=$1.00$ &
    \textsf{CFI}=$.95$;
    \textsf{GFI}=$1.00$&
    \textsf{CFI}=$.96$;
    \textsf{GFI}=$.99$\\
 & \textsf{RMSEA}=$.08$ $[.07, .10]$ &
    \textsf{RMSEA}=$.07$ $[.05, .09]$ &
    \textsf{RMSEA}=$.14$ $[.12, .16]$\\
 & \textsf{SRMR}=$.08$; 
    \textsf{CAIC}=254.6 &
    \textsf{SRMR}=$.08$;
    \textsf{CAIC}=213.3&
    \textsf{SRMR}=$.07$;
    \textsf{CAIC}=404.3\\
\cmidrule(lr){2-2}\cmidrule(lr){3-3}\cmidrule(lr){4-4}
                          & $\uparrow$ & $\uparrow$ & $\uparrow$ \\
\multirow{2}{*}{\emph{Trim} \textsf{ctrl3} \& \textsf{awa3}} & 
                          $\Delta{\mathsf{CFI}} = 0.05$; $\Delta{\mathsf{CAIC}} = -111.6$  & 
                          \multirow{2}{*}{$\Delta{\mathsf{CFI}} = 0.05$; $\Delta{\mathsf{CAIC}} = -82.76$} &
                          \multirow{2}{*}{$\Delta{\mathsf{CFI}} = 0.04$; $\Delta{\mathsf{CAIC}} = -174.5$}\\
                          & Vuong LRT $z = -35.146$, $p < .001$\\
                          & $\downarrow$ & $\downarrow$ & $\downarrow$ \\
\cmidrule(lr){2-2}\cmidrule(lr){3-3}\cmidrule(lr){4-4}
IUIPC-8                  & $\chi^2(17) = 34.532, p = .007$ 
                                       & $\chi^2(17) = 22.04, p = .183$ 
                                       & $\chi^2(10) = 24.844, p = .005$\\
 & \textsf{CFI}=$.99$;
    \textsf{GFI}=$1.00$&
    \textsf{CFI}=$.99$;
    \textsf{GFI}=$1.00$ &
    \textsf{CFI}=$.99$;
    \textsf{GFI}=$1.00$\\
 & \textsf{RMSEA}=$.05$ $[.03, .07]$ &
    \textsf{RMSEA}=$.03$ $[.00, .07]$ &
    \textsf{RMSEA}=$.06$ $[.03, .09]$\\
 & \textsf{SRMR}=$.03$; 
    \textsf{CAIC}=143 &
    \textsf{SRMR}=$.03$;
    \textsf{CAIC}=130.6&
    \textsf{SRMR}=$.03$;
    \textsf{CAIC}=229.8\\
\bottomrule
\end{tabular}
\end{adjustbox}
\\\emph{Note:} $^\ddagger$ Robust estimation with scaled test statistic.
\end{table*}
}
\comparisonVext
\end{DocumentVersionTR}

\subsection{Summary}
The \refterm{construct_validity}{construct validity} of IUIPC according to RQ~\ref{rq:confirm} bears a deliberate discussion. On the one hand, we could confirm the three-dimensionality and second-order model Malhotra et al.~\cite{malhotra2004internet} postulated. On the other hand, factorial validity evidenced in global and local fit---especially the unidimensionality of the first-order factors \textsf{Control} and \textsf{Awareness}---does not seem to be given neither on the main Sample \textsf{B} nor on the validation Sample \textsf{V}. The convergent validity and reliability of these sub-scales is equally in question, our estimates having been lower than Malhotra et al.'s.

\subsection{Estimator Appraisal}
While we have already shown estimator invariance according to RQ~\ref{rq:invariance} throughout confirmation and validation, we are now turning to the question of estimator viability asked in RQ~\ref{rq:estimator}.

Considering the viability of each estimator in their own right, ML without outlier treatment showed the worst performance of the ML-based estimations. We would discourage its use. Let us consider the remaining estimations with outlier treatment. Comparing between ML and MLM, both estimators behaved similarly on the relative changes between the misspecified IUIPC-10 and the respecified IUIPC-8. MLM consistently offered the stronger fit. 

Even though robust WLS is estimating more parameters (loadings, errors, thresholds) than its ML counterparts, we still believe it is fair to say that WLSMVS seemed most sensitive to the misspecification of IUIPC-10. On both samples, we observed a great improvement of fit indices when comparing between the WLSMVS IUIPC-10 and IUIPC-8 models. Hence, the robust WLS estimation is certainly viable in its own right, in fact, we would consider it quite robust against Type I errors.

Having assessed MLM and WLSMVS as face-viable on their global fit, let us further appraise their model interpretations and mean structures.
For this inquiry, we follow the reasoning of Bovaird and Koziol~\cite[pp]{bovaird2012measurement}. Let us train our lens on a single indicator for illustration: \textsf{awa2}. Here, the MLM estimation tells us that individuals with average levels of \textsf{Awareness} will have an expected response on item \textsf{awa2} of $6.62$. Of course, this value does not exist on the 7-point Likert scale. For each increase of one unit of \textsf{Awareness}, we would expect \textsf{awa2} to increase by $1.13$ points. For a single unit of increase from the mean, the prediction $7.75$ is obviously out-of-bounds of the scale, hence, an invalid prediction.
This phenomenon is an artifact of ML-based estimation on heavily non-normal, ordinal data.

Following the same line of inquiry for the robust WLS estimation, these issues do not exist. Here, the loading of \textsf{awa2} $\lambda_{\const{awa2}} = 1.07$ indicates the expected change of the \emph{probit} of an individuals's response to \textsf{awa2}. Under this estimation, an \textsf{Awareness} of $-1.88$ is required to choose response option 6--``Agree'' with a likelihood of 50\%; of $-0.4$ for option 7--``Strongly Agree'' with a likelihood of 50\%. Clearly, the robust WLS prediction of the individual's choice yields a more informative model.

Hence, considering this qualitative appraisal of both estimation methods, MLM and robust WLS, we find that the robust WLS estimation has appreciable advantages in model interpretation. While arguments over probit and likelihoods might be less common in the privacy community than the straight-forward MLM factor loading interpretations, the robust WLS method could offer the community more information how a users really endorse options of a privacy concern scale, such as IUIPC.


\newcommand{\SUCC}{\CIRCLE}
\newcommand{\PART}{\LEFTcircle}
\newcommand{\FAIL}{\Circle}

\begin{table*}
\centering
\footnotesize
\caption{Selected evidence for construct validity and reliability criteria on Samples \textsf{B} and \textsf{V} under MLM estimation%
\explain{.\newline{}The factorial validity shows IUIPC-10 consistently failing fit measures; IUIPC-8 fared better, especially on the validation sample \textsf{V}. The convergent validity of IUIPC-10 is evidenced to be consistently flawed; IUIPC-8 showed suitable results, where some standardized loadings $\beta$ were shown to be border-line without violating $\vari{AVE} > .50$. Divergent validity of IUIPC-10 suffers from the low $\vari{AVE}$; IUIPC-8 fulfills all requirements. IUIPC-10 fails the requirements on internal consistency, especially $\omega > .70$; IUIPC-8 passes them.}}
\label{tab:evidence}
\begin{tabular}{llccccccccc}
\toprule
& & \multicolumn{7}{c}{Construct Validity} & \multicolumn{2}{c}{Reliability}\tabularnewline
\cmidrule(lr){3-9}\cmidrule(lr){10-11}
&& \multicolumn{3}{c}{Factorial} & \multicolumn{2}{c}{Convergent} & \multicolumn{2}{c}{Divergent} & \multicolumn{2}{c}{Internal Consistency}\tabularnewline
\cmidrule(lr){3-5} \cmidrule(lr){6-7} \cmidrule(lr){8-9} \cmidrule(lr){10-11}
&& $H_{\chi^2, 0}$ & $H_{\hat{\varepsilon} \leq .05, 0}$ & $\mathit{HBR}$ & $\beta > .70$ & $\vari{AVE} > .50$ & $\sqrt{AVE} > \forall\bar{r}$ & $\vari{HTMT} < .85$ & $\alpha > .70$  & $\omega > .70$ \tabularnewline
\midrule
\multirow{2}{*}{IUIPC-10} & \textsf{B} & \hyperref[tab:respec]{\FAIL} & \hyperref[tab:respec]{\FAIL} & \hyperref[tab:respec]{\FAIL} &
 \hyperref[tab:loadingsB]{\FAIL} & \hyperref[tab:loadingsB]{\FAIL} &
 \processifversion{DocumentVersionTR}{\hyperref[tab:discriminantB]{\PART}}\processifversion{DocumentVersionConference}{\PART} & \processifversion{DocumentVersionTR}{\hyperref[tab:discriminantB]{\SUCC}}\processifversion{DocumentVersionConference}{\SUCC} & \hyperref[tab:loadingsB]{\FAIL} & \hyperref[tab:loadingsB]{\FAIL} \tabularnewline
                          & \textsf{V} & \hyperref[tab:respec.V]{\FAIL} & \hyperref[tab:respec.V]{\FAIL} & \hyperref[tab:respec.V]{\FAIL}   & \processifversion{DocumentVersionTR}{\hyperref[tab:loadingsV]{\FAIL}}\processifversion{DocumentVersionConference}{\FAIL} & \processifversion{DocumentVersionTR}{\hyperref[tab:loadingsV]{\FAIL}}\processifversion{DocumentVersionConference}{\FAIL} & \processifversion{DocumentVersionTR}{\hyperref[tab:discriminantV]{\FAIL}}\processifversion{DocumentVersionConference}{\FAIL} & \processifversion{DocumentVersionTR}{\hyperref[tab:discriminantV]{\SUCC}}\processifversion{DocumentVersionConference}{\SUCC} & \processifversion{DocumentVersionTR}{\hyperref[tab:loadingsV]{\FAIL}}\processifversion{DocumentVersionConference}{\FAIL} & \processifversion{DocumentVersionTR}{\hyperref[tab:loadingsV]{\FAIL}}\processifversion{DocumentVersionConference}{\FAIL} \tabularnewline
\midrule
\multirow{2}{*}{IUIPC-8}  & \textsf{B} & \hyperref[tab:respec]{\FAIL} & \hyperref[tab:respec]{\FAIL} & \hyperref[tab:respec]{\FAIL} &
 \hyperref[tab:loadingsBredux]{\PART} &  \hyperref[tab:loadingsBredux]{\SUCC} & \processifversion{DocumentVersionTR}{\hyperref[tab:discriminantBredux]{\SUCC}}\processifversion{DocumentVersionConference}{\SUCC} & \processifversion{DocumentVersionTR}{\hyperref[tab:discriminantBredux]{\SUCC}}\processifversion{DocumentVersionConference}{\SUCC} &  \hyperref[tab:loadingsBredux]{\SUCC} &  \hyperref[tab:loadingsBredux]{\SUCC} \tabularnewline
                          & \textsf{V} & \hyperref[tab:respec.V]{\SUCC} & \hyperref[tab:respec.V]{\SUCC} & \hyperref[tab:respec.V]{\SUCC} & \processifversion{DocumentVersionTR}{\hyperref[tab:loadingsVredux]{\PART}}\processifversion{DocumentVersionConference}{\PART} & \processifversion{DocumentVersionTR}{\hyperref[tab:loadingsVredux]{\SUCC}}\processifversion{DocumentVersionConference}{\SUCC} &  \processifversion{DocumentVersionTR}{\hyperref[tab:discriminantVredux]{\SUCC}}\processifversion{DocumentVersionConference}{\SUCC} & \processifversion{DocumentVersionTR}{\hyperref[tab:discriminantVredux]{\SUCC}}\processifversion{DocumentVersionConference}{\SUCC} &  \processifversion{DocumentVersionTR}{\hyperref[tab:loadingsVredux]{\SUCC}}\processifversion{DocumentVersionConference}{\SUCC} & \processifversion{DocumentVersionTR}{\hyperref[tab:loadingsVredux]{\SUCC}}\processifversion{DocumentVersionConference}{\SUCC} \tabularnewline
\bottomrule
\end{tabular}
\\\emph{Note:} \vari{HBR} = Hu and Bentler combination rule used by Malhotra et al.~\cite{malhotra2004internet}; $\beta$ = standardized loading; \vari{AVE} = Average Variance Extracted; $\bar{r}$ = correlation with other factor; \vari{HTMT} = Heterotrait-Monotrait Ratio; $\omega$ = Composite Reliability.
\end{table*}     

\section{Discussion}
\label{sec:discussion}

\subsection{IUIPC-10 could not be confirmed}
While we could attest to the strengths in content validity and the three-dimensionality of IUIPC-10, its CFA revealed a range of weaknesses in its \refterm{construct_validity}{construct validity} (cf. Table~\ref{tab:evidence}):
\begin{inparaenum}[(i)] 
  \item we could not confirm \refterm{factorial_validity}{factorial validity} in terms of global fit, meaning that the corresponding models do not approximate the corresponding observations well;
  \item the inspection of the residuals showed an unsatisfactory local fit for two items of the sub-scales of \textsf{Control} and \textsf{Awareness} calling the the \refterm{unidimensionality}{unidimensionality} of these sub-scales into question.
  \item we found disconfirming evidence on the \refterm{convergent_validity}{convergent validity}.
\end{inparaenum}
We further observed a sub-par \refterm{reliability}{reliability} of the sub-scales \textsf{Control} and \textsf{Awareness}\processifversion{DocumentVersionTR}{ with a signal-to-noise ratio of $1.94$}. As these assessments held true on main and validation samples, irrespective of estimators used, they offer disconfirming evidence against the scale itself.

For privacy researchers, the issues in \refterm{construct_validity}{construct validity} mean that the IUIPC-10 scale shows weaknesses in measuring its hidden construct information privacy concern. The observed sub-par \refterm{reliability}{reliability} means that IUIPC-10 measurements are contaminated by error variance, which entails a low signal-to-noise ratio and can lead to spurious and erratic results. 

\subsection{IUIPC-8 asserts a stronger validity}
With IUIPC-8, we proposed a refined version of IUIPC that performed consistently well in terms of \refterm{construct_validity}{construct validity} and \refterm{reliability}{reliability}. We give an overview of a range of criteria in Table~\ref{tab:evidence}.
In terms of \refterm{factorial_validity}{factorial validity}, we observed good global and local fits. Criteria for \refterm{convergent_validity}{convergent} and \refterm{discriminant_validity}{discriminant} validity were fulfilled consistently.
The respecified scale also showed appreciable improvements in \refterm{reliability}{reliability}, yielding a 33\% to 82\% better signal-to-noise ratio for \textsf{Awareness} and \textsf{Control}, respectively.

We encourage privacy researchers consider carefully: On the one hand, the 10-item version of IUIPC contains more privacy concern information they will care about. It has at least three items per factor and, thereby, creates favorable conditions for CFA model identification and more robust estimation of the true value of the latent factors. However, given that two items seem to misload to a considerable degree and show low reliability, those items may confound the model. The tight fit we obtained for the 8-item version of IUIPC is encouraging: it will approximate the data well and yield sound measurement models for subsequent analysis. The strong concurrent validity with IUIPC-10 further supports using the respecified scale.

Given the evidence in this study, we endorse adopting IUIPC-8 as a brief questionnaire for Internet users' privacy concerns.

\subsection{Questionnaire wording as culprit}
While our reviews of IUIPC in Section~\ref{sec:review} asserted a sound content validity, we equally found evidence for biases rooted in the questionnaire wording. Our analysis of the distributions of IUIPC-10 and its sub-scales from Section~\ref{sec:desc}, especially, the distribution graphs in Figure~\ref{fig:densitySubScales} showed a substantial negative skew and positive kurtosis. This seems to confirm our observation that the use of \refterm{loaded_word}{loaded words} incl.  ``privacy'' and ``autonomy'' may create a systematic bias through priming, further aggravated by \refterm{leading_question}{leading questions}. All these observations point towards the instrument itself influencing the respondents towards agreement. 

IUIPC also suffers from multiple instances of question wording yielding \refterm{nondifferentiation}{nondifferentiation}. While the entire questionnaire can be subject to straightlining due to the absence of reverse-coded items, we believe that items \textsf{ctrl3} and \textsf{awa3} were especially impacted because of the presence of \refterm{double-barreled_question}{double-barreled constructions}. This observation could further explain why these two items showed such a low reliability and why they needed to be removed altogether.

For privacy researchers, these observations stress the importance of inspecting the question wording of instruments carefully and to assess them against commonly known biases~\cite{Oppenheim1992,ChoPak2005}.

\begin{RedundantContent}
We noticed that the two items yielding a low variance explained exhibited a common structure:
\begin{compactdesc}
  \item[\textsf{ctrl3}:] ``I believe that online privacy is invaded when control is lost or unwillingly reduced as a result of
a marketing transaction.''
  \item[\textsf{awa3}:] ``It is very important to me that I am aware and knowledgeable about how my personal information
will be used.''
\end{compactdesc}
They are both double-clause formulations. Item \textsf{ctrl3} exhibits a disjunction (``control is lost \emph{or} unwillingly reduced'')
while item \textsf{awa3} exhibits a conjunction (``I am aware \emph{and} knowledgeable'').
In terms of classical questionnaire formulation biases, they make for \emph{double-barreled questions}~\cite{ChoPak2005}, that is, asking two questions in one. Those constructions make it difficult for the respondent to know what part to answer and how to behave if only one of the clauses is fulfilled. 
In addition, \textsf{ctrl3} contextualizes the item on ``a marketing transaction,'' a profession's term which is not necessarily well-defined for the general public.
Given these observations, we believe that our quantitative results are rooted in a substantive problem in the underlying questions that cause an appreciable bias and low validity.
\end{RedundantContent}

\subsection{How to use IUIPC}

\corrgramScores

Our analysis has a range of implications for privacy researchers, not just for the use of IUIPC but any multi-dimensional privacy concern scale.
First, let us consider judging IUIPC scores for a sample. Given the evidence that the instrument seems to bias responses towards agreement, it is invalid to call a particular sample ``especially privacy-sensitive,'' should the mean IUIPC score be greater than 4--``Neither agree nor disagree.''

Because IUIPC-10 exhibits a sub-standard reliability, privacy researchers need to be conscious a low signal-to-noise ratio. 
The error variance will mask the information about the respondents' true score on IUIPC.

The most important consequence of the low \refterm{reliability}{reliability} for privacy researchers will be the expected attenuation of relations to other latent factors~\cite{RevCon2018}, that is, the magnitude of effects of IUIPC-measured privacy concern on other variables, say behavior, will be reduced. This means that it will be more difficult to show the impact of privacy concern---\emph{even if the true relation between the latent variables is substantive}.

Privacy researchers focusing on simple statistical tools, such as linear regressions on parceled sub-scale scores, are most affected by these shortcomings: the error variance is folded into the scores they use, masking the signal. For these researchers, IUIPC-8 offers considerable advantages by offering stronger validity and reliability. It comes at the price of eliminating two concern-items the researchers might care about.
Considering the comparison of score approximations with the CFA-estimated factor in Figure~\ref{fig:corrgramScores}, we assert that such privacy researchers would be worst off  ignoring the factor structure altogether by simply averaging all items of IUIPC into one ``flat'' score ($r = .74$). It is usually better to take into account the structure, the simplest approach averaging the sub-scale scores into a ``coarse'' factor score. They could further improve their approximation by computing \defterm{factor_scores}{factor scores}\processifversion{DocumentVersionTR}{~\cite{distefano2009understanding,grice2001computing}}, for instance ``weighted'' by factor loadings ($r = .89$) or linearly combined with CFA-derived ``regression'' coefficients. While there is a great deal of discussion on correct and incorrect uses of those factor scores, a common pragmatic approach computes a weighting by the factor loading.

Privacy researchers using advanced tools, such as confirmatory factor analysis or latent-variable structural equation modeling, need to consider a different trade-off. Of course, the analysis of the measurement model will estimate the error variance and establish the loading of each indicator on the corresponding latent factor. Hence, the contamination by error variance is somewhat contained. However, the problematic construct validity of IUIPC-10 will affect their studies, largely visible in sub-par global and local fit. They might discover their measurement model misspecified and their structural models suffering. For them, IUIPC-8 offers a better construct validity, especially useful for investigating sizable latent variable models in relation to investigations of the privacy paradox.

In terms of choice of estimators for CFA and SEM, privacy researchers can viably opt for robust WLS as an estimator tailored to the non-normal, ordinal distributions of IUIPC indicators. As we have shown, a robust WLS estimation on IUIPC-8 offers a good fit at considerable sample sizes. While the privacy community might not be accustomed to the interpretation of these probit models, they offer better model interpretability than the ML or MLM estimations. If privacy researchers decide against robust WLS, we would still advocate using a robust ML estimation (such as MLM or MLR) after a careful data preparation.
\processifversion{DocumentVersionConference}{The extended ArXiv version of this paper includes further considerations on design decisions.}

\begin{DocumentVersionTR}
\subsection{Design decisions matter}
McDonald~\cite{mcdonald2010structural} called structural modeling ``the art of approximation.''
In our analysis, we considered common design decisions researchers must take to gain sound approximations of reality:
\begin{inparaenum}[(i)]
  \item accounting for measurement level and distribution,
  \item accounting for uni- and multi-variate outliers, and
  \item choosing an appropriate estimator for the situation at hand
\end{inparaenum}
are among their numbers.
While we included these dimensions explicitly to ward our analysis against confounding factors, we gained evidence on the sound use of scales like IUIPC for rigorous research.

First, we sound a note of caution: ``Check\dots and check again!''
Just because an instrument shows decent internal consistency measures in its conceptive study (e.g., the popular Cronbach's $\alpha$), this does not guarantee in itself that the instrument will be valid or perform well. Of course, a prudent researcher may refer to independent confirmations of candidate instruments, but those are few and far between in our field. In our opinion, it is sage advise to create complex models in the two-step approach~\cite{anderson1988structural}: diligently evaluating a confirmatory factor analysis of the measurement model first, before continuing to structural or regression analyses.

We could shed light on the fact that the design decision researchers make matter. Were distribution properties checked? Outliers? Were estimators chosen that excel at the given measurement level and distribution?
Fitting a model on non-normal, ordinal, outlier-prone data---such as found in IUIPC-10---relying on Maximum Likelihood (ML) estimation and hoping for the best does not seem to be a prudent way forward.
Not only can this yield a poor fit but these different violations of assumptions and requirements can conspire to create seemingly well-fitting ``results'' that do not offer a sound approximation of reality. 
Those Type-I errors can well haunt a field for years to come.

Given this discussion, we would be so bold to recommend for the evaluation of measurement models:
\begin{compactenum}[1.]
  \item employ a two-step modeling approach~\cite{anderson1988structural},
  \item do your homework in terms of data preparation~\cite{malone2012preparing} and outlier handling~\cite{yuan2001effect},
  \item choose an appropriate estimation method for the measurement level and distribution at hand~\cite{kline2012assumptions},
  \item specifically for ordinal Likert scales with expected non-normality, chose a robust estimator, such as WLSMVS~\processifversion{DocumentVersionConference}{\cite{distefano2002impact,finney2006non,bovaird2012measurement}}\processifversion{DocumentVersionTR}{\cite{muthen1985comparison,distefano2002impact,finney2006non,bovaird2012measurement}},
  \item do not just rely on global fit indices, but explicitly check and report residuals~\cite{kline2015principles}, and finally
  \item let invariable respecification efforts be guided by substantive arguments.
\end{compactenum}
These recommendations are certainly not new. 
However, our experience with the evaluation of IUIPC-10 vouches for applying them consistently to privacy instruments.
\end{DocumentVersionTR}

\subsection{Limitations}
\label{sec:limitations}

In terms of generalizability, this study encountered the usual limitations of self-report instruments, where we sought to ward against straightlining with instructional manipulation checks/attention checks. 
While the sampling aimed at representing a UK population, the sampling process was not randomized and might have been affected by a self-selection bias due to the matchmaking of Prolific.

Naturally, factor analyses like ours were affected by sampling and measurement errors. To ensure our findings gained cross-sample validity, we considered three independent samples. To ensure our findings were valid irrespective of design decisions, we created a direct replication of the IUIPC-10 as well as analyses following contemporary methodological recommendations. The results were valid irrespective of outlier inclusion or estimation method.

\section{Conclusion}
\label{sec:conclusion}

We evaluated the \refterm{construct_validity}{construct validity} and \refterm{reliability}{reliability} of IUIPC-10 in confirmatory factor analyses.
We found that 
\begin{inparaenum}[(i)]
  \item we could confirm the three-dimensionality of IUIPC-10 with the factors \textsf{Control}, \textsf{Awareness}, and \textsf{Collection},
  \item we found disconfirming evidence for the \refterm{factorial_validity}{factorial validity} and \refterm{convergent_validity}{convergent validity}, largely rooted in evidence against the \refterm{unidimensionality}{unidimensionality} of \textsf{Control} and \textsf{Awareness}. These results were consistent in main analysis and validation, irrespective of the estimator used,
  \item we proposed a respecified IUIPC-8 that outperformed the original IUIPC-10 on \refterm{construct_validity}{construct validity} and \refterm{reliability}{reliability} consistently. While we can clearly see its benefits, we are also aware that it only goes so far as the reduced number of items has drawbacks.
  \item we offered empirically grounded recommendations how to use IUIPC and similar privacy concern scales.
\end{inparaenum}

Future work specifically for IUIPC would ideally offer further carefully evaluated revisions to the scale, ideally eliminating identified problems in question wording, aiming for four items per factor, and establishing unassailable \refterm{construct_validity}{construct validity} and an internal consistency $\omega > .80$ for all sub-scales.

We started this paper referring to instruments as measuring stick for privacy concern.
If such a measurement stick is warped, the community's investigation of human factors of PETs and of the privacy paradox may be severely undermined and misled.
We believe that we would benefit from concerted efforts to diligently evaluate standard instruments in the field along similar lines we have pursued in this study.
While reaching consensus on sound measurement instruments on the construct ``information privacy concern'' and its siblings is essential, we would also benefit from following unified recommendations their use. 

\section*{Acknowledgments}
We would like to thank Naresh K. Malhotra, James Agarwal and especially Sung S. Kim for promptly responding with succinct information on the original research process at the conception of IUIPC.
This work was supported by \CASCAde.

\bibliographystyle{abbrv}
\bibliography{sem,privacy,methods_resources,qdesign,./pets_use_IUIPC,validity,reliability}

\begin{appendix}

\section{Materials \& Sample}
\label{app:sample}
\label{app:materials}

We included the used IUIPC-10 questionnaire in Table~\ref{tab:iuipc}. The questionnaire was administered in the first section of a greater survey, which included six instructional manipulation checks (IMCs) as attention checks shown in Table~\ref{tab:imc}.
\begin{table*}[tb]
\centering
\caption{Items of the instrument Internet users' information privacy concerns (IUIPC-10)~\cite{malhotra2004internet}}
\label{tab:iuipc}
\begin{tabular}{llp{.7\textwidth}}
  \toprule
Construct & Item & Question\\ 
  \midrule
\multirow{3}{*}{Control (\textsf{ctrl})} & \textsf{ctrl1} & Consumer online privacy is really a matter of consumers' right to exercise control and autonomy over decisions about how their information is collected, used, and shared.\\
 & \textsf{ctrl2} & Consumer control of personal information lies at the heart of consumer privacy.\\
 & \textsf{ctrl3} & I believe that online privacy is invaded when control is lost or unwillingly reduced as a result of a marketing transaction.\\
\addlinespace
 \multirow{3}{*}{Awareness (\textsf{awa})} & \textsf{awa1} & Companies seeking information online should disclose the way the data are collected, processed, and used.\\
 & \textsf{awa2} & A good consumer online privacy policy should have a clear and conspicuous disclosure.\\
 & \textsf{awa3} & It is very important to me that I am aware and knowledgeable about how my personal information will be used.\\
 \addlinespace
 \multirow{4}{*}{Collection (\textsf{coll})} & \textsf{coll1} & It usually bothers me when online companies ask me for personal information.\\
 & \textsf{coll2} & When online companies ask me for personal information, I sometimes think twice before providing it.\\
 & \textsf{coll3} & It bothers me to give personal information to so many online companies.\\
 & \textsf{coll4} & I'm concerned that online companies are collecting too much personal information about me.\\
\bottomrule
\end{tabular}
\\\emph{Note:} The questionnaire is administered with 7-point Likert items, anchored on 1=``Strongly Disagree'' to 7=``Strongly Agree''
\end{table*}

\begin{table}[tb]
\centering
\caption{Items of our instructional manipulation checks}
\label{tab:imc}
\begin{tabular}{lp{.9\columnwidth}}
  \toprule
Item & Question\\ 
  \midrule
\textsf{A1} & It is important you pay attention to the statements. Please agree by choosing `strongly agree'.\\
\textsf{A2} & To confirm that you are paying attention to the questions in the questionnaire, please select the first option from the left on the scale.\\
\textsf{A3} & I'm paying attention to the questions in this questionnaire. I confirm this by choosing `somewhat agree'. \\
\textsf{A4} & I recognise the importance of paying attention to the questions in this questionnaire. Please select `agree' to confirm your agreement. \\
\textsf{A5} & Paying attention to the questions in this questionnaire is important. I agree by choosing the third option from the left of the scale.\\
\textsf{A6} & When you're responding to the questions in the questionnaire it is important that you're paying attention. Please agree by selecting the second option from the left on the scale.\\
\bottomrule
\end{tabular}
\end{table}

For the reproducibility of the maximum likelihood estimation, \processifversion{DocumentVersionTR}{Table~\ref{tab:inputCorSDBV} contains the correlations and standard deviations (SDs) of samples \textsf{B} and \textsf{V} used in the study}\processifversion{DocumentVersionConference}{Table~\ref{tab:inputCorSDB} contains the correlations and standard deviations (SDs) of Sample~\textsf{B}}. The OSF supplementary materials contain more precise covariance matrices of all samples. \processifversion{DocumentVersionTR}{The Weighted Least Square (WLS) estimation requires the raw data to be computed.}

\begin{DocumentVersionTR} 
ML confirmatory factor analyses (CFAs) can operate on a sample's covariance matrix as long as the sample size $N$ is known, MLM if the NACOV matrix is additionally available.
The covariance matrices, in turn, can be reconstructed from the inputs' correlations and SDs, modulo rounding error:
\[ \mathsf{Cov} := (\mathsf{SD} \cdot \mathsf{t}(\mathsf{SD})) \cdot \mathsf{Cor} \]

Convenience functions for that purpose are readily available, for instance, the functions \textsf{cor2cov}()
present in both the \textsf{R} packages \textsf{lavaan} and \textsf{psych}.
\end{DocumentVersionTR}
\processifversion{DocumentVersionTR}{\inputCorSDBV}
\processifversion{DocumentVersionConference}{\inputCorSDB}

\begin{DocumentVersionConference}
\section{Additional Evidence}

With respect to \refterm{factorial_validity}{factorial validity} in Section~\ref{sec:factor_val}, Table~\ref{tab:comparison.models} shows the comparison of candidate factor solutions. The path model from Figure~\ref{fig:pathPlotCFABref} shows the selected IUIPC-10 model. Table~\ref{tab:residualsB} highlights the residuals founding the assessment of that model's local fit. Table~\ref{tab:loadingsBredux} then offers the loadings and reliability of the respecified model from Section~\ref{sec:respec}. Finally, Table~\ref{tab:respec.V} offers the fit comparison for the validation.


\begin{table*}[tbp]
\centering
\caption{Comparison of different model structures of IUIPC-10 on Sample \textsf{B} with MLM estimation\explain{.\newline{}The models show increasingly better fits in scaled $\chi^2$, \textsf{CFI}, and \textsf{CAIC}, supporting the predicted hierarchical three-factor model.}}
\label{tab:comparison.models}
\begin{adjustbox}{max width=\textwidth}
\begin{tabular}{@{}rrrrrrrrrrrrr@{}}
\toprule
& \multicolumn{3}{c}{One Factor}& \multicolumn{3}{c}{Two Factors}& \multicolumn{3}{c}{Three Factors (1\textsuperscript{st} Order)}& \multicolumn{3}{c}{Three Factors (2\textsuperscript{nd} Order)}\tabularnewline 
\midrule
$\chi^{2} (\mathit{df})$& 481.87 (35)& & & 239.28 (34)& & & 163.69 (32)& &  & 163.69 (32)& & \tabularnewline
$\chi^2/\mathit{df}$ & 13.77 &&& 7.04 &&& 5.12 &&& 5.12 \tabularnewline
\textsf{CFI} & .73& & & .87& & & .92& & & .92& & \tabularnewline
\textsf{GFI}& 1.00& & & 1.00& & & 1.00& & & 1.00& & \tabularnewline
\textsf{RMSEA} & .19& \multicolumn{2}{c}{[.17, .20]} & .13& \multicolumn{2}{c}{[.11, .14]} & .11& \multicolumn{2}{c}{[.09, .12]} & .10& \multicolumn{2}{c}{[.09, .12]} \tabularnewline
\textsf{SRMR}& .14& & & .09& & & .10& & & .10& & \tabularnewline
Scaled $\chi^{2} (\mathit{df})$& 377.87 (35)& & & 189.71 (34)& & & 131.42 (32)& & & 131.42 (32)& & \tabularnewline
\textsf{CAIC} & 600.572 &&& 361.941 &&& 294.264 &&& 294.264 \tabularnewline
Scaled \textsf{CAIC} & 496.570 &&& 312.366 &&& 261.990 &&& 261.990 \tabularnewline
\bottomrule
\end{tabular}
\end{adjustbox}
\end{table*}

\pathPlotCFABref

\residualsB

\loadingsBredux

\discriminantB

\begin{table*}[tbp]
\centering
\caption{Comparison of IUIPC-10 and IUIPC-8 on Validation Sample \textsf{V}%
\explain{.\\The IUIPC-10 models fail the fit tests irrespective of estimator. By Vuong test on the ML estimation and the \textsf{CAIC}, the IUIPC-8 models are better fits than their corresponding IUIPC-10 equivalents. The IUIPC-8 models show very good fit, with the MLM estimation passing even the exact-fit test.}}
\label{tab:respec.V}
\begin{tabular}{llccc}
\toprule
\multirow{2}{*}{Instrument} & \multirow{2}{*}{Respecification} & \multicolumn{3}{c}{Estimator}\\
\cmidrule(lr){3-5}
&& ML & MLM$^\ddagger$ & WLSMVS$^\ddagger$ \\
\midrule
\multirow{4}{*}{IUIPC-10} &            & $\chi^2(32) = 122.015, p < .001$ 
                                       & $\chi^2(32) = 80.69, p < .001$ 
                                       & $\chi^2(16) = 151.118, p < .001$\\ 
 && \textsf{CFI}=$.94$; 
    \textsf{GFI}=$1.00$ &
    \textsf{CFI}=$.95$;
    \textsf{GFI}=$1.00$&
    \textsf{CFI}=$.96$;
    \textsf{GFI}=$.99$\\
 && \textsf{RMSEA}=$.08$ $[.07, .10]$ &
    \textsf{RMSEA}=$.07$ $[.05, .09]$ &
    \textsf{RMSEA}=$.14$ $[.12, .16]$\\
 && \textsf{SRMR}=$.08$; 
    \textsf{CAIC}=254.6 &
    \textsf{SRMR}=$.08$;
    \textsf{CAIC}=213.3&
    \textsf{SRMR}=$.07$;
    \textsf{CAIC}=404.3\\
\cmidrule(lr){3-3}\cmidrule(lr){4-4}\cmidrule(lr){5-5}
                          & & $\uparrow$ & $\uparrow$ & $\uparrow$ \\
                          & \multirow{2}{*}{Trim \textsf{ctrl3} \& \textsf{awa3}} & 
                          $\Delta{\mathsf{CFI}} = 0.05$; $\Delta{\mathsf{CAIC}} = -111.6$  & 
                          \multirow{2}{*}{$\Delta{\mathsf{CFI}} = 0.05$; $\Delta{\mathsf{CAIC}} = -82.76$} &
                          \multirow{2}{*}{$\Delta{\mathsf{CFI}} = 0.04$; $\Delta{\mathsf{CAIC}} = -174.5$}\\
                          & & Vuong LRT $z = -35.146$, $p < .001$\\
                          & & $\downarrow$ & $\downarrow$ & $\downarrow$ \\
\cmidrule(lr){3-3}\cmidrule(lr){4-4}\cmidrule(lr){5-5}
IUIPC-8                  & & $\chi^2(17) = 34.532, p = .007$ 
                                       & $\chi^2(17) = 22.04, p = .183$ 
                                       & $\chi^2(10) = 24.844, p = .005$\\
 && \textsf{CFI}=$.99$;
    \textsf{GFI}=$1.00$&
    \textsf{CFI}=$.99$;
    \textsf{GFI}=$1.00$ &
    \textsf{CFI}=$.99$;
    \textsf{GFI}=$1.00$\\
 && \textsf{RMSEA}=$.05$ $[.03, .07]$ &
    \textsf{RMSEA}=$.03$ $[.00, .07]$ &
    \textsf{RMSEA}=$.06$ $[.03, .09]$\\
 && \textsf{SRMR}=$.03$; 
    \textsf{CAIC}=143 &
    \textsf{SRMR}=$.03$;
    \textsf{CAIC}=130.6&
    \textsf{SRMR}=$.03$;
    \textsf{CAIC}=229.8\\
\bottomrule
\end{tabular}
\\\emph{Note:} $^\ddagger$ Robust estimation with scaled test statistic.
\end{table*}

\end{DocumentVersionConference}

\begin{DocumentVersionConference}
\section{Factor Analysis}
\label{sec:fa}

\end{DocumentVersionConference}

\end{appendix}

\processifversion{Supplements}{\clearpage\newpage\onecolumn\section*{\LARGE{Supplementary Materials} for \\\Large{\emph{Revisiting the Factor Structure of IUIPC-10}}}}
\begin{Supplements}
\section*{Correlation Matrices}
For the reproducibility of the maximum likelihood estimation, Table~\ref{tab:inputCorSD} contains the correlations and standard deviations (SDs) of all samples used in the study, along with their sample sizes. The OSF supplementary materials contain more precise covariance matrices of all samples. 

ML confirmatory factor analyses (CFAs) can operate on a sample's covariance matrix as long as the sample size $N$ is known, MLM if the \textsf{NACOV} matrix is additionally available.
The covariance matrices, in turn, can be reconstructed from the inputs' correlations and SDs, modulo rounding error:
\[ \mathsf{Cov} := (\mathsf{SD} \cdot \mathsf{t}(\mathsf{SD})) \cdot \mathsf{Cor} \]

Convenience functions for that purpose are readily available, for instance, the functions \textsf{cor2cov}()
present in both the \textsf{R} packages \textsf{lavaan} and \textsf{psych}.

\inputCorSD

\section*{Exploratory Factor Analyses}
Figure~\ref{fig:screePlotB} shows the corresponding Scree plot to illustrate the different
factor solutions possible.
\screePlotB

Because of the factor correlations we chose an oblique transformation, namely \textsf{Oblimin}, for the exploratory factor analyses. We display the loadings for Samples \textsf{A} and \textsf{B} in Table~\ref{tab:efaloadingspoly}.

\efaloadingspoly

In Figure~\ref{fig:biplotB} we display the biplot of the exploratory factor analysis on Sample~\textsf{B}. Therein, one can clearly see the misloading of \textsf{ctrl3} and \textsf{awa3}.

\biplotB

\clearpage
\newpage

\section*{CFA of IUIPC-10}

We display the overview of the different estimations of IUIPC-10 on Sample~\textsf{B} in Table~\ref{tab:comparison.estimators}. It includes the factor loadings as well as scaled and unscaled fit statistics.

\begin{table*}[tbp]
\centering
\caption{Overview of IUIPC-10 CFAs with different estimators on Sample~\textsf{B}}
\label{tab:comparison.estimators}
\begin{adjustbox}{max width=\textwidth}
\begin{tabular}{@{}rrrrrrrrrrrrr@{}}
\toprule
& \multicolumn{4}{c}{ML}& \multicolumn{4}{c}{MLM}& \multicolumn{4}{c}{WLSMVS}\tabularnewline
\cmidrule(lr){2-5}\cmidrule(lr){6-9}\cmidrule(lr){10-13}
& \multicolumn{1}{c}{Estimate}& \multicolumn{1}{c}{\vari{SE}}& \multicolumn{1}{c}{$z$}& \multicolumn{1}{c}{$p$}& \multicolumn{1}{c}{Estimate}& \multicolumn{1}{c}{\vari{SE}}& \multicolumn{1}{c}{$z$}& \multicolumn{1}{c}{$p$}& \multicolumn{1}{c}{Estimate}& \multicolumn{1}{c}{\vari{SE}}& \multicolumn{1}{c}{$z$}& \multicolumn{1}{c}{$p$}\tabularnewline
\midrule
& \multicolumn{9}{c}{\underline{Factor Loadings}}\tabularnewline \multicolumn{1}{l}{\underline{ctrl}}\tabularnewline
\textsf{ctrl1} & 1.00$^+$& & & & 1.00$^+$& & & & 1.00$^+$& & & \tabularnewline
\textsf{ctrl2} & 1.00& 0.12& 8.50& $< .001$& 1.00& 0.11& 8.76& $< .001$& 1.00& 0.09& 11.50& $< .001$\tabularnewline
\textsf{ctrl3} & 0.59& 0.09& 6.33& $< .001$& 0.59& 0.11& 5.36& $< .001$& 0.95& 0.08& 11.59& $< .001$\tabularnewline
 \multicolumn{1}{l}{\underline{aware}}\tabularnewline
\textsf{awa1} & 1.00$^+$& & & & 1.00$^+$& & & & 1.00$^+$& & & \tabularnewline
\textsf{awa2} & 1.13\phantom{$^+$}& 0.11& 10.26& $< .001$& 1.13\phantom{$^+$}& 0.13& 8.53& $< .001$& 1.07\phantom{$^+$}& 0.10& 11.08& $< .001$\tabularnewline
\textsf{awa3} & 0.90\phantom{$^+$}& 0.12& 7.39& $< .001$& 0.90\phantom{$^+$}& 0.14& 6.64& $< .001$& 0.92\phantom{$^+$}& 0.08& 11.55& $< .001$\tabularnewline
 \multicolumn{1}{l}{\underline{collect}}\tabularnewline
\textsf{coll1} & 1.00$^+$& & & & 1.00$^+$& & & & 1.00$^+$& & & \tabularnewline
\textsf{coll2} & 0.76\phantom{$^+$}& 0.05& 16.50& $< .001$& 0.76\phantom{$^+$}& 0.05& 14.86& $< .001$& 0.96\phantom{$^+$}& 0.03& 37.00& $< .001$\tabularnewline
\textsf{coll3} & 1.06\phantom{$^+$}& 0.05& 21.75& $< .001$& 1.06\phantom{$^+$}& 0.04& 23.63& $< .001$& 1.14\phantom{$^+$}& 0.02& 47.74& $< .001$\tabularnewline
\textsf{coll4} & 0.95\phantom{$^+$}& 0.05& 19.64& $< .001$& 0.95\phantom{$^+$}& 0.05& 18.14& $< .001$& 1.06\phantom{$^+$}& 0.02& 43.02& $< .001$\tabularnewline
 \multicolumn{1}{l}{\underline{iuipc}}\tabularnewline
\textsf{collect} & 0.41\phantom{$^+$}& 0.08& 5.14& $< .001$& 0.41\phantom{$^+$}& 0.08& 5.42& $< .001$& 0.46\phantom{$^+$}& 0.04& 10.78& $< .001$\tabularnewline
\textsf{ctrl} & 0.42\phantom{$^+$}& 0.07& 5.88& $< .001$& 0.42\phantom{$^+$}& 0.07& 6.05& $< .001$& 0.52\phantom{$^+$}& 0.05& 10.35& $< .001$\tabularnewline
\textsf{aware} & 0.36\phantom{$^+$}& 0.06& 6.54& $< .001$& 0.36\phantom{$^+$}& 0.06& 6.40& $< .001$& 0.73\phantom{$^+$}& 0.07& 10.39& $< .001$\tabularnewline
& \multicolumn{9}{c}{\underline{Intercepts}}\tabularnewline
\textsf{ctrl1} & 5.97\phantom{$^+$}& 0.05& 123.59& $< .001$& 5.97\phantom{$^+$}& 0.05& 123.59& $< .001$& 0.00$^+$& & & \tabularnewline
\textsf{ctrl2} & 5.96\phantom{$^+$}& 0.05& 122.96& $< .001$& 5.96\phantom{$^+$}& 0.05& 122.96& $< .001$& 0.00$^+$& & & \tabularnewline
\textsf{ctrl3} & 5.96\phantom{$^+$}& 0.05& 115.20& $< .001$& 5.96\phantom{$^+$}& 0.05& 115.20& $< .001$& 0.00$^+$& & & \tabularnewline
\textsf{awa1} & 6.68\phantom{$^+$}& 0.03& 234.36& $< .001$& 6.68\phantom{$^+$}& 0.03& 234.36& $< .001$& 0.00$^+$& & & \tabularnewline
\textsf{awa2} & 6.62\phantom{$^+$}& 0.03& 226.48& $< .001$& 6.62\phantom{$^+$}& 0.03& 226.48& $< .001$& 0.00$^+$& & & \tabularnewline
\textsf{awa3} & 6.25\phantom{$^+$}& 0.04& 146.99& $< .001$& 6.25\phantom{$^+$}& 0.04& 146.99& $< .001$& 0.00$^+$& & & \tabularnewline
\textsf{coll1} & 5.26\phantom{$^+$}& 0.07& 74.70& $< .001$& 5.26\phantom{$^+$}& 0.07& 74.70& $< .001$& 0.00$^+$& & & \tabularnewline
\textsf{coll2} & 5.76\phantom{$^+$}& 0.06& 100.28& $< .001$& 5.76\phantom{$^+$}& 0.06& 100.28& $< .001$& 0.00$^+$& & & \tabularnewline
\textsf{coll3} & 5.69\phantom{$^+$}& 0.06& 88.30& $< .001$& 5.69\phantom{$^+$}& 0.06& 88.30& $< .001$& 0.00$^+$& & & \tabularnewline
\textsf{coll4} & 5.73\phantom{$^+$}& 0.06& 90.65& $< .001$& 5.73\phantom{$^+$}& 0.06& 90.65& $< .001$& 0.00$^+$& & & \tabularnewline
& \multicolumn{9}{c}{\underline{Fit Indices}}\tabularnewline
$\chi^{2}(\vari{df})$& 163.69 (32)& & & $< .001$& 163.69& & & $< .001$ & 213.83& & & $< .001$ \tabularnewline
\textsf{CFI} & .92& & & & .92& & & & .98& & & \tabularnewline
\textsf{GFI} & 1.00& & & & 1.00& & & & .99& & & \tabularnewline
\textsf{RMSEA}& .11&  \multicolumn{2}{c}{$[.09, .12]$}  & $< .001$ & .11&  \multicolumn{2}{c}{$[.09, .12]$}  & $< .001$ & .12&  \multicolumn{2}{c}{$[.11, .14]$}  & $< .001$\tabularnewline
\textsf{SRMR}& .10& & & & .10& & & & .10& & & \tabularnewline
Scaled $\chi^{2} (\vari{df})$& & & & & 131.42 (32)& & & $< .001$& 181.15 (14.95)& & & $< .001$\tabularnewline
\textsf{CAIC} & 294.264 & & & & 294.264 & & & & 447.278 & & & \tabularnewline
Scaled \textsf{CAIC} &  & & & & 261.990 & & & & 414.598 & & & \tabularnewline
\bottomrule
\end{tabular}
\end{adjustbox}
\\\emph{Note:} $N_{\mathsf{B}}^\prime = 371$; $^+$Fixed parameter; 
the ML CFA model without outlier correction with $N_{\mathsf{B}} = 380$ yielded $\chi^2(32) = 202.789, p < .001$; $\mathsf{CFI} = .90$; $\mathsf{GFI} = 1.00$; $\mathsf{RMSEA} = .12 [.10, .14]$; $\mathsf{SRMR} = .11$; $\mathsf{CAIC} = 333.758$.
\end{table*}

In the following we also offer the residuals, that is, correlation and (standardized, if possible) covariance residuals for all CFAs computed. Table~\ref{tab:residualsBasIUIPC} includes the residuals of the replication of the approach of Malhotra et al.~\cite{malhotra2004internet}, that is, computing an ML-estimated CFA on ordinal, non-normal data without outlier treatment.

Table~\ref{tab:residualsBml} contains the residuals of the ML-estimation on our outlier-treated Sample~\textsf{B}, corresponding to the left column on Table~\ref{tab:comparison.estimators}. We can see that as a tendency the residuals of the replication in Table~\ref{tab:residualsBasIUIPC} and the outlier-treated ML-estimation in Table~\ref{tab:residualsBml} are largely aligned.

able~\ref{tab:residualsB} displays the residuals of the MLM estimation on Sample~\textsf{B}, which is the touch stone of the paper (center column of Table~\ref{tab:comparison.estimators}. 

Finally, Table~\ref{tab:residualsBwls} shows the residuals of the WLSMVS estimation of IUIPC-10. Note that \textsf{lavaan} cannot compute standardized covariance residuals for a WLSMVS estimation and that we have included the raw covariance residuals instead.

\residualsBasIUIPC

\residualsBml

\residualsB

\residualsBwls

\section*{Respecification: IUIPC-8}

We computed the models for the respecification IUIPC-8 on Sample~\textsf{B}. 


Again, we include the residuals for the corresponding models. Table~\ref{tab:residualsBreduxml} displays the residuals for the ML estimation on IUIPC-8. Already in this estimation, we observe a far lower number of statistically significant covariance residuals than in the IUIPC-10 model.

The MLM estimation with robust standard errors is shown in Table~\ref{tab:residualsBredux}. Here we have two statistically significant residuals with concerning correlations left. 
We inspected the questionnaire itself for evidence of method correlation. We found specifically that \textsf{coll1} and \textsf{coll3} follow a parallel construction with the stem ``It bothers me\dots,'' which could lead to correlated error variances on substantive grounds. Testing the this variant with an LRT difference test on nested models, we found that the difference is statistically significant in the ML estimation, but not under Satorra-Bentler correction, $\chi^2(1) = 3.975, p = .046$ and $\chi^2(1) = 2.548, p = .110$, respectively. The difference between the variants is even less pronounced under WLSMVS estimation. Hence, we choose to keep the model without correlated errors.

Table~\ref{tab:residualsBreduxwls}, then, shows the residuals for the WLSMVS estimated model on Sample~\textsf{B}.
We observe that the WLSMVS estimation offers a perfect fit, passing even the exact-fit hypothesis at a sample size of $N^\prime_{\mathsf{B}} = 391$.

\pathPlotCFABreduxref

\residualsBreduxml

\residualsBredux

\residualsBreduxwls

\discriminantBredux

\clearpage
\newpage

\section*{Validation}
The validation CFAs were computed on Sample \textsf{V}. 
\subsection*{IUIPC-10 Validation}
First, we consider the performance of IUIPC-10 on this independent sample.

We also provide the comparison with the standardized solution for the MLM estimation in Table~\ref{tab:loadingsV}.
Again, we observe that the variance explained by the indicators \textsf{ctrl3} and \textsf{awa3} is intolerably low, with $R^2 = .22$ and $R^2 = .15$ respectively. Hence, the validation CFA of IUIPC-10 indicates the same convergent validity problem seen in the main analysis on base Sample~\textsf{B}.

\loadingsV

Considering the residuals on Sample~\textsf{V}, we find the same pattern as in the main analysis in the Tables~\ref{tab:residualsVml} (ML estimation) and~\ref{tab:residualsV} (MLM estimation). The WLSMVS estimation in Table~\ref{tab:residualsVwls} yield related correlation patterns, perhaps not quite as pronounced.

\residualsVml

\residualsV

\residualsVwls

\discriminantV

\subsection*{IUIPC-8 Validation}

Considering the standardized loadings in Table~\ref{tab:loadingsVredux}, we find that the indicator variables explain more consistently more than $50\%$ of the variance. Indicator \textsf{coll2} is an exception to note.

\loadingsVredux

The residuals displayed in Tables~\ref{tab:residualsVreduxml} (ML estimation), \ref{tab:residualsVredux} (MLM estimation), and~\ref{tab:residualsVreduxwls} are all in good shape, indicating a good local fit throughout. 

\residualsVreduxml

\residualsVredux

\residualsVreduxwls

\discriminantVredux

\clearpage
\newpage

\section*{Used Software}
The statistics were computed in \textsf{R} (v. 3.6.2) using the package \textsf{lavaan} (v. 0.6-5) for the confirmatory factor analyses.
We included a truncated version of the \textsf{R} \textsf{sessionInfo()} below, including the attached packages and their version numbers.

\begin{verbatim}
R version 3.6.2 (2019-12-12)
Platform: x86_64-apple-darwin15.6.0 (64-bit)
Running under: macOS High Sierra 10.13.6

Matrix products: default

Random number generation:
 RNG:     L'Ecuyer-CMRG 
 Normal:  Inversion 
 Sample:  Rejection 
 
locale:
 en_GB.UTF-8

attached base packages:
 stats     graphics  grDevices utils     datasets  methods   base     

other attached packages:
 semTable_1.7          semPower_1.0.0        pastecs_1.3.21        matrixcalc_1.0-3      
 GPArotation_2014.11-1 semTools_0.5-2        semPlot_1.1.2         lavaan_0.6-5
 dplyr_0.8.4           weights_1.0.1         mice_3.7.0            gdata_2.18.0
 Hmisc_4.3-1           ggplot2_3.2.1         Formula_1.2-3         survival_3.1-8
 lattice_0.20-38       xtable_1.8-4          psych_1.9.12.31 
\end{verbatim}

\end{Supplements}

\clearpage
\newpage
\listoftodos

\end{document}